\documentclass[epj]{svjour}
\usepackage{blindtext}
\usepackage{amsfonts}
\usepackage{tensor}
\usepackage{graphicx}
\usepackage{pict2e}
\usepackage{array}

\usepackage{makecell}
\usepackage{longtable}
\usepackage{float,placeins}
\usepackage{url}

\usepackage{amssymb,amsmath,environ}

\usepackage{color}


\newcommand{\bse}{\begin{subequations}}
	\newcommand{\ese}{\end{subequations}}
\newcommand{\be}{\begin{equation}}
\newcommand{\ee}{\end{equation}}

\NewEnviron{eqsplit}{%
	\begin{equation}\begin{split}
	\BODY
	\end{split}\end{equation}
}

\newcommand{\bea}{\begin{eqnarray}}
\newcommand{\eea}{\end{eqnarray}}
\newcommand{\ba}{\begin{array}}
	\newcommand{\ea}{\end{array}}

\newcommand{\C}{\mathcal{C}}
\newcommand{\A}{\mathcal{A}}
\newcommand{\K}{\mathcal{K}}

\newcommand{\D}{\mathcal{D}}

\newcommand{\la}{\langle}
\newcommand{\ra}{\rangle}
\newcommand{\ve}{\epsilon}

\begin{document}
\title{A Fourier-Cumulant Analysis for Multiharmonic Flow Fluctuation}
\subtitle{by employing a multidimensional generating function approach }
\author{Seyed Farid Taghavi% etc
%\thanks{\emph{Present address:} Physik Department E62, Technische Universit\"{a}t M\"{u}nchen, James Franck Str.~1, 85748 Garching, Germany}%
}
\institute{Physik Department E62, Technische Universit\"{a}t M\"{u}nchen, James Franck Str.~1, 85748 Garching, Germany}
\date{Received: date / Revised version: date}
\abstract{The Fourier analysis of the final particle distribution followed by cumulant study of the Fourier coefficient event-by-event fluctuation is one of the main approaches for testing the collective evolution in the heavy-ion collision.  Using a multidimensional generating function, we propose a method to extract any possible cumulant of multiharmonic flow fluctuations and classify them in terms of the order of cumulants and harmonics involved in them. In particular, we show that there are 33 distinct cumulants with order $2,3,4,5$ and harmonics $2,3,4,5$. We compute the normalized version of these cumulants from hydrodynamic simulation for Pb--Pb collisions based on {\tt T$_\text{R}$ENTo}+{\tt VISH2+1}+{\tt UrQMD}. We compare the simulation with those normalized cumulants that the LHC has measured and predict the unmeasured ones.  Comparing the initial and final state fluctuation normalized cumulants, we compute the linear and nonlinear hydrodynamic response couplings.  We finally introduce the genuine three-particle correlation function containing information of all third-order cumulants.
\PACS{
      {25.75.Ld}{Collective flow, relativistic collisions}   \and
      {25.75.Gz}{Particle correlations, relativistic collisions}
     } % end of PACS codes
} %end of abstract
\maketitle
\section{Introduction}

It is widely accepted in the heavy-ion community that QCD matter in the deconfined phase, quark-gluon plasma, is produced in ultrarelativistic heavy-ion collisions. The collective models equipped with QCD as an underlying physics can successfully explain the majority of experimental observations. However, measuring the accurate transport coefficients such as shear and bulk viscosity over entropy density and finding QCD critical point are still one of the main challenges in the heavy-ion community \cite{Heinz:2013th,Braun-Munzinger:2015hba,Busza:2018rrf}. Moreover,  observing similar collectivity signals in small system collisions such as p--p or p--Pb \cite{Abelev:2014mda,Khachatryan:2010gv,Khachatryan:2016txc,Aaboud:2016yar,Aaboud:2017blb,Adamczyk:2015xjc,PHENIX:2018lia} raise debates about relation between the observations and the collective processes. These indicate that the ultimate goal has not been achieved, and we need to deepen our understanding of heavy-ion collision physics. To this end, introducing new observables to probe various aspects of the full image is crucial. The present manuscript tries to introduce a systematic procedure to generate observables related to the anisotropic flow and classify them.  Some of these observables are studied before, and some others are introduced for the first time.

The anisotropic particle emission in the azimuthal direction \cite{Ollitrault:1992bk,Barrette:1994xr,Barrette:1996rs} is one of the most important evidences of collectivity in ultrarelativistic heavy-ion collisions. The final particle momentum distribution in the azimuthal direction can be expanded in a Fourier series,
\bea\label{Fourier}
f(\varphi)=\frac{1}{2\pi}\left[1+2 \sum_{n=1}^\infty v_{n} \cos n(\varphi-\psi_n) \right],
\eea  
with Fourier coefficients $v_ne^{in\psi_n}$, called flow harmonics. The imprints of different stages of a heavy-ion collision, pre-equilibrium, initial state, collective evolution, and freezeout are cumulatively encoded in the flow harmonics. On account of the quantum mechanical nature of partons inside the nucleons, the initial geometry has a complicated structure and more importantly fluctuates from one event to the other (event-by-event fluctuation) leading to nonvanishing and fluctuating $v_n$ and $\psi_n$ for any $n$ \cite{Abelev:2014mda,Aad:2014vba,Chatrchyan:2013nka,Adamczyk:2015obl,Adam:2016izf,Sirunyan:2017pan,Sirunyan:2017fts,Acharya:2019vdf,Aaboud:2018syf,Aaboud:2019sma,Alver:2010gr}. Practically, we measure the statistical properties of the $v_ne^{in\psi_n}$ fluctuation in the experiment. Several attempts focus on the moments  \cite{Jia:2017hbm,Teaney:2013dta,Jia:2012ma,Jia:2012ju,Bhalerao:2011ry,Jia:2012sa,Aad:2014fla,Bilandzic:2020csw,Acharya:2017zfg,Acharya:2020taj} or on cumulants of fluctuations to achieve information from flow harmonic fluctuations. Single harmonic cumulants \cite{Borghini:2000sa,Borghini:2001vi}, symmetric cumulants \cite{Bilandzic:2013kga,ALICE:2016kpq,Acharya:2017gsw,STAR:2018fpo}, generalized (or higher-order) symmetric cumulants \cite{Mordasini:2019hut,Acharya:2021afp}, asymmetric cumulant \cite{Jia:2017hbm} are some examples of these attempts.

 It is noteworthy that each cumulant of a probability density function (p.d.f.) contains independent information about the underlying fluctuation. Therefore, a complete study of all possible cumulants up to a given order of cumulant expansion is crucial to gain a comprehensive insight into the fluctuations. In the present work, we start with the standard definition of cumulants based on generating functions for an arbitrary number of variables. We extract and classify all possible cumulants depending on their order and the involved number of harmonics. Considering the space limitation, we focus on cumulants of harmonics $n=2,3,4,5$ up to fifth order, which contains 33 distinct cumulants. Among them, there are already known cumulants together with new cumulants containing symmetry plane correlations. To ease the future application, we have prepared a {\tt Mathematica} \cite{mathematica} package accessible as an ancillary file for the present manuscript or in the GitHub repository \cite{github}. The package returns the cumulants in three different forms: in terms of $v_n$ and $\psi_n$ moment, final particle azimuthal angle correlations, and $Q$-vectors (see Eq.~\eqref{QvecDef}) notations. For cases that the {\tt Mathematica} software is not available, a list of few first multiharmonic cumulants in terms of $v_n$ and $\psi_n$ symbolic moments are tabulated in the appendix~\ref{allCumulApp}.  

Regarding multiharmonic cumulants, other studies have been done previously \cite{Jia:2014jca,DiFrancesco:2016srj} with some similarities and differences compared to the present work. In Ref.~\cite{Jia:2014jca}, the underlying p.d.f. of the fluctuations are classified into three categories, flow-amplitude (only $v_n$ fluctuations), event-plane-correlation  (only $\psi_n$ fluctuations) and mixed-correlations (both $v_n$ and $\psi_n$ fluctuations). We find that, instead of classifying the p.d.f. into three types, considering one general p.d.f. is more suitable way to extract all possible cumulants up to a given degree.
 %while in our study the p.d.f. is considered as a whole. In our realization, there is no event-plane-correlation p.d.f.  Also, in this realization, for some combination of harmonics, $\psi_n$ appears in flow-amplitude cumulants as well \cite{Moravcova:2020wnf} which is not compatible with a p.d.f. as a function of flow-amplitude only (see appendix~\ref{allCumulApp}, and compare cumulants $c_{3,4,6}^{\{0,0\}}\{2,2,2\}$ and $c_{2,3,5}^{\{0,0\}}\{2,2,2\}$ as one particular example). 
The author believes that the cumulants extracted in the present paper are achievable with the method introduced earlier in Ref.~\cite{DiFrancesco:2016srj}. In this reference, however, the explicit calculations are limited to the simple few known observables, and the dependence of the cumulant expressions to the harmonics is not presented. Our explicit calculations show that the cumulants' form is related to the involved harmonics in the cumulant (see also Refs.~\cite{Jia:2014jca,Moravcova:2020wnf}). For instance, we see that the first cumulant involving simultaneous harmonics $n=2,3,5$ appears in the third order, and the next cumulants  appear in the fifth order in three different forms. For the case involving simultaneous harmonics $n=2,4,5$, however, cumulants starts from fifth order with two different forms (see Table~\ref{firstCumulTable}). Besides, our study is equipped with Monte Carlo hydrodynamic simulation as well ({\tt T$_\text{R}$ENTo} \cite{Moreland:2014oya}+{\tt VISH2+1} \cite{Shen:2014vra,Song:2007ux}+{\tt UrQMD} \cite{Bass:1998ca,Bleicher:1999xi}  with parameter calibrated by a Global Bayesian analysis \cite{Bernhard:2019bmu}) to see which unmeasured cumulants would have larger signals in future measurements. Comparing this simulation with future measurements can also validate the heavy-ion collision parameters tuned with Global Bayesian analysis. We compare the simulation with normalized cumulants that are measured at the LHC, which reveals a rather good agreement.

We study the initial state fluctuation cumulants to see how the final state inherits the fluctuations from the initial state. The initial energy density anisotropy translation to the final state momentum anisotropy is formulated by a set of equations so-called hydrodynamic response \cite{Teaney:2010vd}. By comparing the initial and final state fluctuations, we obtain the hydrodynamic response coefficients. Moreover, our systematic cumulant study guide us to extend the two-particle correlation function (2PC) notion  \cite{Khachatryan:2010gv,Aad:2015gqa,Aad:2012gla,Zhu:2013zgd} to multiparticle correlation functions, $q$PC (see the definition in section~\ref{threeParticle}). We specifically study the flow-induced 3PC for harmonics 2 to 5  via our hydrodynamic simulation. This quantity contains information on all third-order cumulants.

This paper is structured as follows: The section \ref{generatingFunction} is dedicated to introduce generating function method for multiharmonic cumulants. In section \ref{package}, the functions available in the {\tt Mathematica} package are introduced. A realistic Monte Carlo study of the first Fourier-cumulant expansion terms is presented in section \ref{MonteCarlo}. The linear and nonlinear hydrodynamic response couplings are extracted by comparing the initial state, and final state fluctuation normalized cumulants in section \ref{phase}. Finally, we study genuine three-particle correlation functions in section \ref{threeParticle}. In the appendix \ref{allCumulApp}, a list of few first cumulants is tabulated. In appendix \ref{appB}, a technical study about the statistical fluctuation of multiharmonic correlations is presented. Detail of some derivations in the text is presented in the appendices  \ref{appBp} and \ref{appC}.

\section{Multiharmonic generating function}\label{generatingFunction}

Despite the simplicity of Eq.~\eqref{Fourier},  experimental measurement of flow harmonics is a challenging task. The number of final produced particles per event in a given collision is not enough to extract statistically accurate flow harmonics values. Many events average resolves the problem of low statistics while leads to the convolution of event-by-event fluctuation of the initial state into the measurement. Strictly speaking, the measurable quantities are the moments or cumulants of the  following gigantic probability density function (p.d.f.)
\bea\label{pdf}
p(v_{n_1,x},v_{n_1,y},\ldots,v_{n_k,x},v_{n_k,y}),
\eea
where we have used the Cartesian coordinate notation for the flow harmonics, $v_{n,x}=v_n\cos n\psi_n$ and $v_{n,y}=v_n\sin n\psi_n$.
Noting the fact that $f(\varphi)$ (Eq.~\eqref{Fourier}) is assigned to each initial state, the p.d.f.~\eqref{pdf} is tightly related to the event-by-event fluctuation of the initial state, which is, accordingly, related to the underlying physics of nucleons/nuclei. Moreover, the map which connects the above p.d.f. to the initial state fluctuation is governed by the collective evolution process. Indeed, there are stochastic processes in the collective evolution that can be convoluted into the above p.d.f. as well. Besides the experiment's statistical uncertainty issues, the reaction plane angle is not experimentally measurable conveniently, which leads to loss of one more degree of freedom in p.d.f.~\eqref{pdf}. In the polar coordinate the p.d.f.~\eqref{pdf} depends on $k$ flow amplitudes $v_{n_i}$s and $k$ symmetry plane angles (or event-planes) $\psi_{n_i}$s. After imposing the randomness of the reaction plane angle, from $k$ variable $\psi_{n_i}$, only $k-1$ independent variables $\psi_{n_i}-\psi_{n_1}$ remain 
\bea\label{pdf2}
p(v_{n_1},v_{n_2},\ldots,v_{n_{k}},\psi_{n_2}-\psi_{n_1},\ldots,\psi_{n_k}-\psi_{n_1}).
\eea
Here, $\psi_{n_1}$ is chosen conventionally to compute the angle differences. One could choose any other combination of two symmetry plane angle differences. The final cumulants are independent of this choice.

In the present section, we introduce the method of generating function to extract all possible cumulants of the p.d.f.~\eqref{pdf2}. To this end, we start with a standard definition of generating function of a 
generic multivariate p.d.f. $g(\vec{x})$ (see for instance \cite{Kubo}),
\bea\label{gf}
G(\vec{k})\equiv \la e^{i\vec{x}\cdot \vec{k}}\ra_g =\int d\vec{x} \,g(\vec{x})\, e^{i\vec{x}\cdot \vec{k}},
\eea
where $\vec{x}$ is a $m$-dimensional random vector. The cumulants of the distribution function $g(\vec{x})$ (shown by $\K_{a_1,\ldots,a_m}$) are the Taylor expansion coefficients of the cumulant generating function $\log G(\vec{k})$,
\begin{equation} \label{nDcumul}
\begin{split}
\log G(\vec{k})=\sum_{a_1,\ldots,a_k}\left( \prod_{i=1}^k\frac{(i k_i)^{a_i}}{a_i!}\right)\K_{a_1,\ldots,a_k}.
\end{split}
\end{equation}

The above generating function returns the cumulants of  p.d.f. \eqref{pdf}, while in heavy-ion physics, we are interested in cumulants of p.d.f. \eqref{pdf2}. To clarify the difference between cumulants of p.d.f.~\eqref{pdf} compared to p.d.f.~\eqref{pdf2}, let us start with the most simple example. Focusing on one-harmonics $v_n e^{in\psi_n}$, the p.d.f. of the flow fluctuation in the Cartesian coordinate has the form $p(v_{n,x},v_{n,y})$, and its cumulants can be obtained via the definition presented in Eq.~\eqref{nDcumul} for $k=2$. For instance, 
\begin{equation} \label{kapa02kapa30}
\begin{split}
\mathcal{K}_{0,2}&=\la v_{n,y}^2\ra-\la v_{n,y}\ra^2,\\
\mathcal{K}_{3,0}&=\la v_{n,x}^3\ra-3\la v_{n,x}\ra\la v_{n,x}\ra^2+2\la v_{n,x}\ra^3,\\
\end{split}
\end{equation}
are the width of the distribution in the $v_{n,y}$ and the skewness in the $v_{n,x}$ direction, respectively. In the experiment, however, we lose one of the degrees of freedom in $p(v_{n,x},v_{n,y})$ due to the randomness of the reaction plane angle.  After imposing event-by-event random reaction plane angle to $p(v_{n,x},v_{n,y})$, we obtain a rotationally symmetric p.d.f. $\tilde{p}(v_{n,x},v_{n,y})$. The information of the latter p.d.f. is encoded in a radial distribution $p(v_n)$. Traditionally, we assume the cumulants of $p(v_n)$ are $c_n\{2k\}$ \cite{Borghini:2001vi} where the two first orders of them are given by
 \begin{equation} \label{cn2k}
 \begin{split}
 c_n\{2\}&= \la v_n^2 \ra,\\
 c_n\{4\}&= \la v_n^4 \ra -2\la v_n^2 \ra^2.
 \end{split}
 \end{equation}
 These quantities, however, are specific combinations of two-dimensional cumulants of $p(v_{n,x},v_{n,y})$ that survived after imposing the randomness of the reaction plane angle. For instance, it turns out \cite{Abbasi:2017ajp,Mehrabpour:2018kjs},
\begin{equation} \label{2Dcumul}
\begin{split}
c_n\{2\}&= \K_{20}+\K_{02},\\
c_n\{4\}&=\K_{40}+2\K_{22}+\K_{04}.
\end{split}
\end{equation}
Here, we have assumed that $\bar{v}_n=\la v_{n,x} \ra=0$ and $\la v_{n,y} \ra=0$ for simplicity. This condition is satisfied when the flow harmonics are sourced purely from fluctuations, namely ellipticity in central Pb-Pb collisions, or triangularity in round nuclei collisions. For nonvanishing $\bar{v}_n$, similar relations with more terms in the right-hand side of Eq.~\eqref{2Dcumul} are obtained.   
In any case, the cumulants $c_n\{2k\}$ are the maximum information we can achieve from the cumulants of the original distribution $p(v_{n,x},v_{n,y})$.

Before proceeding, let us briefly review the generating functional method to find $c_n\{2k\}$ \cite{Giacalone:2016eyu}.  The generating function in Eq.~\eqref{gf} corresponds to the rotationally symmetric  $\tilde{p}(v_{n,x},v_{n,y})$ reads as
\begin{equation} \label{1dcn2kGF}
\begin{split}
G(k_x,k_y)&=\int dv_{n,x}dv_{n,y} \tilde{p}(v_{n,x},v_{n,y}) e^{i v_{n,x}k_x +i v_{n,y}k_y}\\
&=\int v_n dv_n d\psi_n \tilde{p}(v_n) e^{i v_n k \cos\psi_n}\\
&=\int dv_n \left[v_n \tilde{p}(v_n)\right] J_0(k \,v_n)\\
&=\la J_0(k \,v_n) \ra_p,
\end{split}
\end{equation}
where in the above $J_0(k \,v_n)$ is the Bessel function of the first kind and the averaging $\la J_0(k \,v_n) \ra_p$ is performed with respect to $p(v_n)\equiv v_n \tilde{p}(v_n) $. One can expand the logarithm of the generating function in terms of $k$,
\bea\label{1dExpand}
\log \la J_0(k\,v_n)\ra_{\tilde{p}}=\sum_m k^m \, \ell(m)\,c_n\{m\},
\eea
with
\bea\label{coeff1}
\ell(m)= \frac{i^m}{2^{m}(m/2!)^2},
\eea
to find the cumulants $c_n\{2k\}$.

We extend the above procedure into the cases with an arbitrary number of harmonics. To simplify the notation, we rewrite the p.d.f.~\eqref{pdf2} as $p(\vec{v},\vec{\delta\psi})$ where
\bea\label{16}
\vec{v}=(v_{n_1},\ldots,v_{n_k}),\quad \vec{\delta\psi}=(\delta\psi_{n_1},\ldots,\delta\psi_{n_{k-1}}),
\eea
and 
\bea\label{deltaPsi}
\delta\psi_{n_i}=\psi_{n_{i+1}}-\psi_{n_1}.
\eea
Considering the rotational symmetry imposed by the randomness of the reaction plane angle, one can first average the characteristic function over the azimuthal angle to find a symmetric characteristic function. Then the nonvanishing combination of cumulants of $p(v_{n_1,x},v_{n_1,y},\ldots,v_{n_k,x},v_{n_k,y})$ can be obtained directly from the symmetric characteristic function,
\bea\label{newG}
\mathcal{G}(\vec{k},\vec{\delta\phi})=\la \mathcal{J}(\vec{v},\vec{k},\vec{\delta\psi},\vec{\delta\phi}) \ra_{p},
\eea
where
\begin{equation}\label{J1}
\begin{split}
\mathcal{J}&\left(\vec{v},\vec{k},\vec{\delta\psi},\vec{\delta\phi}\right)=\frac{1}{2\pi}\int_{0}^{2\pi} d\psi_{n_1} e^{i\,L},
\end{split}
\end{equation}
and
\begin{equation}\label{J2}
\begin{split}
L&=\,v_{n_1} k_{n_1}\cos n_1\psi_{n_1}\\
&+\sum_{i=2}^k v_{n_i} k_{n_i}\cos \left[n_i\left(\psi_{n_1}+\delta\psi_{n_{i-1}}-\delta\phi_{n_{i-1}}\right)\right].
\end{split}
\end{equation}
For the derivation refer to the appendix \ref{appC}. Having found $\mathcal{G}(\vec{k},\vec{\delta\phi})$, the cumulants of flow fluctuations can be obtained from the following expansion:
\begin{equation}\label{cumulativeFuncI}
\begin{split}
\log\mathcal{G}(\vec{k},\vec{\delta\phi})&=\sum_{\substack{m_1,\ldots,m_k\\\alpha_1,\ldots\alpha_{k-1}}} k_{n_1}^{m_1}\cdots k_{n_k}^{m_k}\,e^{i\alpha_1 \delta\phi_{n_1}}\cdots e^{i\alpha_{k-1} \delta\phi_{n_{k-1}}}\\
& \times \ell(\vec{m},\vec{\alpha})\; c_{n_1,\ldots,n_k}^{\{\alpha_1,\ldots\alpha_{k-1}\}}\{m_1,\ldots,m_k\},
\end{split}
\end{equation}
where the coefficients
\be\label{newNotation}
\begin{aligned}
	&c_{n_1,\ldots,n_k}^{\{\alpha_1,\ldots\alpha_{k-1}\}}\{m_1,\ldots,m_k\}\equiv\\
	&\quad\la v_{n_1}^{m_1}\cdots v_{n_k}^{m_k} \cos(\alpha_1 \delta\psi_{n_1}+\cdots+\alpha_k \delta\psi_{n_k}) \ra_c
\end{aligned}
\ee
are the cumulants ($\delta\psi_{n_i}$ are defined in Eq.~\eqref{deltaPsi}). In the right-hand side of the above equation, we have represented the cumulant with its highest rank moment together with a subscript $c$. Here, $n_i=1,2,\ldots$ stand for the involving harmonics, $m_i=0,1,2,\ldots$ are the power of flow amplitudes and $\alpha_i=0,\pm 1, \pm 2,\ldots$ are the coefficients of the symmetry plane angle differences. The order of the cumulant is given by
\bea
q =m_1+\cdots+m_k.
\eea
By definition, $n_i$s are all distinguished while, without loss of generality, we impose a strictly ascending order convention to them, $n_1<\cdots< n_k$.\footnote{In this manuscript, we use harmonic index $a_i$ for those  repeated in the sequence, while $n_i$ is reserved for strictly ascending indices. } The coefficient $\ell(\vec{m},\vec{\alpha})$ is a numerical factor that does not depend on the moments. We fix the coefficient so that the numerical factor of the highest rank moment in cumulant $c_{n_1,\ldots,n_k}^{\{\alpha_1,\ldots\alpha_{k-1}\}}\{m_1,\ldots,m_k\}$ turns to be equal to unity. As we will see in the next section, some combinations of harmonics $n_i$, $m_i$, and $\alpha_i$ lead to vanishing cumulant. A trivial example is $k=1$ where the Eq.~\eqref{newNotation} reduces to $c_n\{m\}\equiv \la v_n^m \ra_c$. It is known that this cumulant is nonvanishing only for even $m$ (see Eq.~\eqref{cn2k}).

Each distinguished cumulant $c_{n_1,\ldots,n_k}^{\{\alpha_1,\ldots\alpha_{k-1}\}}\{m_1,\ldots,m_k\}$ contains a piece of independent information about the p.d.f.~\eqref{pdf2}. To numerically compute these cumulants for a given p.d.f., we need to find the explicit form of them written in terms of the moments.
Although the cumulants $c_{n_1,\ldots,n_k}^{\{\alpha_1,\ldots\alpha_{k-1}\}}\{m_1,\ldots,m_k\}$ can be found analytically using Eq.~\eqref{cumulativeFuncI}, its computation is cumbersome, except for the case $k=1$. By considering one flow harmonic ($k=1$), the generating function reduces to that mentioned in Eq.~\eqref{1dcn2kGF}. Keeping two flow harmonics, the flow fluctuation distribution contains three degrees of freedom $v_{n_1}$, $v_{n_2}$, and $\delta\psi_{n_1}$.  In this case, the function $\mathcal{J}$ can be written in terms of generalized Bessel function \cite{generalBessel}, and by expansion, one obtains the cumulants $c_{n_1,n_2}^{\{\alpha_{1}\}}\{m_1,m_2\}$. However, extracting them is arithmetically more involved. For more general cases, the complexity increases, forcing us to choose a more practical and efficient way for the computation. 

\section{One package for all cumulants }\label{package}

Using the cumulant generating function in Eq.~\eqref{cumulativeFuncI}, one can find the cumulants $c_{n_1,\ldots,n_k}^{\{\alpha_1,\ldots\alpha_{k-1}\}}\{m_1,\ldots,m_k\}$  written in terms of symbolic moments of variables $v_n$ and $\psi_n$. These results can be immediately used in theoretical studies where $v_n$ and $\psi_n$ of every single event are accurately accessible. In the experiment as well as some simulations, however, the outcomes are particles azimuthal angles. As a result, one needs to invest an extra effort to translate the azimuthal angle of final particles into the averages containing $v_n$ and $\psi_n$. For that, we use the multiparticle techniques to compute the flow harmonic cumulants. We rewrite the moments in terms of $Q$-vectors (see Eq.~\eqref{QvecDef}) which can be calculated from final particle azimuthal angles at every single event and employed in computing the cumulant \cite{Bilandzic:2013kga,Bilandzic:2010jr}. This section presents a practical way to find cumulants and their statistical uncertainties in terms of symbolic moments, correlation of particle azimuthal angles, and $Q$-vectors.

\subsection{Cumulants from generating function}

As it is mentioned in the previous section, the analytical computation of a generic cumulant $c_{n_1,\ldots,n_k}^{\alpha_1,\ldots\alpha_{k-1}}\{m_1,\ldots,m_k\}$ is cumbersome. For that reason, we do it symbolically in {\tt Mathematica}. We encapsulated this computation into different functions (see below) and implemented them  into a {\tt Mathematica} package. One can load the package in a separate {\tt Mathematica} notebook and recall the functions. This package is available as an ancillary file of this manuscript, or in the GitHub repository \cite{github}.

In the package, we first compute the Taylor expansion of the exponential function in Eq.~\eqref{J1} in terms $k_{n_1},k_{n_2},\ldots$ up to order $m_1,m_2,\ldots$. Then we perform the integral in Eq.~\eqref{J1}. After that, we replace the combinations $ v_{n_1}^{w_1}\cdots v_{n_k}^{w_k}e^{\beta_1 \delta\psi_{n_1}}\cdots e^{\beta_k \delta\psi_{n_k}}$ in the expansion with a symbolic variable as a moment. Computing the logarithm of the result, we read the coefficients of the Taylor series for the variables $k_1,k_2,\ldots$, and the Fourier series for the variables $\alpha_1,\alpha_2,\ldots$ (see Eq.~\eqref{cumulativeFuncI}). We extract the cumulants in terms of our symbolic moment variables up to a numerical factor $\ell(\vec{m},\vec{\alpha})$ by comparing the result with Eq.~\eqref{cumulativeFuncI}. Finally, we single out the highest rank moment in the result and fix the coefficient $\ell(\vec{m},\vec{\alpha})$ such that the numerical factor of this moment turns to unity.

We listed the available functions in a short manual at the header of the package file. One of these functions is
\be
\begin{aligned}
	&\verb|c[{| \text{m}_1\verb|,...,|\text{m}_k\verb|},{| \alpha_1\verb|,...,|\alpha_{k-1}\verb|},{| \text{n}_1\verb|,...,|\text{n}_k\verb|},| \text{v}\verb|,| \psi \verb|]|  
\end{aligned}
\ee
that returns the associated cumulants written in terms of moments of variables $v_n$ and $\psi_n$. For instance, 
\be
\begin{aligned}
	&\scriptsize\verb|In[1]:=|\normalsize \verb| c[{4},{},{2},| \text{v}\verb|,| \psi \verb|]|\\
	&\scriptsize\verb|Out[1]= |\normalsize \la v_2^4  \ra-2\la v_2^2\ra^2\\
	&\scriptsize\verb|In[2]:=|\normalsize \verb| c[{2,1},{4},{2,4},| \epsilon\verb|,| \phi \verb|]|\\
	&\scriptsize\verb|Out[2]= |\normalsize   \la  \epsilon_2^2 \epsilon_4 \cos \left(4 \left(\phi _2-\phi _4\right)\right)\ra \\		
	&\scriptsize\verb|In[3]:=|\normalsize\verb|c[{2,2},{0},{2,3},| \text{v}\verb|,| \psi \verb|]|\\
	&\scriptsize\verb|Out[3]= |\normalsize\la v_2^2 v_3^2  \ra-\la v_2^2\ra \la v_3^2\ra.
\end{aligned}
\ee
By using {\tt cMean} instead of {\tt c}, the angle brackets are replaced by {\tt Mean[]}. For instance, {\tt cMean[\{2\},\{\},\{2\},v,$\psi$]} returns {\tt Mean[v[2]$^2$]}.

The above representation of cumulant is useful when flow harmonics are accurately accessible in a single event, typically in hydrodynamic simulations. If the final state is the particle azimuthal angles, one needs to employ the particle correlations,
\begin{equation} \label{QVecsIIpp}
\begin{split}
&\la k \ra_{a_1,\ldots,a_k}\equiv\la e^{i\,a_1 \varphi_{i_1}+\cdots+i \,a_k \varphi_{i_k}} \ra,\\
&=\frac{(M-k)!}{M!} \sum_{i_1\neq \cdots \neq i_k} e^{i\,a_1 \varphi_{i_1}+\cdots+i \,a_k \varphi_{i_k}} ,\quad \sum_i a_i=0.\\
\end{split}
\end{equation}
Here, $\varphi_i$ is the azimuthal angle of the $i$th particle in an event, and $M$ is the multiplicity of the event \cite{Borghini:2000sa,Borghini:2001vi,Bilandzic:2013kga,Bilandzic:2010jr}.
A generic flow harmonic moment $\la V_{a_1,\ldots,a_k} \ra$ with
\begin{equation}\label{23}
\begin{split}
V_{a_1,\ldots,a_k} &\equiv v_{a_1} \cdots v_{a_k}\;\cos(a_1 \psi_{a_1}+\cdots+a_k \psi_{a_k}), 
\end{split}
\end{equation}
 can be written in terms of particle correlations as $\la V_{a_1,\ldots,a_k} \ra=\la \la k \ra_{a_1,\ldots,a_k} \ra$. The function
 \bea\label{24}
 {\tt cCorr[\{m_1,\ldots,m_k\},\{\alpha_1,\ldots,\alpha_{k-1}\},\{n_1,\ldots,n_k  \}]}
 \eea
returns the cumulant written in terms of correlations. For instance,
\be
\begin{aligned}
&\scriptsize\verb|In[1]:= |\normalsize {\tt cCorr[\{2,2\},\{0\},\{2,4\}] }\\
&\scriptsize\verb|Out[1]= |\normalsize \la \la 4\ra_{-4,-2,2,4}\ra-\la \la 2\ra_{-2,2}\ra \la \la 2\ra_{-4,4}\ra.
\end{aligned}
\ee
For external applications, it is easier if the output can be simply copied to the codes outside of {\tt Mathematica}. This can be obtained via 
\be\label{32}
\begin{aligned}
	\verb|cCorr[{| \text{m}_1\verb|,...,|\text{m}_k\verb|},{| \alpha_1&\verb|,...,|\alpha_{k-1}\verb|}|\\
	&\verb|,{| \text{n}_1\verb|,...,|\text{n}_k\verb|},| \text{corr} \verb|]|.  
\end{aligned}
\ee
Using function $\verb|CForm|$  ($\verb|FortranForm|$) in {\tt Mathematica}, one immediately convert the output of {\tt cCorr} to an expression applicable in a $\verb|C++|$ ($\verb|Fortran|$) code. For instance,
\be
\begin{aligned}
	&\scriptsize\verb|In[1]:=|\normalsize \verb| cCorr[{2,2},{0},{2,4}},| \text{corr} \verb|]|\\
	&\scriptsize\verb|Out[1]= |\normalsize \text{corr}[-4,-2,2,4]-\text{corr}[-2,2] \text{corr}[-4,4],
\end{aligned}
\ee
where,
\be\label{QVecsIIp}
\begin{aligned}
&\text{corr}[a_1,\ldots,a_k]\equiv \la\la k \ra_{a_1,\ldots,a_k}\ra.
\end{aligned}
\ee

From the practical point of view, computing $\la\la k \ra_{a_1,\ldots,a_k}\ra$ contains several nested loops depending on the value of $k$ which are computationally expensive. A technique is introduced in Refs.~\cite{Bilandzic:2013kga,Bilandzic:2010jr} where the moments $ \la k \ra_{a_1,\ldots,a_k}$ are obtained in terms of $Q$-vectors,\footnote{We ignored the particle weights $w_k$ in the $Q$-vector definition for simplicity.}
\bea\label{QvecDef}
Q_{n}=\sum_{k=1}^{M} e^{in\varphi_k}.
\eea
Using this technique, only one loop over particles in an event is needed to compute the correlations.

%The function \eqref{32} together with the techniques explained in Refs.~\cite{Bilandzic:2010jr,Bilandzic:2013kga}, one can find the cumulants directly from particle azimuthal angles. 

For those who want to analyze anisotropic flow inside {\tt Mathematica}, we implemented the recursive algorithm mentioned in Ref.~\cite{Bilandzic:2013kga} into the package to find $ \la k \ra_{a_1,\ldots,a_k}$ in terms of $Q$-vectors. Substituting $v_n$ and $\psi_n$ symbolic averages with moments written in terms of $Q$-vectors, we find cumulants in terms $Q$-vectors (so-called $Q$-cumulants),
\be\label{cQvec}
\begin{aligned}
	\verb|cQvec[{| \text{m}_1\verb|,...,|\text{m}_k\verb|},{| \alpha_1&\verb|,...,|\alpha_{k-1}\verb|}|\\
	&\verb|,{| \text{n}_1\verb|,...,|\text{n}_k\verb|},M,Q]|,  
\end{aligned}
\ee
where {\tt M} and {\tt Q} are the multiplicity and $Q$-vector symbols, respectively. For example,
\be
\begin{aligned}\label{28}
	&\scriptsize\verb|In[1]:=|\normalsize \verb| cQvec[{2},{},{2},M,Q]|\\
	&\scriptsize\verb|Out[1]= | {\tt Mean\Big[}\frac{{\tt M-Q[-2]Q[2]}}{{\tt M(1-M)}}{\tt \Big]},
\end{aligned}
\ee
which is nothing but
\bea\label{c22}
c_2\{2\}=\left\langle \frac{M-Q_2^* Q_2}{M(1-M)} \right\rangle,
\eea
as it is already obtained in Refs.~\cite{Bilandzic:2013kga,Bilandzic:2010jr}. Having defined {\tt M}, {\tt Q[-2]}, and {\tt Q[2]} as a list in {\tt Mathematica} containing multiplicity, $Q_2$, and $Q_2^*$ of many events, the function in Eq.~\eqref{28} returns the numerical value of the cumulant $c_2\{2\}$. The function works accordingly for all other cumulants discuss in the present work.

 The other function we advertise here is 
\be
\begin{aligned}
	& \verb|cTable[{| \text{n}_1\verb|,...,|\text{n}_k\verb|},minOrd,maxOrd,| \text{v}\verb|,| \psi \verb|]|
\end{aligned}
\ee
that returns a table of all nonvanishing cumulants for flow harmonics $n_1,\ldots,n_k$ and orders between $\verb|minOrd|$ and $\verb|maxOrd|$.   It is worth mentioning that considering more flow harmonics and computing higher-order cumulants makes computations more time-consuming.
For future studies and cases that {\tt Mathematica} software is not available, we tabulate the cumulants of flow harmonics $n=2,\ldots,6$ up to four harmonics combinations in the appendix \ref{allCumulApp}. 
In Tables~\ref{one-harmonic} to \ref{four-harmonics} in the appendix \ref{allCumulApp}, we present the one- and two-harmonics cumulants up to eight, and three- and four-harmonics up to six orders, concerning the length of the final expressions.   The interested reader can extract the cases which are not appeared in the tables by using the {\tt Mathematica} package.

Apart from  $c_n\{2m\}$ cumulant which is obviously a special case of $c_{n_1,\ldots,n_k}^{\{\alpha_1,\ldots\alpha_{k-1}\}}\{m_1,\ldots,m_k\}$, several multiharmonic cumulants are known before, such as symmetric cumulants \cite{Bilandzic:2013kga},
\bea
\text{SC}(k,l)\equiv c_{k,l}^{\{0\}}\{2,2\}=\la v_k^2 v_l^2 \ra-\la v_k^2\ra \la v_l^2 \ra,
\eea
generalized (or higher-order) symmetric cumulants for some combination of harmonics \cite{Mordasini:2019hut} (the actual form of the cumulant in terms of moments can be found in the appendix~\ref{allCumulApp}),\footnote{The equivalence of generalized symmetric cumulants and $c_{k,l,m}^{\{0,0\}}\{2,2,2\}$ is not exact. As a different approach in Ref.~\cite{Mordasini:2019hut}, the cumulants of flow amplitude squared fluctuations are studied, which for most of the cases it is equivalent to what is presented here.} $\text{SC}(k,l,m)\equiv c_{k,l,m}^{\{0,0\}}\{2,2,2\}$, 
and asymmetric cumulant \cite{Aaboud:2018syf,Aaboud:2019sma,Jia:2017hbm}, 
\begin{equation} \label{newcumulants1}
\begin{split}
\text{ac}_n\{3\}&\equiv c_{n,2n}^{\{2n\}}\{2,1\} = \la v_n^2 v_{2n} \cos 2n \left(\psi_{n}-\psi_{2n}\right) \ra,
\end{split}
\end{equation}
as special cases of what we present here. There are other cumulants, however, missed in other studies so far. For instance, 
\begin{equation} \label{newcumulants3}
\begin{split}
c_{2,4}^{\{4\}}\{2,3\}&= \la  v_2^2 v_4^3 \cos \left(4 \left(\psi _2-\psi _4\right)\right)\ra \\
& -2 \la 
v_4^2\ra   \la  v_2^2 v_4 \cos \left(4 \left(\psi _2-\psi
_4\right)\right)\ra
\end{split}
\end{equation}
is an example of a cumulant which have not been studied before. 

\subsection{Statistical uncertainty of cumulants }\label{statisticalSec}

Having computed the flow harmonic cumulants, we discuss the statistical fluctuation of the cumulants. In case the flow harmonics $v_n e^{in\psi_n}$ are accurately accessible in  a single event, one can follow a standard procedure (see for instance Ref.~\cite{kendallBook}) to find the covariance matrix of the moments $\la V_{a_1,\ldots,a_k}\ra$ (see Eq.~\eqref{23}). The covariance matrix of two generic moments are simply given by 
\bea
\text{cov}(V_i,V_j)=\frac{1}{N}\left(\la V_iV_j \ra - \la V_i\ra \la V_j \ra\right),
\eea
where $N$ is the number of events and the subscript $i$ stands for a generic collective index $a_1,\ldots,a_k$.
The variance of any function of moments, $f(\la V_1 \ra,\la V_2 \ra,\ldots)$ is obtained by \cite{kendallBook},
\bea\label{36}
\sigma_f^2 =\sum_{i,j}\frac{\partial f}{\partial \la V_i \ra}\frac{\partial f}{\partial \la V_j \ra} \text{cov}( V_i , V_j).
\eea
For instance, the variance of $c_{2}\{4\}=\la v_2^4 \ra-2\la v_2^2 \ra^2$ is given by
\begin{equation}\label{UncertaintyExample}
\begin{split}
\sigma_{c_{2}\{4\}}^2 &=\frac{1}{N}\Big[\text{cov}( v_2^4 , v_2^4 )-8\la v_2^2\ra\text{cov}( v_2^4 , v_2^2 )\\
&+16 \la v_2^2\ra^2 \text{cov}(v_2^2 , v_2^2 )\Big],\\
&=\frac{1}{N}\Big[\la v_2^8\ra-8\la v_2^2 \ra \la v_2^6\ra-\la v_2^4\ra^2+24\la v_2^2\ra^2 \la v_2^4\ra\\
&-16\la v_2^2\ra^4\Big].
\end{split}
\end{equation}

If the azimuthal angle of particles in the final state are available, the flow harmonic $V_{a_1,\ldots,a_k}$ should be replaced by particle correlations $\text{Re}\la k \ra_{a_1,\ldots,a_k}$. Although $\text{Re}\la k \ra_{a_1,\ldots,a_k}$ is not an accurate estimation for $V_{a_1,\ldots,a_k}$, in the ultimate many events average it approaches to an accurate estimation. By replacing $V_{a_1,\ldots,a_k}$ with $(\la k \ra_{a_1,\ldots,a_k}+\la k \ra_{-a_1,\ldots,-a_k} )/2$ in Eq.~\eqref{36}, we obtain the variance of the function of correlations in terms of correlations. For instance, in this notation, we have $c_2\{4\}=\la \la 4\ra_{-2,-2,2,2} \ra-2 \la \la 2\ra_{-2,2}\ra$, and accordingly, the example in Eq.~\eqref{UncertaintyExample} turns to the following form:
\begin{equation}\label{41}
\begin{split}
\sigma_{c_{2}\{4\}}^2 &=\frac{1}{N}\Big[\la \la 4\ra^2_{-2,-2,2,2}\ra-8\la \la 2\ra_{-2,2} \ra \la \la 2 \ra_{-2,2} \la 4 \ra_{-2,-2,2,2}\ra\\
&-\la \la 4 \ra_{-2,-2,2,2}\ra^2+16\la \la 2\ra_{-2,2}^2\ra\la \la 2\ra_{-2,2}\ra^2\\
&+8\la \la 4\ra_{-2,-2,2,2}\ra\la \la 2\ra_{-2,2}\ra^2-16\la \la 2\ra_{-2,2}\ra^4\Big],
\end{split}
\end{equation}
where we have used the explicit form of covariance matrix in Eq.~\eqref{36}.
One can immediately substitute the quantity $\la k \ra_{a_1,\ldots,a_k}$ in terms of $Q$-vectors for practical computations by using Refs.~\cite{Bilandzic:2013kga,Bilandzic:2010jr}. 

The procedure explained above is implemented into the following functions in the package:
\begin{subequations}\label{42}
	\begin{align}
	&{\tt Nsigma2[func]}\label{42a},\\
	&{\tt Nsigma2Mean[func,v,}\psi{\tt ]}\label{42c},\\
	&{\tt Nsigma2Qvec[func,M,Q]}\label{42d}.
\end{align}
\end{subequations}
The output of the above functions is $N\sigma_f^2$. The input function ({\tt func}) can be any function of correlations but the correlations should be always  written in the form {\tt corr[$a_1,\ldots,a_k$]} as the output of Eq.~\eqref{24}.\footnote{It does not matter how the correlations are named, {\tt correl[$a_1,\ldots,a_k$]}, {\tt f[$a_1,\ldots,a_k$]}, or anything else.} For instance, to find $N \sigma^2_{c_2\{2\}}$ by using {\tt Nsigma2}, we should call the function as follows:
\be\label{43}
\begin{aligned}
	&\scriptsize\verb|In[1]:= |\normalsize {\tt Nsigma2[cCorr[\{2\},\{\},\{2\},corr]]} \\
	&\scriptsize\verb|Out[1]= |\normalsize \la \la 2 \ra_{-2,2}^2 \ra-\la \la 2 \ra_{-2,2}\ra^2.
\end{aligned}
\ee
The difference between functions in Eq.~\eqref{42} is in their output presentations. The output of function {\tt Nsigma2} is shown in Eq.~\eqref{43}, the function {\tt Nsigma2Mean} output is similar to {\tt cMean} function, and {\tt Nsigma2Qvec} returns the variance in terms of $Q$-vectors similar to {\tt cQvec} function (see Eq.~\eqref{28}). As an example, the statistical error of $c_2\{2\}=\la \la 2 \ra_{-2,2} \ra$ in terms of $Q$-vectors can be found below:  
\be\label{44}
\begin{aligned}
	&\scriptsize\verb|In[1]:= |\normalsize {\tt Nsigma2Qvec[cCorr[\{2\},\{\},\{2\},M,Q]]} \\
	&\scriptsize\verb|Out[1]= |\normalsize  {\tt Mean\Big[}\Big(\frac{{\tt M-Q[-2]Q[2]}}{{\tt M(1-M)}}{\tt \Big)^2 \Big]}- {\tt Mean\Big[}\frac{{\tt M-Q[-2]Q[2]}}{{\tt M(1-M)}}{\tt \Big]}.
\end{aligned}
\ee

Referring to Eqs.~\eqref{41} and \eqref{43} (and to any other {\tt Nsigma2} outcomes), we find that the statistical uncertainty has no explicit multiplicity dependence. Let us consider two equal-size sets of events when the multiplicity of events in one set is smaller than in the other set. We expect that the statistical uncertainty of cumulants computed from the set of events with smaller multiplicity is larger than the statistical uncertainty obtained from the other set with higher multiplicity events. This apparent contradiction with intuition is due to the presence of correlations $\la \la k\ra_{a_1,\ldots,a_k} \la \ell\ra_{a_1,\ldots,a_\ell} \ra$ in the statistical uncertainty (for instance $\la \la 2 \ra_{-2,2} \la 4 \ra_{-2,-2,2,2}\ra$ in Eq.~\eqref{41}). These forms of correlations are the consequence of $\text{cov}(V_i,V_j)$ in Eq.~\eqref{36} and have not appeared in previous studies in Refs.~\cite{Bilandzic:2013kga,Bilandzic:2010jr}. In fact, the correlations $\la \la k\ra_{a_1,\ldots,a_k} \la \ell\ra_{a_1,\ldots,a_\ell} \ra$ depend on multiplicity due to the presence of autocorrelations remained in them. To be more specific, we focus on the first terms in the output in Eqs.~\eqref{43},
\be\begin{aligned}
\la 2 \ra_{-2,2}^2=\frac{1}{M^2(M-1)^2}\sum_{i_1\neq i_2,i_3\neq i_4}e^{2i(\varphi_{i_1}-\varphi_{i_2}+\varphi_{i_3}-\varphi_{i_4})}.
\end{aligned}
\ee
The above relation is a four-particle correlation with some remaining autocorrelations.
In the technical appendix~\ref{appB}, a more direct approach is employed to extract the statistical uncertainties. This approach is implemented in the function {\tt Nsigma2P[func,M]}. The result of this function for $c_2\{2\}$ is as follows:
\begin{equation} \label{46}
\begin{split}
&N \sigma_{c_2\{2\}}^2=\Bigg\la \frac{1+\la 2\ra_{-4,4}}{M(M-1)}\\
&+\frac{(M-2)}{M(M-1)}\bigg[\la 3\ra_{-4,2,2}+\la 3 \ra_{-2,-2,4}+2\la 2\ra_{-2,2}\bigg]\\
&+\frac{(M-2)(M-3)}{M(M-1)}\la  4 \ra_{-2,-2,2,2}\Bigg\ra-\la \la 2 \ra_{-2,2}\ra^2.
\end{split}
\end{equation}
As seen from the above equation, on the one hand, there is no correlation with the form $\la \la k\ra_{a_1,\ldots,a_k} \la \ell\ra_{a_1,\ldots,a_\ell} \ra$, and on the other hand, there are explicit $M$ dependences as we expect. One can explicitly show that the terms inside the first bracket in Eq.~\eqref{46} are exactly equal to $\la 2 \ra_{-2,2}^2$. It becomes more apparent when we replace all $\la k \ra_{a_1,\ldots,a_\ell}$ in Eq.~\eqref{46} with $Q$-vectors using the results of Refs.~\cite{Bilandzic:2010jr,Bilandzic:2013kga} where one finds that  Eq.~\eqref{46} in terms of $Q$-vectors is identical with that mentioned in Eq.~\eqref{44}. In general, the result of function \eqref{42c} is identical with that obtained from function {\tt Nsigma2P}. 

As a final remark in this section, we would like to compare our statistical uncertainty results with previous studies. By ignoring the event-by-event fluctuation, all events would be identical. Therefore, we can ignore the outer angle brackets in Eq.~\eqref{46}. Also, by collecting many events we can increase the accuracy such that each $\la k\ra_{a_1,\ldots,a_k} $ estimates the true value of the flow harmonic $V_{a_1,\ldots,a_k}$ (see Eq.~\eqref{23}). In such an ideal scenario, Eq.~\eqref{46} turns into the following form:
\begin{equation}
\begin{split}
N\sigma_{c_2\{2\}}^2&=\frac{v_{4}^2+1}{M(M-1)}+\\
&\frac{2(M-2)}{M(M-1)}\bigg[ v_2^2 v_{4}\cos(4(\psi_2-\psi_{4}))+ v_2^2 \bigg]\\
&+\frac{(M-2)(M-3)}{M(M-1)} v_2^4- v_2^4,
\end{split}
\end{equation}
which has been obtained before in Refs.~\cite{Bilandzic:2013kga,Bilandzic:2014qga} for the case with $\psi_2=\psi_4$. 

\section{Fourier - cumulant expansion to study flow distribution}\label{MonteCarlo}

\FloatBarrier
\begin{table*}
	\centering
	\setlength{\arrayrulewidth}{1.2pt}
	\caption{List of all cumulants with order $q=2,3,4,5$ and harmonics $n=2,3,4,5$.}
	\scriptsize
	\begin{tabular}[t]{p{0.7cm} p{3.2cm} p{1.9cm} p{10cm}}
		\hline
		& cumulant & order  & cumulant expression \\
		\hline
		1 & $c_2\{2\}$ & $2$   & $\la v_2^2 \ra$ \\
		\Xhline{0.3\arrayrulewidth}					
		2 &   $c_3\{2\}$ & $2$   & $\la v_3^2 \ra$ \\
		\Xhline{0.3\arrayrulewidth}					
		3 &   		$c_4\{2\}$ & $2$   & $\la v_4^2 \ra$ \\
		\Xhline{0.3\arrayrulewidth}			
		4 &   $c_5\{2\}$ & $2$   & $\la v_5^2 \ra$ \\
		\hline		
		5 &   $c_{2,4}^{\{4\}}\{2,1\}$ & $3$   & $  \la  v_2^2 v_4 \cos \left(4 \left(\psi _2-\psi _4\right)\right)\ra  $\\
		\Xhline{0.3\arrayrulewidth}			
		6 &   $c_{2,3,5}^{\{-3,5\}}\{1,1,1\}$ & $3$   & $ \la  v_2 v_3 v_5 \cos \left(2 \psi _2+3 \psi _3-5 \psi _5\right)\ra  $\\	
		\hline		
		7 &   $c_2\{4\}$ & $4$   & $\la v_2^4 \ra-2\la v_2^2 \ra^2$ \\
		\Xhline{0.3\arrayrulewidth}			
		8 &   $c_3\{4\}$ & $4$   &  $\la v_3^4 \ra-2\la v_3^2 \ra^2$ \\
		\Xhline{0.3\arrayrulewidth}			
		9 &   $c_4\{4\}$ & $4$   &  $\la v_4^4 \ra-2\la v_4^2 \ra^2$\\
		\Xhline{0.3\arrayrulewidth}			
		10 &   $c_5\{4\}$ & $4$   &  $\la v_5^4 \ra-2\la v_5^2 \ra^2$\\		
		\Xhline{0.3\arrayrulewidth}			
		11 &   $c_{2,3}^{\{0\}}\{2,2\}$ & $4$   &  $\la v_2^2 v_3^2 \ra -\la v_2^2 \ra \la v_3^2 \ra$ \\		
		\Xhline{0.3\arrayrulewidth}			
		12 &   $c_{2,4}^{\{0\}}\{2,2\}$ & $4$   &  $\la v_2^2 v_4^2 \ra -\la v_2^2 \ra \la v_4^2 \ra$ \\
		\Xhline{0.3\arrayrulewidth}			
		13 &   $c_{2,5}^{\{0\}}\{2,2\}$ & $4$   &   $\la v_2^2 v_5^2 \ra -\la v_2^2 \ra \la v_5^2 \ra$\\
		\Xhline{0.3\arrayrulewidth}			
		14 &   $c_{3,4}^{\{0\}}\{2,2\}$ & $4$   &   $\la v_3^2 v_4^2 \ra -\la v_3^2 \ra \la v_4^2 \ra$\\
		\Xhline{0.3\arrayrulewidth}			
		15 &   $c_{3,5}^{\{0\}}\{2,2\}$ & $4$   &   $\la v_3^2 v_5^2 \ra -\la v_3^2 \ra \la v_5^2 \ra$\\
		\Xhline{0.3\arrayrulewidth}			
		16 &   $c_{4,5}^{\{0\}}\{2,2\}$ & $4$   &   $\la v_4^2 v_5^2 \ra -\la v_4^2 \ra \la v_5^2 \ra$\\		 
		\Xhline{0.3\arrayrulewidth}			
		17 &   $c_{2,3,4}^{\{6,-4\}}\{1,2,1\}$ & $4$   &  $ \la  v_3^2 v_2 v_4 \cos \left(2 \left(\psi _2-3 \psi _3+2 \psi _4\right)\right)\ra  $ \\
		\Xhline{0.3\arrayrulewidth}			
		18 &   $c_{3,4,5}^{\{8,-5\}}\{1,2,1\}$ & $4$   & $ \la  v_4^2 v_3 v_5 \cos \left(3 \psi _3-8 \psi _4+5 \psi _5\right)\ra  $ \\
		\Xhline{0.3\arrayrulewidth}			
		19 &   $c_{2,3,4,5}^{\{3,4,-5\}}\{1,1,1,1\}$  & $4$   & $ \la  v_2 v_3 v_4 v_5 \cos \left(2 \psi _2-3 \psi _3-4 \psi _4+5 \psi _5\right)\ra  $ \\	
		\hline
		20 &   $c_{2,3}^{\{6\}}\{3,2\}$ & $5$   &  $ \la  v_2^3 v_3^2 \cos \left(6 \left(\psi _2-\psi _3\right)\right)\ra  $ \\	
		\Xhline{0.3\arrayrulewidth}			
		21 &   $c_{2,4}^{\{4\}}\{2,3\}$ & $5$   &  $ \la  v_2^2 v_4^3 \cos \left(4 \left(\psi _2-\psi _4\right)\right)\ra  -2 \la 
		v_4^2\ra   \la  v_2^2 v_4 \cos \left(4 \left(\psi _2-\psi
		_4\right)\right)\ra   $ \\	
		\Xhline{0.3\arrayrulewidth}			
		22 &   $c_{2,4}^{\{4\}}\{4,1\}$ & $5$   &  $ \la  v_2^4 v_4 \cos \left(4 \left(\psi _2-\psi _4\right)\right)\ra  -3 \la 
		v_2^2\ra   \la  v_2^2 v_4 \cos \left(4 \left(\psi _2-\psi
		_4\right)\right)\ra   $ \\	
		\Xhline{0.3\arrayrulewidth}			
		23 &   $c_{2,3,4}^{\{-6,8\}}\{1,2,2\}$ & $5$   & $ \la  v_4^2 v_3^2 v_2 \cos \left(2 \left(\psi _2+3 \psi _3-4 \psi _4\right)\right)\ra  $ \\	
		\Xhline{0.3\arrayrulewidth}			
		24 &   $c_{2,3,4}^{\{0,4\}}\{2,2,1\}$ & $5$   &  $ \la  v_3^2 v_2^2 v_4 \cos \left(4 \left(\psi _2-\psi _4\right)\right)\ra  -\la  v_3^2\ra   \la  v_2^2 v_4 \cos
		\left(4 \left(\psi _2-\psi _4\right)\right)\ra  $ \\	
		\Xhline{0.3\arrayrulewidth}			
		25 &   $c_{2,3,5}^{\{-3,5\}}\{1,1,3\}$ & $5$   & $ \la  v_5^3 v_2 v_3 \cos \left(2 \psi _2+3 \psi _3-5 \psi _5\right)\ra  -2 \la  v_5^2\ra  \la  v_2 v_3 v_5 \cos \left(2 \psi _2+3 \psi _3-5 \psi
		_5\right)\ra    $  \\	
		\Xhline{0.3\arrayrulewidth}			
		26 &   $c_{2,3,5}^{\{-3,5\}}\{1,3,1\}$ & $5$   &   $ \la  v_3^3 v_2 v_5 \cos \left(2 \psi _2+3 \psi _3-5 \psi _5\right)\ra  -2 \la  v_3^2\ra   \la  v_2 v_3 v_5 \cos
		\left(2 \psi _2+3 \psi _3-5 \psi _5\right)\ra  $\\	
		\Xhline{0.3\arrayrulewidth}			
		27 &   $c_{2,3,5}^{\{-3,5\}}\{3,1,1\}$ & $5$   &  $ \la  v_2^3 v_3 v_5 \cos \left(2 \psi _2+3 \psi _3-5 \psi _5\right)\ra  -2 \la  v_2^2\ra   \la  v_2 v_3 v_5 \cos
		\left(2 \psi _2+3 \psi _3-5 \psi _5\right)\ra  $ \\	
		\Xhline{0.3\arrayrulewidth}			
		28 &   $c_{2,4,5}^{\{-8,10\}}\{1,2,2\}$ & $5$   &  $ \la  v_5^2 v_4^2 v_2 \cos \left(2 \left(\psi _2+4 \psi _4-5 \psi _5\right)\right)\ra  $ \\	
		\Xhline{0.3\arrayrulewidth}			
		29 &   $c_{2,4,5}^{\{4,0\}}\{2,1,2\}$ & $5$   & $ \la  v_5^2 v_2^2 v_4 \cos \left(4 \left(\psi _2-\psi _4\right)\right)\ra  -\la  v_2^2 v_4 \cos \left(4 \left(\psi _2-\psi
		_4\right)\right)\ra   \la  v_5^2\ra  $ \\	
		\Xhline{0.3\arrayrulewidth}			
		30 &   $c_{3,4,5}^{\{-4,10\}}\{2,1,2\}$ & $5$   &  $ \la  v_5^2 v_3^2 v_4 \cos \left(6 \psi _3+4 \psi _4-10 \psi _5\right)\ra  $\\	
		\Xhline{0.3\arrayrulewidth}			
		31 &   $c_{3,4,5}^{\{4,5\}}\{3,1,1\}$ & $5$   &   $ \la  v_3^3 v_4 v_5 \cos \left(9 \psi _3-4 \psi _4-5 \psi _5\right)\ra  $\\	
		\Xhline{0.3\arrayrulewidth}			
		32 &   $c_{2,3,4,5}^{\{-3,0,5\}}\{1,1,2,1\}$ & $5$   & $ \la  v_4^2 v_2 v_5 v_3 \cos \left(2 \psi _2+3 \psi _3-5 \psi _5\right)\ra  -\la  v_4^2\ra   \la  v_2 v_3 v_5 \cos
		\left(2 \psi _2+3 \psi _3-5 \psi _5\right)\ra  $ \\	
		\Xhline{0.3\arrayrulewidth}			
		33 &   $c_{2,3,4,5}^{\{3,-4,5\}}\{2,1,1,1\}$ & $5$   &  $ \la  v_2^2 v_3 v_4 v_5 \cos \left(4 \psi _2-3 \psi _3+4 \psi _4-5 \psi _5\right)\ra  $\\					
		\hline
	\end{tabular}
	\label{firstCumulTable}
\end{table*}%

An important property of cumulant analysis is that the lowest order cumulants capture the global features of the p.d.f., and by moving toward higher-orders, more detailed features become relevant. The same is true for Fourier expansion of  $f(\varphi)$ in Eq.~\eqref{Fourier} where lower harmonics contain more coarse-grained pictures than higher harmonics. As a result, we consider a double expansion for studying p.d.f.~\eqref{pdf2}: first, harmonic expansion, and second cumulant order expansion. According to this approach, $c_2\{2\}$ contains the most coarse-grained information (we ignore direct flow in the present study). The harmonic $n=2$ is the first dominant flow, and $c_2\{2\}$ is the only possible second-order cumulant for $n=2$. It is well-known that $c_2\{2\}$ contains information about the initial state ellipticity together with elliptic flow fluctuation. The first third-order cumulant is $c_{2,4}^{\{4\}}\{2,1\}$ which means, in addition to second harmonic $n=2$, the harmonic $n=4$ should be involved as well. Keeping more harmonics and higher-order cumulants, more detailed information about the p.d.f.~\eqref{pdf2} and $f(\varphi)$ can be achieved. The number of cumulants increases rapidly by including more harmonics and keeping higher-order cumulants. 

Henceforth, we study harmonics $n=2,3,4,5$ and cumulants up to fifth order, which includes 33 different cumulants as they are tabulated in Table~\ref{firstCumulTable}. Apart from those cumulants that contain only one moment, there are cumulants containing more than one event-plane moment (see lines 21, 24, 25, 26, 27, 29, and 32 in Table~\ref{firstCumulTable}) which are studied here for the first time.  We should point out that three different five-particle cumulants have been measured by ALICE \cite{ALICE:2011ab,Bilandzic:2012an} two of them (see \cite{Bilandzic:2012an}) contain more than one event-plane moment. The direct flow (flow harmonics with $n=1$) is involved in all these cumulants. The cumulant $c_{2,4}^{\{4\}}\{4,1\}$ (see Ref.~\cite{Jia:2014jca} as well) in line 22 is equivalent with $v_4\{5\}$ measured by CMS up to a normalization factor \cite{Chatrchyan:2013kba}.  

\FloatBarrier
\begin{figure*}
	%\vspace*{-1.6cm}
	\hspace*{-1.0cm}
	\includegraphics[width=19cm]{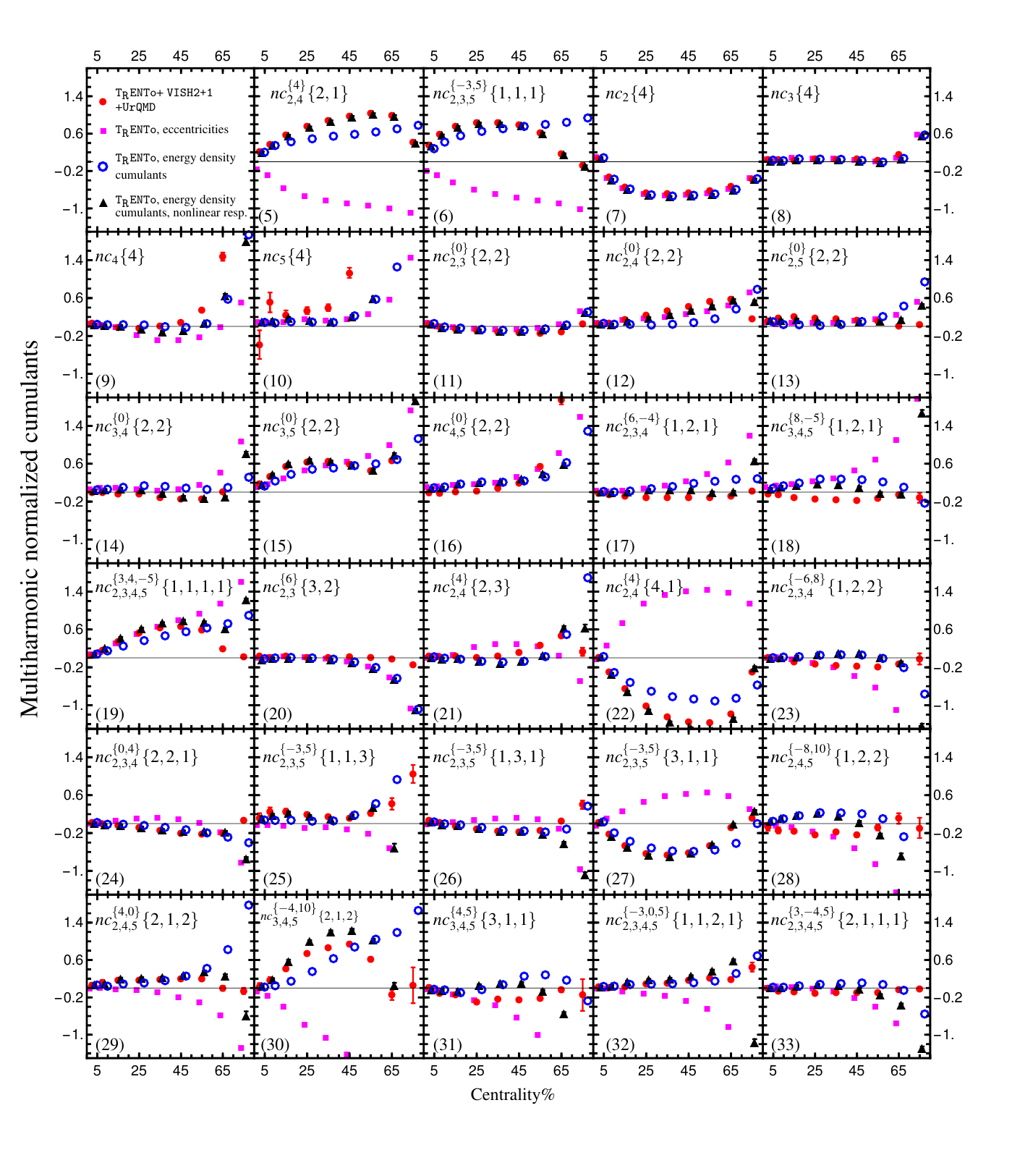} 	
	\caption{All  $2,3,4,5$-order normalized cumulants for harmonics $n=2,3,4,5$. Magenta squares and open blue circles are eccentricity and energy density cumulant fluctuations from {\tt T$_\text{R}$ENTo} initial state model, red filled circles are flow fluctuations after hydrodynamic evolution ({\tt VISH2+1}+{\tt UrQMD}). The black triangles are the initial state with a linear and nonlinear hydrodynamic response. In the lower-left part of each panel, the corresponding cumulant line number in Table~\ref{firstCumulTable} are presented.}
	\label{fig1}
\end{figure*}

\subsection{Normalized cumulants from initial and final states }\label{4.1}

To understand the observed cumulants' origin, let us recall that the central part of the flow fluctuations comes from the initial state. To study the initial state fluctuation, we need to quantify the shape of the initial state with quantities similar to those done for flow harmonics, namely eccentricities \cite{Gardim:2011xv},
\bea\label{polarMoments}
\ve_{n}e^{i n \phi_{n} }\equiv -\frac{\{r^n e^{in\varphi}\}}{\{r^n\}},\qquad n>1,
\eea
where $\{\cdots\}=\int r dr d\varphi \varepsilon(r,\varphi)$ is the averaging with respect to the energy density $\varepsilon(r,\varphi)$. 
The true initial energy density shape is  captured by the cumulants of the distribution $\varepsilon(r,\varphi)$ not its moments (eccentricities) in Eq.~\eqref{polarMoments} \cite{Teaney:2010vd}. The cumulants of the energy density have been obtained by Teaney and Yan in the seminal paper \cite{Teaney:2010vd} and follow-up papers \cite{Teaney:2013dta,Teaney:2012ke}. Employing the convention in Ref.~\cite{Teaney:2012ke}, the two-dimensional energy density cumulants in the polar coordinate are given by
\begin{subequations}\label{ieCumul}
	\begin{align}
			&\mathcal{C}_{2}e^{i2\Phi_2}=\epsilon_{2}e^{i2\phi_2},\label{ieCumulA}\\
			&\mathcal{C}_{3}e^{i3\Phi_3}=\epsilon_{3}e^{i3\phi_3},\label{ieCumulB}\\
		&\mathcal{C}_{4}e^{i4\Phi_4}=\epsilon_4e^{i4\phi_4}+3\left(\frac{\{r^2\}^2}{\{r^4\}}\right) \epsilon_2^2\, e^{i4\phi_2}, \label{ieCumulC}\\
	&\mathcal{C}_{5}e^{i5\Phi_5}=\epsilon_5 e^{i5\phi_5}+10\left(\frac{\{r^2\} \{r^3\}}{\{r^5\}}\right) \epsilon_2 \epsilon_3 e^{i2\phi_2} e^{i3\phi_3}\label{ieCumulD}.
	\end{align}
\end{subequations}
One notes that the eccentricities for $n=2$ and $3$ are cumulants as well, while for $n>3$ there are contributions from lower harmonics eccentricities and radial shape of the energy density (see Eqs.~(2.9) and (2.11) in Ref.~\cite{Teaney:2012ke}).

The anisotropic flow in the final state is a collective response to the initial anisotropic geometry. This process is formulated via the following response relation \cite{Teaney:2010vd},
\bea\label{nonLinearOrginal}
v_n e^{i\,n\,\psi_n}=w_n \, \mathcal{C}_n\,e^{i\,n\,\Phi_n}+\text{nonlinear terms}.
\eea 
Up to the leading order, the physics of collectivity and freezeout is encoded in the linear coupling $w_n$. 
Ignoring the nonlinear terms, the flow harmonic cumulants of the p.d.f \eqref{pdf2} should be proportional to the cumulants of the  p.d.f. $
p_\text{ini}(\C_{n_1},\ldots,\Phi_{n_2}-\Phi_{n_1},\ldots)
$. The cumulants of the latter p.d.f. are obtained precisely similar to those studied so far by replacing $v_ne^{in\psi_n}$ with $\mathcal{C}_{n}e^{in\Phi_n}$.\footnote{One notes that $\mathcal{C}_{n}e^{in\Phi_n}$ is the cumulant of the initial energy density $\varepsilon(r,\varphi)$ in a single event. We call it energy density cumulants in the text. We also study $\mathcal{C}_{n}e^{in\Phi_n}$ event-by-event fluctuation cumulants (the cumulants of the p.d.f. $
	p_\text{ini}(\C_{n_1},\ldots,\Phi_{n_2}-\Phi_{n_1},\ldots)
	$) which are called initial state fluctuation cumulants in the manuscript.  } 
To be more precise, let us recall the important properties of cumulants, homogeneity. Based on this property, by scaling the random variable $x_i$ as $x_i \to w_i x_i$, the cumulants defined in Eq.~\eqref{nDcumul} are scaled as
\begin{equation} \label{KappaScal}
\begin{split}
\mathcal{K}_{a_1,\ldots,a_k} \to w_1^{a_1}\cdots w_k^{a_k} \, \mathcal{K}_{a_1,\ldots,a_k}.
\end{split}
\end{equation}
One can examine the above relation in two explicit examples shown in Eq.~\eqref{kapa02kapa30}. 
By ignoring the nonlinear parts in Eq.~\eqref{nonLinearOrginal}, we can relate the cumulants $c_{n_1,\ldots,n_k}^{\{\alpha_1,\ldots\alpha_{k-1}\}}\{m_1,\ldots,m_k\}$ obtained from eccentricities and flow harmonics as
\begin{equation} \label{cnScal}
\begin{split}
&c_{n_1,\ldots,n_k}^{\{\alpha_1,\ldots\alpha_{k-1}\}}\{m_1,\ldots,m_k\} =\\ &\left(w_{n_1}^{m_1}\cdots w_{n_k}^{m_k}\right)c_{n_1,\ldots,n_k}^{\{\alpha_1,\ldots\alpha_{k-1}\}(\text{init})}\{m_1,\ldots,m_k\}.
\end{split}
\end{equation}
The above relation indicates that the initial state and final state fluctuation cumulants differ by a numerical factor, similar to that mentioned in Eq.~\eqref{KappaScal}.

For comparing the initial and final state cumulants in the linear response approximation, the values of $w_n$ are needed. However, it is still possible to modify our cumulant definition such that it is independent of $w_n$ numerical value. To this end, we define \textit{normalized} (\textit{standardized}) cumulants,
\begin{equation} \label{normalCumu}
\begin{split}
&nc_{n_1,\ldots,n_k}^{\{\alpha_1,\ldots\alpha_{k-1}\}}\{m_1,\ldots,m_k\}=\frac{c_{n_1,\ldots,n_k}^{\{\alpha_1,\ldots\alpha_{k-1}\}}\{m_1,\ldots,m_k\}}{\sqrt{c_{n_1}^{m_1}\{2\}\cdots c_{n_k}^{m_k}\{2\}}}.
\end{split}
\end{equation}
Referring to Eqs.~\eqref{cnScal} and \eqref{normalCumu}, one finds that the normalized cumulants of the initial anisotropy and flow harmonics fluctuations must be precisely the same at the linear approximation. Any deviation between two normalized cumulants should be sourced from nonlinear terms.

The definition of normalized cumulants in Eq.~\eqref{normalCumu} is
compatible with the suggestion in Ref.~\cite{Bhalerao:2011yg,Teaney:2013dta} made for event-plane correlations. According to the above definition, for $k=1$, we have
\bea
nc_n\{2m\}=\frac{c_n\{2m\}}{c_n^m\{2\}}\propto \left(\frac{v_n\{2m\}}{v_n\{2\}}\right)^{2m},
\eea
compatible with Ref.~\cite{Giacalone:2017uqx}. Also for $nc_{m,n}^{\{0\}}\{2,2\}$, we obtain
\begin{equation} \label{normalSC}
\begin{split}
nc_{m,n}^{\{0\}}\{2,2\}&= \frac{c_{m,n}^{\{0\}}\{2,2\}}{\la v_m^2 \ra \, \la v_n^2 \ra} \\
&\equiv \text{NSC}(m,n)
\end{split}
\end{equation}
where $\text{NSC}(m,n)$ is the normalized symmetric cumulant. 

There is an alternative way of defining the normalized cumulant,
\begin{equation} \label{alterNormalCumu}
\begin{split}
&n'c_{n_1,\ldots,n_k}^{\{\alpha_1,\ldots\alpha_{k-1}\}}\{m_1,\ldots,m_k\}=\frac{c_{n_1,\ldots,n_k}^{\{\alpha_1,\ldots\alpha_{k-1}\}}\{m_1,\ldots,m_k\}}{\sqrt{\la v_{n_1}^{2m_1} \ra \cdots \la v_{n_k}^{2m_k} \ra}}.
\end{split}
\end{equation}
This definition is compatible with the scalar product method mentioned in Ref.~\cite{Aad:2014fla}. For cumulants with all $m_i=1$, two approaches coincide. In the present study, we use the first convention in Eq.~\eqref{normalCumu} when all cumulants are normalized with the first single-harmonics cumulants. In comparison between the LHC data and simulation, we employ the alternative approach only when the first approach is not measured.

\subsection{Normalized cumulants from simulation and the LHC}

\FloatBarrier
\begin{table*}
	\centering
	\setlength{\arrayrulewidth}{1.2pt}
	\caption{Measured normalized cumulants at the LHC for Pb--Pb collisions. The far left column indicates the line number of the corresponding cumulant in Table~\ref{firstCumulTable}. In the references shown by [...]$'$, the alternative version of normalized cumulants (see Eq.~\eqref{alterNormalCumu}) are presented. The method of the measurements, scalar product (SP), standard multiparticle (ST), and subevent multiparticle (SE) methods, are indicated for each reference. }
	\scriptsize
	\begin{tabular}{l l  c c c c c c }
		\hline
		\multicolumn{2}{c}{normalized cumulant}   & \multicolumn{2}{c}{ALICE} & \multicolumn{2}{c}{ATLAS} & \multicolumn{2}{c}{CMS}  \\
		&  & 2.76\,TeV & 5.02\,TeV & 2.76\,TeV & 5.02\,TeV & 2.76\,TeV & 5.02\,TeV  \\	
		\hline
		5 & $nc_{2,4}^{\{4\}}\{2,1\}$ & & & \cite{Aad:2014fla}$'$(SP), \cite{Aaboud:2018syf}$'$(SE) & \cite{Aaboud:2019sma}$'$(ST,SE) & &  \\ 
		\Xhline{0.3\arrayrulewidth}
		6 & $nc_{2,3,5}^{\{-3,5\}}\{1,1,1\}$ & & & \cite{Aad:2014fla}(SP)  & & & \\
		\Xhline{0.3\arrayrulewidth}
		7 & $nc_{2}\{4\}$ & &  &   & \cite{Aaboud:2019sma}(ST,SE)  & & \\
		\Xhline{0.3\arrayrulewidth}
		8 & $nc_{3}\{4\}$ & &  & & \cite{Aaboud:2019sma}(ST,SE) & & \\
		\Xhline{0.3\arrayrulewidth}
		9 & $nc_{4}\{4\}$ & &  &  & \cite{Aaboud:2019sma}(ST,SE) &  & \\
		\Xhline{0.3\arrayrulewidth}
		11 & $nc_{2,3}^{\{0\}}\{2,2\}$ & \cite{ALICE:2016kpq}(ST)   & \cite{Acharya:2019vdf}(SE)  &  \cite{Aaboud:2018syf}(SE) & \cite{Aaboud:2019sma}(ST,SE) & & \cite{Sirunyan:2017uyl}(ST) \\ 
		\Xhline{0.3\arrayrulewidth}
		12 & $nc_{2,4}^{\{0\}}\{2,2\}$ & \cite{ALICE:2016kpq}(ST)  &  \cite{Acharya:2019vdf}(SE)  & \cite{Aaboud:2018syf}(SE)  & \cite{Aaboud:2019sma}(ST,SE) & & \cite{Sirunyan:2017uyl}(ST) \\
		\Xhline{0.3\arrayrulewidth}
		13 & $nc_{2,5}^{\{0\}}\{2,2\}$ & \cite{Acharya:2017gsw}(ST)  &  & & &    &  \\
		\Xhline{0.3\arrayrulewidth}
		14 & $nc_{3,4}^{\{0\}}\{2,2\}$ & \cite{Acharya:2017gsw}(ST)  &  & & &    &  \\
		\Xhline{0.3\arrayrulewidth}
		15 & $nc_{3,5}^{\{0\}}\{2,2\}$ & \cite{Acharya:2017gsw}(ST) &    &  &  &  &  \\
		\Xhline{0.3\arrayrulewidth}
		17 & $nc_{2,3,4}^{\{6,-4\}}\{1,2,1\}$ & &  & \cite{Aad:2014fla}$'$(SP) & & & \\
		\Xhline{0.3\arrayrulewidth}
		20 & $nc_{2,3}^{\{6\}}\{3,2\}$   & &  & \cite{Aad:2014fla}$'$(SP) & & &  \\
		\hline
	\end{tabular}
	\label{measuredCumulTable}
\end{table*}%

%\FloatBarrier
\begin{figure*}
	%\vspace*{-1.6cm}
	%\hspace*{-1.0cm}
	\centering
	\includegraphics[width=17cm]{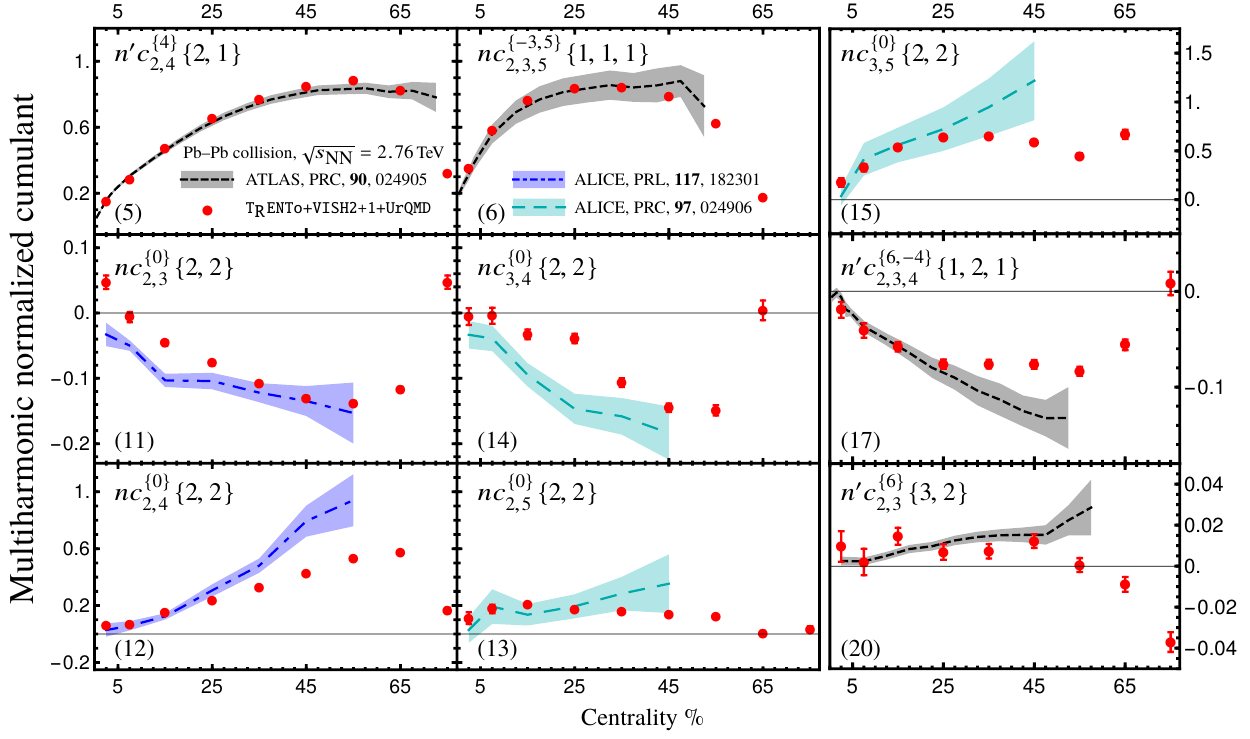} 	
	\caption{The (alternative) normalized cumulants that measured By ALICE \cite{ALICE:2016kpq,Acharya:2017gsw} and ATLAS \cite{Aad:2014fla} compared with {\tt T$_\text{R}$ENTo}+{\tt VISH2+1}+{\tt UrQMD}. }
	\label{expFig}
\end{figure*}

We study the first 29 normalized cumulants in the Fourier-cumulant expansion by using a hybrid hydrodynamic model, {\tt T$_\text{R}$ENTo} \cite{Moreland:2014oya}+{\tt VISH2+1} \cite{Shen:2014vra,Song:2007ux}+{\tt UrQMD} \cite{Bass:1998ca,Bleicher:1999xi} ($nc_n\{2\}=1$ by definition). We simulate events for Pb--Pb collisions at $\sqrt{s_{\text{NN}}}=2.76\;$TeV. The global Bayesian analysis calibrates the model's free parameters (consist of the temperature dependence of shear and bulk viscosity over entropy density) to explain the measurements by ALICE experiment \cite{Bernhard:2019bmu}. In Fig.~\ref{fig1}, the normalized version of flow harmonic cumulants listed in Table~\ref{firstCumulTable} are plotted with red filled circles. We kept a fixed range for the vertical axes in all panels to simplify the magnitude comparison of different normalized cumulants. In the calibration, the cumulants $c_n\{2\}$ for $n=2,3,4$, and $c_2\{4\}$ are used. This means that only the information of normalized cumulant $nc_2\{4\}$ (in panel (7)) has been used for calibration and the rest of the simulations are predictions. 

 A series of normalized cumulants in Fig.~\ref{fig1} has already been measured at the LHC (see Table~\ref{measuredCumulTable} for a summary). The normalized symmetric cumulants, $nc_{m,n}^{\{0\}}\{2,2\}$, are measured by ALICE \cite{Acharya:2019vdf,ALICE:2016kpq,Acharya:2017gsw}, ATLAS \cite{Aaboud:2018syf,Aaboud:2019sma}, and CMS \cite{Sirunyan:2017uyl} experiments. The event-plane correlations via scalar product method for Pb--Pb collisions at $\sqrt{s_{\text{NN}}}=2.76\,$TeV are measured by ATLAS experiment \cite{Aad:2014fla}. The measurement of the normalized version (with alternative convention in Eq.~\eqref{alterNormalCumu}) of cumulants $c_{2,4}^{\{4\}}\{2,1\}$, $c_{2,3,5}^{\{-3,5\}}\{1,1,1\}$, $c_{2,3,4}^{\{6,-4\}}\{1,2,1\}$, and $c_{2,3}^{\{6\}}\{2,3\}$ can be found in this analysis.\footnote{The normalization conventions in Eq.~\eqref{alterNormalCumu} and \eqref{normalCumu} are equivalent for  $c_{2,3,5}^{\{-3,5\}}\{1,1,1\}$.} The alternative normalized version of $c_{2,4}^{\{4\}}\{2,1\}$ is also measured by ATLAS experiment for Pb--Pb collisions at $\sqrt{s_{\text{NN}}}=2.76\,$TeV \cite{Aaboud:2018syf} and $\sqrt{s_{\text{NN}}}=5.02\,$TeV \cite{Aaboud:2019sma}. And finally, the normalized cumulants  $nc_n\{4\}$ for $n=2,3,4$ are measured at $\sqrt{s_{\text{NN}}}=5.02\,$TeV by ATLAS experiment \cite{Aaboud:2019sma}.  
 
 Although the simulation is calibrated with ALICE experiment kinematics ($0.2<p_T<5\,\text{GeV}$ and $|\eta|<0.8$), the normalized versions of cumulants are less sensitive to these kinematics. In particular, the transverse momentum range dependence of the normalized cumulants has been examined in Refs.~\cite{Aaboud:2019sma,Acharya:2017gsw}. As a result, in addition to the ALICE experiment measurements, we compare our simulation with ATLAS measurements ($p_T>0.5\,$GeV and $|\eta|<2.5$) for Pb--Pb collisions at $\sqrt{s_{\text{NN}}}=2.76\,$TeV. The results are depicted in Fig.~\ref{expFig}. In general, the simulations and experimental measurements are in agreement.  As seen in the figure, the the simulation for cumulants $nc_{2,4}^{\{4\}}\{2,1\}$ and $n'c_{2,3,5}^{\{-3,5\}}\{1,1,1\}$ can accurately explain the ATLAS measurements. The ALICE measurement for normalized symmetric cumulant $nc_{2,3}^{\{0\}}\{2,2\}$ is overestimated by simulation at centralities below $\sim 25\%$. The same is true for $nc_{3,4}^{\{0\}}\{2,2\}$ around centrality $\sim 25\%$. However, the normalized cumulant $nc_{2,4}^{\{0\}}\{2,2\}$ is underestimated for centralities above $35\,\%$. The simulation at centralities below 10\% reveals a sign change for $nc_{2,3}^{\{0\}}\{2,2\}$ that is not observed by ALICE measurement.
 
 Except $nc_2\{4\}$ which is used for calibration, the rest of the normalized cumulants shown in Fig.~\ref{fig1} and not displayed in Fig.~\ref{expFig} contain an independent piece of information that can be examined in the future experiment.  Here, we only focus on the Pb--Pb collisions at fixed center-of-mass energy. 
  The cumulants' system size and energy dependence would lead to interesting information about the underlying initial state fluctuation and collective evolution in the future. 
 Referring to Fig.~\ref{fig1}, we find that the normalized symmetric cumulants $nc_{m,n}^{\{0\}}\{2,2\}$ are not the most pronounced ones compared to the others. For instance, the fourth-order normalized cumulant $nc_{2,3,4,5}^{3,4,-5}\{1,1,1,1\}$ (panel (19)), and fifth-order normalized cumulants $nc_{2,4}^{\{4\}}\{4,1\}$,\footnote{The measured quantity $v_4\{5\}$ in Ref.~\cite{Chatrchyan:2013kba} is equivalent with $c_{2,4}^{\{4\}}\{4,1\}$ with normalization factor $-2v_2^4\{4\}$.}  $nc_{2,3,5}^{\{-3,5\}}\{3,1,1\}$ and $nc_{3,4,5}^{\{-4,10\}}\{2,1,2\}$ (panels (22), (27), and (30)) are the largest cumulants up to $n=5$ harmonics that have not been measured yet.

 Concerning the feasibility of the cumulant measurements, there is a resent analysis by ALICE collaboration measuring higher-order normalized symmetric cumulants (shown by NSC$(m,n,\ell)$) consisting flow amplitudes $v_2$, $v_3$, $v_4$ and $v_2$, $v_3$, $v_5$ \cite{Acharya:2021afp}. To calculate these sixth-order cumulants, the moments as $\la v_2^2 v_3^2 v_4^2\ra$ and $\la v_2^2 v_3^2 v_5^2\ra$ are needed to be measured. Given that these cumulants are one order higher than those we presented in Table~\ref{firstCumulTable}, we would expect that all the unmeasured normalized cumulants in Table~\ref{firstCumulTable} are experimentally accessible in the near future. The cumulant $nc_{2,3,5}^{-3,5}\{1,1,3\}$ (see line25 in Table~\ref{firstCumulTable}) contains moments with $v_3 v_5^3$ combination. Compared to NSC$(2,3,5)$, it has one power higher for flow amplitude for $v_5$ and one power lower for $v_3$. Considering its large signal from the simulation, we expect that an experimental observation for this cumulant as well.

So far, we have studied the normalized cumulants of Pb--Pb collisions and investigate to what extend the Monte Carlo simulation can explain the experimental data. To better understand both initial state and collective evolution, we have also compared the initial and final state fluctuation cumulants. In the next section, we investigate it in more detail.
As we have discussed earlier, up to the linear order of hydrodynamic response, the normalized cumulants of flow harmonics and energy density cumulant fluctuations must be the same. As a result, the observed difference in Fig.~\ref{fig1} is due to the nonlinear hydrodynamic response.

\section{Linear and nonlinear hydrodynamic response coefficients}\label{phase}

In Sec.~\ref{4.1}, it has been discussed that by comparing the normalized cumulants of the initial and final state fluctuation, we can obtain information about the collective hydrodynamic evolution. In the present section, we quantitatively study the hydrodynamic response coefficients.

In the present study, we keep only the linear terms for $n=2,3$, and the first subleasing nonlinear terms for $n=4,5$. In particular, we consider the following explicit form of Eq.~\eqref{nonLinearOrginal},
\begin{subequations}\label{linearAndNonlinearCn}
	\begin{align}
	&v_2 e^{i2\psi_2}\simeq w_2\, \C_{2} e^{i2\Phi_2}\\
	&v_3 e^{i3\psi_3}\simeq  w_3\, \C_{3} e^{i3\Phi_3}\\
	&v_4 e^{i4\psi_4}\simeq  w_4\, \C_{4} e^{i4\Phi_4}+w_{4(22)}\,\C_2^2 e^{i4\Phi_2}\label{57a},\\
	&v_5 e^{i5\psi_4}\simeq  w_5\, \C_{5} e^{i5\Phi_5}+w_{5(23)}\,\C_2 \C_3 e^{i2\Phi_2+i3\Phi_3}.
	\end{align}
\end{subequations}
There is another convention in studying hydrodynamic response (see Ref.~\cite{Gardim:2011xv}) in which the eccentricities are employed to present the hydrodynamic response,
\begin{subequations}\label{linearAndNonlinear}
	\begin{align}
	&v_2 e^{i2\psi_2}\simeq k_2 \, \epsilon_{2} e^{i2\phi_2},\\
	&v_3 e^{i3\psi_3}\simeq k_3 \,\epsilon_{3} e^{i3\phi_3},\\
&v_4 e^{i4\psi_4}\simeq k_4 \,\epsilon_{4} e^{i4\phi_4}+k_{4(22)}\,\epsilon_2^2 e^{i4\phi_2},\\
&v_5 e^{i5\psi_4}\simeq k_5 \, \epsilon_{5} e^{i5\phi_5}+k_{5(23)}\,\epsilon_2 \epsilon_3 e^{i2\phi_2+i3\phi_3}.
	\end{align}
\end{subequations}
We will discuss the interpretation of these two conventions later.

Different approaches can be employed to extract the couplings $w_n$, $w_{4(22)}$, and $w_{5(23)}$ (or $k_n$, $k_{4(22)}$, and $k_{5(23)}$). We can start from a Gaussian geometrical initial energy density and deform it with one (or a few numbers) of nonvanishing $\C_{n} e^{in\Phi_n}$. Then we change the value of the energy density cumulant to probe the hydrodynamic response from this ``single shot'' simulation \cite{Teaney:2013dta,Teaney:2012ke}. It is shown that (Marcinkiewciz theorem\footnote{Marcinkiewicz has proved that only Gaussian distribution has a polynomial generating function with a finite number of nonzero cumulants \cite{MarcinkiewcizTheorem}.}) reproducing a distribution (except Gaussian distribution) with a finite number of cumulants leads to negative values at some parts of distribution domain. These negative values should be regulated, which consequently produces spurious cumulants \cite{Teaney:2012ke}. Until spurious cumulants are small, we can neglect their effect in the coupling estimations. The initial state with larger deformation would have more such a problem.

As another approach, one can generate many events with complicated initial geometries and study the hydrodynamic response. For instance, to extract the response coefficient $k_n$ in Refs.~\cite{Gardim:2011xv,Noronha-Hostler:2015dbi}, the authors multiply both sides of Eq.~\eqref{nonLinearOrginal} (ignore nonlinear terms) with $\epsilon_n e^{-i n\phi_n}$ and then average over many simulated events in a given centrality class. This approach leads to the following estimation: $k_n^{\text{est}}=\la\epsilon_n v_n \cos(n(\psi_n-\phi_n)) \ra/\la \epsilon_n^2 \ra$. By writing this estimator in the Cartesian coordinate, $\epsilon_n e^{in\phi_n}=\epsilon_{n,x}+i\epsilon_{n,y}$ and  $v_n e^{in\psi_n}=v_{n,x}+iv_{n,y}$, we see that this relation is the Pearson correlation between eccentricities and flow harmonics. This estimator is related to the covariance matrix of  $p_{\epsilon|v}(\epsilon_{n,x},\epsilon_{n,y},v_{n,x},v_{n,y})$. To estimate the nonlinear hydrodynamic couplings similar to what is done in Ref.~\cite{Gardim:2011xv}, we need to include more harmonics in p.d.f. $p_{\epsilon|v}$. Using this approach, we have ignored the non-Gaussian effects encoded in $p_{\epsilon|v}$. 

\begin{figure}
	%\vspace*{-1.6cm}
	%	\hspace*{-1.0cm}
	\begin{tabular}{c}
		\includegraphics[width=8.3cm]{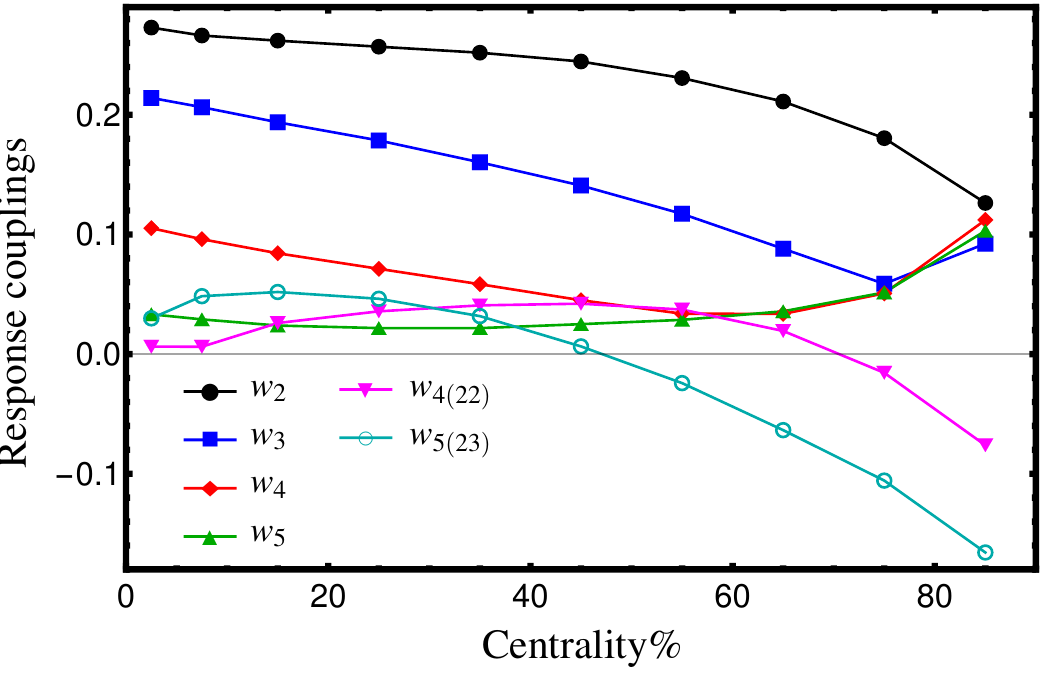}\\
		
	\end{tabular}	
	\caption{Linear and nonlinear hydrodynamic response couplings obtained by comparing initial and final state normalized cumulants.}
	\label{HydroCouplingFigP}
\end{figure}

Comparing the event-by-event fluctuation of the initial and final states in a simulation is an alternative method (the one we employ here) to study the hydrodynamic response couplings. For that, we technically compare $p(v_{n_1},\ldots,\psi_{n_2}-\psi_{n_1},\ldots)$ with $
p_\text{ini}(\C_{n_1},\ldots,\Phi_{n_2}-\Phi_{n_1},\ldots)
$. By substituting $v_n e^{in\psi_n}$ in the flow harmonic fluctuation cumulant (moments) with the corresponding expression for the hydrodynamic response Eqs.~\eqref{linearAndNonlinearCn}, we obtain a function written in terms of initial state fluctuation cumulant (moments). For instance, by employing Eq.~\eqref{57a}, we obtain 
\begin{equation} \label{58}
\begin{split}
\la v_4^2 \ra &\simeq \la| w_4\, \C_{4} e^{i4\Phi_4}+w_{4(22)}\C_2^2 e^{i4\Phi_2} |^2 \ra\\
&=w_4^2\la \C_4^2 \ra+w_{4(22)}^2\la \C_2^4 \ra\\
&\hspace*{1.5cm}+2w_4w_{4(22)}\la \C_2^2\C_4\cos(4\Phi_4-4\Phi_2)\ra.
\end{split}
\end{equation}
Similarly, we work out the following flow harmonic cumulants in terms of $\C_n e^{in\Phi_n}$,
\begin{equation} \label{59}
\begin{split}
&\la v_2^2 \ra \simeq w_2^2 \la \C_2^2 \ra,\\
&\la v_3^2 \ra \simeq w_3^3 \la \C_3^2 \ra,\\
&\la v_5^2 \ra \simeq w_5^2\la \C_5^2 \ra+w_{5(23)}^2\la \C_2^2 \C_3^2 \ra\\
&\hspace*{0.5cm}+2w_5 w_{5(23)}\la \C_2 \C_3\C_5\cos(5\Phi_5-3\Phi_3-2\Phi_2)\ra,\\
&\la v_2^2 v_4 \cos(4\psi_4-4\psi_2)\ra\simeq  \\
&\hspace*{1cm} w_2^2 w_{4(22)} \la \C_2^4\ra+ w_2^2w_4\la\C_2^2 \C_4 \cos(4\Phi_4-4\Phi_2) \ra,\\
&\la v_2 v_3 v_5 \cos(5\psi_5-3\psi_3-2\psi_2)\ra\simeq  w_2 w_3 w_{5(23)} \la \C_2^2 \C_3^2 \ra\\
&\hspace*{0.7cm}+w_2 w_3 w_5 \la \C_2 \C_3\C_5\cos(5\Phi_5-3\Phi_3-2\Phi_2)\ra.
\end{split}
\end{equation}
We can rewrite the right-hand sides of the above equations in terms of cumulants as well. 

There are six equations (Eqs.~\eqref{58} and \eqref{59}) and six unknown response coupling constants that we can find numerically. We do a na\"ive analysis by ignoring the statistical errors in our Monte Carlo simulation to extract the unknown coefficients. At the linear level, the initial state's pressure gradient enforces that the event-plane angle $\psi_n$  to be the same as the participant plane $\Phi_n$.  As a result, we expect that all $w_n$s are real and positive. By demanding that the linear response coupling constants $w_n$, $n=2,\ldots,5$ are positive, we obtain only one set of solutions for couplings at each centrality class.  The results are depicted in Fig.~\ref{HydroCouplingFigP}. The couplings $k_2$ and $k_3$ (which are identical with $w_2$ and $w_3$ in our study) are explicitly reported in Ref.~\cite{Noronha-Hostler:2015dbi}. Considering that a different hydrodynamic model (with different tuning) has been used, and a different method (mentioned in Ref.~\cite{Gardim:2011xv}) is employed to extract the couplings, our results for $w_2$ and $w_3$  are compatible with those computed in Ref.~\cite{Noronha-Hostler:2015dbi}. The couplings $w_4$ and $w_5$ are computed in Ref.~\cite{Teaney:2013dta}. The values of these couplings approach zero and change the sign at mid-central collisions. In our method, there are valid positive solutions for $w_4$ and $w_5$ with no sign change. We have also found larger values for these couplings. However, our computations have rather similar behavior as those reported in Ref.~\cite{Teaney:2013dta} for the nonlinear coefficients. The observed differences between the two studies could be due to the different hydrodynamic models or different approaches of extracting the coefficients (single-shot approach has been employed in Ref.~\cite{Teaney:2013dta}). One notes that our estimations depend on the number of nonlinear response terms that we have considered in our hydrodynamic response estimation. By adding more terms, we need to employ more moments (cumulants) as input. This modification can lead to slightly different values for the couplings.

As it has been mentioned in Sec.~\ref{4.1}, to understand that how much flow harmonic fluctuation is originated from the initial state, one can compute the initial state fluctuation normalized cumulants and compare them with those obtained from flow harmonic. In Fig.~\ref{fig1},  $\C_n e^{in\Phi_n}$ fluctuation normalized cumulants (blue empty circles) and $\epsilon_n e^{in\phi_n}$ fluctuation normalized cumulants (magenta squares) are depicted. The observed difference between $\epsilon_n e^{in\phi_n}$ and $v_n e^{in\psi_n}$ cumulants is interpreted as the presence of nonlinear terms similar to what is mentioned in Eqs.~\eqref{linearAndNonlinear}. The comparison of $\C_n e^{in\Phi_n}$ and $v_n e^{in\psi_n}$ gives us a hint about the nonlinear terms similar to what has been shown in Eqs.~\eqref{linearAndNonlinearCn}. 

We see in the figure that some of the normalized cumulants that are computed from $\epsilon_n e^{in\phi_n}$ have a different sign from those calculated from $v_n e^{in\psi_n}$ (see for instance panels (5), (6), (22), and (30), in Fig.~\ref{fig1}). In fact, the ``wrong'' sign of $\epsilon_n e^{in\phi_n}$ fluctuation in participant plane correlation has been observed in the previous studied \cite{Aad:2014fla,Teaney:2013dta,Teaney:2012ke,Qiu:2012uy,Qiu:2013wca,Qian:2016fpi}. In particular, the quantities $\chi_{422}$ and $\chi_{523}$ in \cite{Qian:2016fpi} are similar to $nc_{2,4}^{\{4\}}\{2,1\}$ and $nc_{2,3,5}^{\{-3,5\}}\{1,1,1\}$, up to a normalization factor. A sign difference between initial and final state fluctuations has been observed for these quantities. The authors of Ref.~\cite{Qian:2016fpi} conclude that this sign difference is a signature of hydrodynamic response to the initial state. Up to a normalization factor, the same quantities have been studied in \cite{Teaney:2013dta}. Again a sign change has been observed between eccentricity and flow harmonic fluctuations while the sign change is resolved by replacing $\epsilon_n e^{in\phi_n}$ with $\C_n e^{in\Phi_n}$. However,  we cannot conclude that replacing $\epsilon_n e^{in\phi_n}$ with $\C_n e^{in\Phi_n}$ always leads to a compatible correlation sign with final state fluctuation. For instance, one can find cases in panels (18), (23), and (28) of Fig.~\ref{fig1} that $\C_n e^{in\Phi_n}$ correlations have opposite sign compared to flow harmonic fluctuations. Few examples can be found in \cite{Teaney:2013dta} as well.\footnote{We have observed that if we \textit{effectively} assume the hydrodynamic couplings $k_4$ and $k_5$ are negative in many event averages, \textit{all} the observables in Fig.~\ref{fig1} reveals a compatible correlation sign with $v_n e^{in\psi_n}$ fluctuation at the linear hydrodynamic response approximation. Further investigations are needed in this regard.}

The cumulants calculated from hydrodynamic nonlinear response estimation are shown with black filled triangles in Fig.~\ref{fig1}, where the response couplings are those displayed in Fig.~\ref{HydroCouplingFigP}. The figure shows that the two first cumulants ($c_{2,4}^{\{4\}}\{2,1\}$ and $c_{2,3,5}^{\{-3,5\}}\{1,1,1\}$) have a perfect agreement with the nonlinear response estimation. These two cumulants are inputs to estimate the response couplings. Since normalized cumulants are plotted in Fig.~\ref{fig1}, only two ratios $w_{4(22)}/w_4$ and $w_{5(23)}/w_5$ play a role. As a result, two normalized cumulants are enough to fix these ratios. The rest of the normalized cumulants computed from the nonlinear response are, in fact, the predictions. Some of these predictions perfectly match  with hydrodynamic computations. For instance, $nc_{3,5}^{\{0\}}\{2,2\}$, $nc_{2,4}^{\{4\}}\{4,1\}$ and $nc_{3,4,5}^{\{-4,10\}}\{2,1,2\}$ has shown in panels (15), (22), and (30) in Fig.~\ref{fig1}. There are cases with a poor agreement between hydrodynamic simulation and the nonlinear response estimation, namely $nc_4\{4\}$ 
and $nc_5\{4\}$ (panels (9) and (10) in Fig.~\ref{fig1}). The nonlinear terms could not cure the initial and final correlation sign differences in panels (18), (23), and (28). Including more nonlinear terms would increase the accuracy of the latter cases.

\begin{figure}
	%\vspace*{-1.6cm}
	%	\hspace*{-1.0cm}
	\begin{tabular}{c}
		\includegraphics[width=8.3cm]{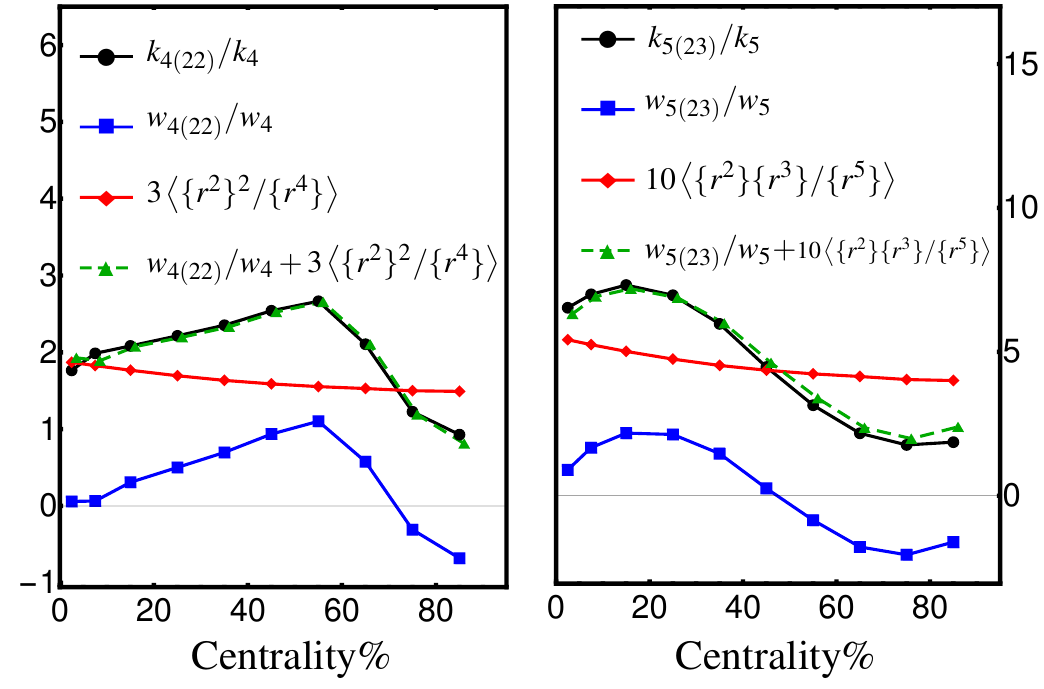}\\	
	\end{tabular}	
	\caption{The {\tt T$_\text{R}$ENTo} initial state radial shape (diamond points), and the nonlinear over linear hydrodynamic response couplings from {\tt VISH2+1}+{\tt UrQMD} (circle and square points).}
	\label{CouplingRatios}
\end{figure}

 Concerning the interpretation of the nonlinear hydrodynamic response coefficients, we note that the true deformation of the initial energy density is quantified by cumulants $\C_{n} e^{in\Phi_n}$. One expects that the final anisotropy to be proportional to the true deformation \cite{Teaney:2010vd}. Assuming $w_{4(22)}$ and $w_{5(23)}$ have solely collective evolution contributions, the nonlinear couplings $k_{4(22)}$ and $k_{5(23)}$ in Eq.~\eqref{linearAndNonlinear} have contributions from both initial shape and collective evolution. One can relate these two couplings  by substituting cumulants \eqref{ieCumul} into Eqs.~\eqref{linearAndNonlinearCn} and compare them with Eqs.~\eqref{linearAndNonlinear}. The linear couplings are identical in both conventions, $w_n=k_n$. However, we obtain the following relation for the nonlinear couplings:
\begin{subequations}\label{couplingRelation}
	\begin{align}
	\frac{k_{4(22)}}{k_4}&=\frac{w_{4(22)}}{w_4}+3\frac{\{r^2\}^2}{\{r^4\}},\\
	\frac{k_{5(23)}}{k_4}&=\frac{w_{5(23)}}{w_5}+10\frac{\{r^2\} \{r^3\}}{\{r^5\}}.
	\end{align}
\end{subequations}  
The above relations mean that the coefficients $k_{4(22)}$ and $k_{5(23)}$ receive contributions from the radial shape of the initial energy density as well. In Fig.~\ref{CouplingRatios}, the coupling ratios $w_{4(22)}/w_4$ and $w_{5(23)}/w_5$ are plotted with square points (blue curves). The couplings $k_{4(22)}/k_4$ and $k_{5(23)}/k_5$ (shown by circles, black curves) are estimated by solving six equations in Eqs.~\eqref{58} and \eqref{59} where energy density cumulants are replaced by eccentricities. Since the coupling ratio estimations are estimated from many events in a given centrality, we compute the right-hand side of Eqs.~\eqref{couplingRelation}, where the initial energy density contributions are averaged over events in the given centrality class. The quantities $3\left\la\{r^2\}^2/\{r^4\}\right\ra$ and $10\left\la\{r^2\} \{r^3\}/\{r^5\}\right\ra$ (shown by diamonds, red curve) are directly computed from {\tt T$_\text{R}$ENTo} events. The estimated $k_{4(22)}/k_4$ and $k_{5(23)}/k_5$ from Eqs.~\eqref{couplingRelation} are shown by triangles (green curves) in Fig.~\ref{CouplingRatios}. We see a rather perfect match between black curves and green curves, as we expected. From the figure, we see a nontrivial centrality dependence of quantities $3\left\la\{r^2\}^2/\{r^4\}\right\ra$ and $10\left\la\{r^2\} \{r^3\}/\{r^5\}\right\ra$. This observation indicates that there are contributions from the radial shape of the initial energy density in the values of  $k_{4(22)}$ and $k_{5(23)}$. This contribution should be taken into account in interpreting the nonlinear couplings $k_{4(22)}$ and $k_{5(23)}$ as hydrodynamic response couplings.

\section{Flow-induced genuine three-particle correlation }\label{threeParticle}

The observation of long-range correlations between particles at $\Delta\varphi\simeq0$ and nonzero $\Delta\eta$ is one of the first and most important fluid(-like) signals in large (small) system collisions \cite{Khachatryan:2010gv,Aad:2015gqa,Aad:2012gla,Zhu:2013zgd,Aamodt:2011by}. In particular, function $C(\Delta\eta,\Delta\varphi)$, the two-particle correlation function (2PC),\footnote{In the present work, we refer to the quantity in Eq.~\eqref{QVecsIIp} as correlations, while $C(\Delta\eta,\Delta\phi)$ is called correlation functions. } quantifies the correlation between two particles in the final state. In the present section, we introduce a generalized version of this function and study its relation with flow harmonic cumulants.

Ignoring the experimental complications, in principle, the function $C(\Delta\eta,\Delta\phi)$ is measured as follows: we choose all distinguished pairs of particles in an event and compute $\Delta\eta$ and $\Delta\varphi$ for each pair and fill a histogram from a collection of pairs in many events. To find the connection between 2PC with cumulants, we focus only on the $\Delta\varphi$ part of the correlation. In other words, we choose all the distinguished particles and compute $\Delta\varphi$ irrespective of their position in the $\eta$ direction. One can use $\eta$-gaps (see Refs.~\cite{Jia:2017hbm,Adler:2002pu}) to decrease the contamination of nonflow effects in the correlation function estimation. For two-particle correlation, we find the well-known relation between 2PC and second-order cumulants $c_n\{2\}$,
\bea\label{12p}
C_2(\Delta \varphi)=\frac{1}{(2\pi)^2}\left[1+2\sum_{n=1}^{\infty} c_n\{2\}\cos(n\Delta\varphi)\right].
\eea
Here, we have used the notation $C_2(\Delta\varphi)$ instead of commonly used notation $C(\Delta\varphi)$ for future generalization. The advantage of measuring $C_2(\Delta \varphi)$ compared to $c_n\{2\}$ is that it contains a cumulative information of all second order cumulants $c_n\{2\}$ with $n>0$. 

\begin{figure}
	%\vspace*{-1.6cm}
	%	\hspace*{-1.0cm}
	\begin{tabular}{c}
		\includegraphics[width=8.5cm]{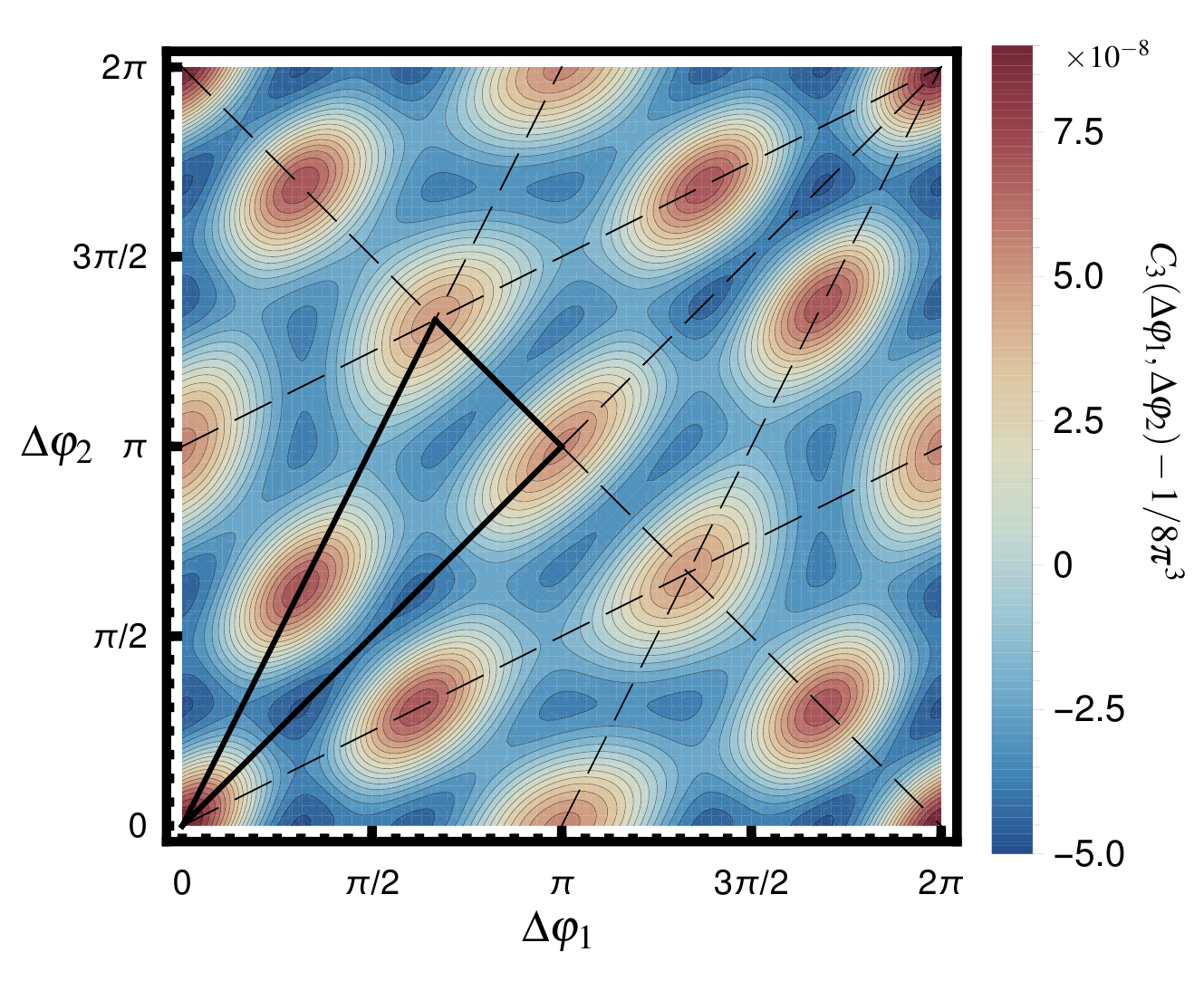}\\
	\end{tabular}	
	\caption{ Genuine three-particle correlation function, 3PC. The correlation is obtained from the {\tt T$_\text{R}$ENTo}+{\tt VISH2+1}+{\tt UrQMD} simulation for Pb--Pb collision at $\sqrt{s_{\text{NN}}}=2.76\,$TeV in the 10$-$20\% centrality class.}
	\label{threeCorrFig1}
\end{figure}

We can extend the notion of two-particle correlation function into $q$-particle correlation function ($q$PC). Specifically, in the following, we focus on correlations of three particles in the final state, $C_3(\Delta \varphi_1,\Delta \varphi_2)$ because it is a function of two variables
\bea\label{fiLable}
\Delta\varphi_1=\varphi_{2}-\varphi_{1},\qquad \Delta\varphi_2=\varphi_{3}-\varphi_{1},
\eea
and easy to visualize. For the same reason (and finding a clear connection to the flow harmonic cumulants), we ignore the $\eta$ dependence similar to Eq.~\eqref{12p}. The $\eta$-gap method can also be used in this case to decrease the nonflow effects as well.
Employing the systematic study of multiharmonic cumulants presented in this manuscript,
we can find an expansion of any $q$PC in terms $q$-order cumulants, similar to what has been written in Eq.~\eqref{12p}. The technical details of finding the relation between $q$PC and $q$ order cumulants can be found in appendix \ref{appBp}. Here, we show the final result for 3PC where harmonics $n=2,\ldots,5$ are involved,  
\begin{equation} \label{c9p}
\begin{split}
&C_3(\Delta \varphi_1,\Delta \varphi_2)=\\
&\frac{1}{(2\pi)^3}\bigg[1+2c_{2,4}^{\{4\}}\{2,1\}\,\Phi_{2,4}^{\{4\}}(\Delta\varphi_1,\Delta\varphi_2)\\
&\hspace*{1cm}+2c_{2,3,5}^{\{-3,5\}}\{1,1,1\}\,\Phi_{2,3,5}^{\{-3,5\}}(\Delta\varphi_1,\Delta\varphi_2)\bigg].
\end{split}
\end{equation}
The above relation is the correlation function of three particles when all lower-order correlations are removed (see appendix \ref{AppC2}).
The functions $\Phi_{2,4}^{4}$ and $\Phi_{2,3,5}^{\{-3,5\}}$ can be found in Eq.~\eqref{c5pp}. 
By employing $c_{2,4}^{\{4\}}\{2,1\}$ and $c_{2,3,5}^{\{-3,5\}}\{1,1,1\}$ from {\tt T$_\text{R}$ENTo}+{\tt VISH2+1}+{\tt UrQMD} simulation, we have plotted $C_3(\Delta \varphi_1,\Delta \varphi_2)$ in Fig.~\ref{threeCorrFig1}. Here, we have subtracted $1/8\pi^3$ to focus only on the nontrivial correlation. 

 As seen in Fig.~\ref{threeCorrFig1}, the correlation reveals repeating patterns which is a consequence of symmetries. Before explaining these symmetries in three-particle correlations, let us discuss them in a more simple case, two-particle correlation functions. To measure two-particle correlation functions, two (charged) particles in a given event are chosen and compute $\Delta\varphi=\varphi_2-\varphi_1$. The signal distribution is obtained by measuring $\Delta\varphi$ for many events.\footnote{To suppress the combinatorial backgrounds and acceptance effects, a background distribution is constructed from  $\Delta\varphi=\varphi_2-\varphi_1$ when one particle is chosen from the given even and the second particle from several randomly selected other events. The ratio of signal and background distribution is proportional to $C_2(\Delta\varphi)$ \cite{Chatrchyan:2011eka}.} Let us call two distinguished particles in a given event as $\alpha$ and $\beta$ with azimuthal angles $\varphi_\alpha$ and $\varphi_\beta$. Choosing all pairs of distinguished particles in an event and call them as particle 1 and 2, the particle $\alpha$ is labeled as particle 1 and $\beta$ as particle 2 once and the  particle $\alpha$ as particle 2 and $\beta$ as particle 1 again. The former labeling leads to $\Delta\varphi=\varphi_2-\varphi_1=\varphi_\beta-\varphi_\alpha$ and the latter to $\Delta\varphi'=\varphi_2-\varphi_1=\varphi_\alpha-\varphi_\beta=-\Delta\varphi$. As a result, in the signal distribution both $\Delta\varphi$ and $-\Delta\varphi$ have contribution from a single configuration of particles $\alpha$ and $\beta$.
This labeling brings us to the following symmetry, $C_2(\Delta\varphi)=C_2(-\Delta\varphi)$. One notes that the mentioned ``double counting'' is also considered in $c_n\{2\}= \la \la 2 \ra_{-n,n} \ra$  (see Eq.~\eqref{QVecsIIpp}). As a result, this symmetry is manifestly true in Eq.~\eqref{12p}. We also have an obvious periodic symmetry $C_2(\Delta\varphi)=C_2(\Delta\varphi+2n\pi)$ for any integer $n$. Using these two symmetries, we find that we have independent information in $C_2(\Delta\varphi)$ in the range $0<\Delta\varphi<\pi$.  The correlation $C_2(\Delta\varphi)$ (or $C_2(\Delta\varphi,\Delta\eta)$) is mostly reported in the range $-\frac{\pi}{2} \lesssim \Delta\varphi\lesssim \frac{3\pi}{2}$ to see a clear ridge and shoulder structures (see for instance \cite{Aamodt:2011by}). Due to the symmetries explained above, both ridge and shoulder structures are symmetric with respect to the axes pass through their peaks at $\Delta\varphi=0$ and $\Delta\varphi=\pi$ and only half of these structures contain independent information in the range $-\frac{\pi}{2} < \Delta\varphi< \frac{3\pi}{2}$.

When we have three particles in the final state, there are $3!$ possible ways to label three distinguished particles, leading to six different values for $(\Delta\varphi_1,\Delta\varphi_2)$ with a single configuration of particles. Let us call three particles as $\alpha$, $\beta$, and $\gamma$ with azimuthal angles
$\varphi_\alpha$, $\varphi_\beta$, and $\varphi_\gamma$. If particles $\alpha$, $\beta$, and $\gamma$ are labeled by 1, 2, and 3 (see Eq.~\eqref{fiLable}), we obtain $\Delta\varphi_1=\varphi_\beta-\varphi_\alpha$, and $\Delta\varphi_2=\varphi_\gamma-\varphi_\alpha$. The particles $\alpha$, $\beta$ and $\gamma$ can be labeled as 2, 1, and 3. With this labeling, we have $\Delta\varphi'_1=\varphi_\alpha-\varphi_\beta=-\Delta\varphi_1$ and $\Delta\varphi'_2=\varphi_\gamma-\varphi_\beta=\Delta\varphi_2-\Delta\varphi_1$. Both values for $(\Delta\varphi_1,\Delta\varphi_2)$ and $(\Delta\varphi'_1,\Delta\varphi'_2)$ refer to the same configuration of particles and exist in the signal distribution. As a result, the final distributions have the following symmetry: $C_3(\Delta\varphi_1,\Delta\varphi_2)=C_3(-\Delta\varphi_1,\Delta\varphi_2-\Delta\varphi_1)$. Another case is labeling $\alpha$, $\beta$, and $\gamma$ particles with 1, 3, and 2. This labeling leads to the values $\Delta\varphi''_1=\varphi_\gamma-\varphi_\alpha=\Delta\varphi_2$ and $\Delta\varphi''_2=\varphi_\beta-\varphi_\alpha=\Delta\varphi_1$. Consequently, the distribution has symmetry $C_3(\Delta\varphi_1,\Delta\varphi_2)=C_3(\Delta\varphi_2,\Delta\varphi_1)$. There are three other permutations, but all of them lead to a combination of symmetries explained above. 
More than the above symmetries, for any configuration of particles, we expect the same probability for the mirrored configurations, $\varphi_i\to -\varphi_i$, which leads to the symmetry $C_3(\Delta\varphi_1,\Delta\varphi_2)= C_3(-\Delta\varphi_1,-\Delta\varphi_2)$. Finally, we have rotational symmetry $\varphi_i\to\varphi_i+2n_i\pi$ for any integer $n_i$, which eventually leads to the periodic condition for $C_3(\Delta\varphi_1,\Delta\varphi_2)$ with periodicity $2\pi$. We summarize all the symmetries of 3PC as follows,
\bea\label{symmetries}
\begin{cases} 
	C_3(\Delta\varphi_1,\Delta\varphi_2)= C_3(\Delta\varphi_2,\Delta\varphi_1),\\
	C_3(\Delta\varphi_1,\Delta\varphi_2)= C_3(-\Delta\varphi_1,\Delta\varphi_2-\Delta\varphi_2),\\
	C_3(\Delta\varphi_1,\Delta\varphi_2)= C_3(-\Delta\varphi_1,-\Delta\varphi_2),\\
	C_3(\Delta\varphi_1,\Delta\varphi_2)= C_3(\Delta\varphi_1+2n_1\pi,\Delta\varphi_2+2n_2\pi).
\end{cases}
\eea

\begin{figure}
	%\vspace*{-1.6cm}
	%	\hspace*{-1.0cm}
	\begin{tabular}{c}
		\includegraphics[width=7.5cm]{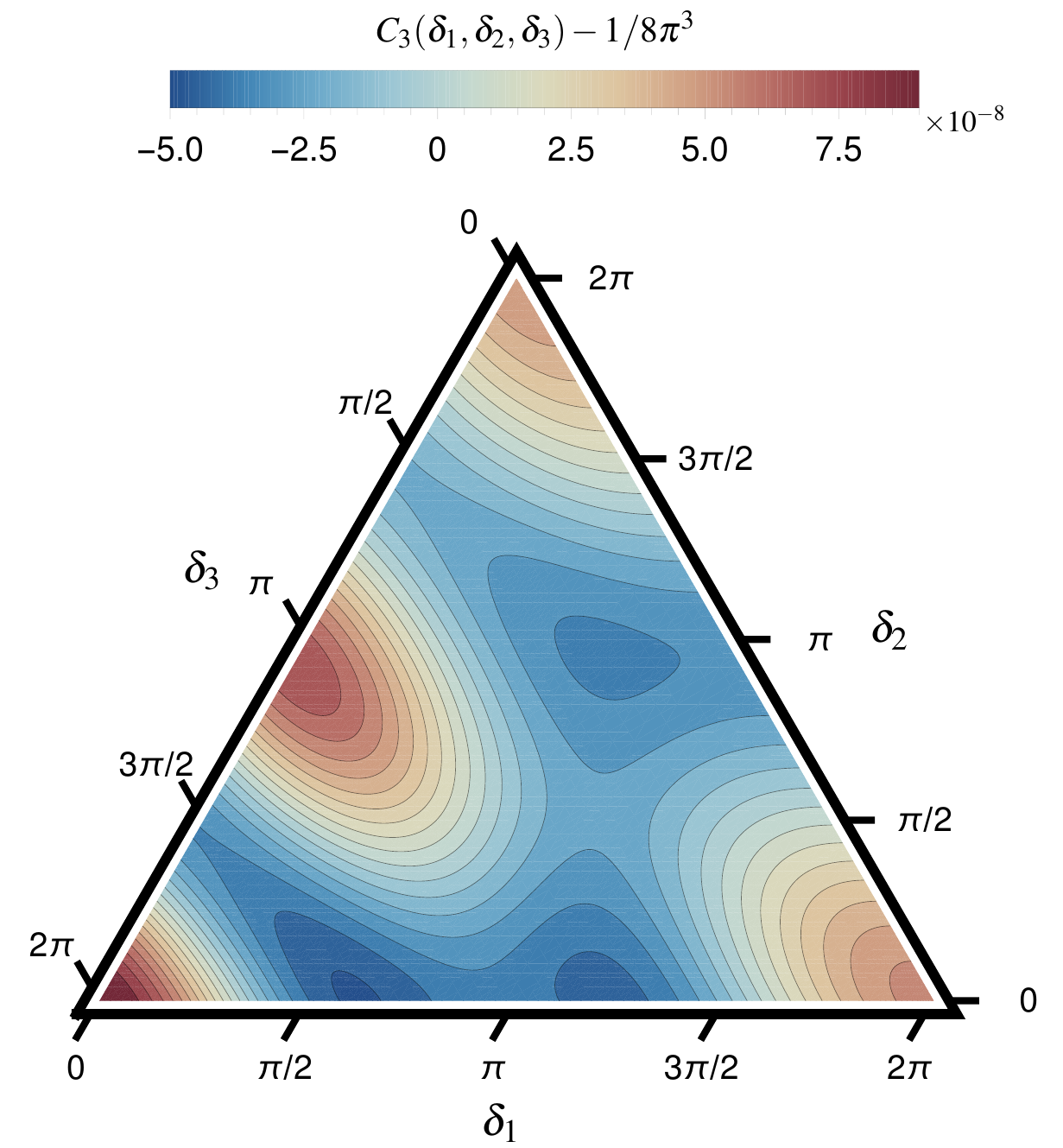}  
	\end{tabular}	
	\caption{Dalitz-like plot of the ``unit-cell'' (black triangle in Fig.~\ref{threeCorrFig1}) of genuine three-particle correlation function. }
	\label{threeCorrFig}
\end{figure}

The symmetries in Eq.~\eqref{symmetries} are responsible of repeating patterns observed in Fig.~\ref{threeCorrFig1}. Each dashed line in the figure refers to one (or combination of more than one) symmetry(ies). As seen in the figure, the range $0<\Delta\varphi_1,\Delta\varphi_2<2\pi$ is divided into twelve repeating triangular regions. In the experiment and the simulations, we lose the statistics twelve times by considering the full region  without gaining any more information. 
For that reason, we confine ourselves into a ``unit cell'' of $\Delta \varphi_1$ and $\Delta \varphi_2$ containing all the nontrivial non-repeating patterns. Here, we conventionally choose the following region, 
\begin{equation} \label{48}
\begin{cases} 
\Delta\varphi_2<2\Delta\varphi_1,\\
\Delta\varphi_2>\Delta\varphi_1,\\
\Delta\varphi_2<-\Delta\varphi_1+2\pi,
\end{cases}
\end{equation}
as it is shown by a black triangle in Fig.~\ref{threeCorrFig1}. If any choice of particles leads to $(\Delta\varphi_1,\Delta\varphi_2)$ outside this triangle, the symmetries mentioned in Eq.~\eqref{symmetries} can be used to map that point inside it. 
The edges of the black triangle in Fig.~\ref{threeCorrFig1} refer to the combinations of the angles, $\Delta\varphi_1$ and $\Delta\varphi_2$ with unequal lengths. 
It will have more symmetric visualization if we use the following variables,
\begin{equation} \label{49}
\begin{split}
\delta_1&=4 \Delta\varphi_1-2\Delta\varphi_2,\\
\delta_2&= 3\Delta\varphi_2- \Delta\varphi_1,\\
\delta_3&=2\pi-\Delta\varphi_2- \Delta\varphi_1,
\end{split}
\end{equation} 
where $0<\delta_1,\delta_2,\delta_2<2\pi$ and  $\delta_1+\delta_2+\delta_3=2\pi$. Using this variables, we can plot the unit cell of $C_3(\Delta \varphi_1,\Delta \varphi_2)$ in a Dalitz-like plot as it is shown in Fig.~\ref{threeCorrFig}.

 Similar to $C_2(\Delta\varphi)$ which has cumulative information from all second-order cumulants, experimental measurement of $C_3(\Delta\varphi_1,\Delta\varphi_2)$ (or equivalently $C_3(\delta_1,\delta_2,\delta_3)$) contains a cumulative information of all third-order cumulants at the same time.  Measuring these correlations in large and small systems can be used as an independent method of testing the collectivity and event-by-event fluctuation in heavy-ion experiments. We leave more investigation of this measurable quantity to future studies.

\section{Conclusion}\label{Conclusion}

 A multidimensional generating function method was introduced to extract a large class of cumulants related to the flow harmonic fluctuations. We proposed an ordering based on the Fourier-cumulant expansion for these observables to systematically capture the most dominant features of the flow harmonic fluctuations. Using this method and reproducing the already known cumulants, we have found new cumulants consisting of symmetry plane correlations that have not been studied before. We defined the normalized cumulants to compare the cumulant's magnitude with each other and with initial state fluctuation. We employed hydrodynamic simulation for Pb--Pb collisions ({\tt T$_\text{R}$ENTo}+{\tt VISH2+1}+{\tt UrQMD}) calibrated by a Global Bayesian analysis to predict the unmeasured normalized cumulants' value. The observables introduced in this study can be used as inputs for calibration of the Bayesian analysis or to validate the already tuned parameters. 
  
We extract the linear and nonlinear hydrodynamic response by comparing the initial anisotropy and flow harmonic fluctuation cumulants. This method can be extended to obtain higher-order nonlinear terms by comparing more initial and final state fluctuations cumulants.
  
Based on the Fourier-cumulant expansion, we also introduced a general way to find the genuine $q$-particle distribution function, $q$PC. In particular, we studied 3PC containing information of all third-order cumulants for all harmonics. The flow-induced 3PC for harmonics $n=2,\cdots,5$ was presented by using  {\tt T$_\text{R}$ENTo}+{\tt VISH2+1}+{\tt UrQMD} simulation. This measurable helps to study flow and nonflow effects in large and small systems in the future. 

\section*{Acknowledgment}

The author thanks Ante Bilandzic, Cindy Mordasini, Marcel Lesch and other members of multiparticle correlations group (MuPa) at the Technical University of Munich, also Jiangyong Jia, Jing Qian and Ulrich Heinz for useful discussions. This project has received funding from the European Research Council (ERC) under the European Unions Horizon 2020 research and innovation programme (grant agreement No 759257).

\appendix

\onecolumn

\section{List of few first order cumulants for harmonics $2$ to $6$}\label{allCumulApp}

By employing the {\tt Mathematica} package introduced in section~\ref{package}, we tabulate the cumulants containing one and two harmonics up eight order. The cumulants harmonics containing three and four harmonics are presented in up to six orders. Except for the one-harmonic, the cumulants' final expression is mostly dependent on the values of $n_1,\ldots,n_k$. There are still repeating patterns in the expressions, like $c_{n_1,n_2}^{\{0\}}\{2,2\}=\la v_{n_1}^2 v_{n_2}^2 \ra-\la v_{n_1}^2\ra \la v_{n_2}^2 \ra$. However, we sacrifice brevity for the sake of clarity and dedicate a separate table for each combination of flow harmonics regardless of repeating patterns.

%\FloatBarrier
\begin{table*}[!ht]
	\raggedright
	%	\centering
	\setlength{\arrayrulewidth}{1.2pt}
	\caption{One-harmonic cumulants up to order eight}
	\scriptsize
\begin{tabular}[t]{p{1cm} p{16.5cm}}
		\hline
		$\{m\}$ & $c_n\{m\}$\\
		\hline
		$\{2\}$ & $\la v_n^2 \ra $\\
		\Xhline{0.3\arrayrulewidth}
		$\{4\}$ & $\la v_n^4 \ra-2\la v_n^2 \ra^2 $\\
		\Xhline{0.3\arrayrulewidth}
		$\{6\}$ & $\la v_n^6\ra-9 \la v_n^4\ra   \la v_n^2\ra  + 12 \la v_n^2\ra^3 $\\
		\Xhline{0.3\arrayrulewidth}
		$\{8\}$ & $\la  v_n^8\ra  -18 \la  v_n^4\ra^2 -16 \la  v_n^6\ra   \la  v_n^2\ra +144 \la  v_n^4\ra   \la  v_n^2\ra^2-144 \la  v_n^2\ra^4  $\\
		\hline
	\end{tabular}
	\label{one-harmonic}
\end{table*}%
\FloatBarrier
\begin{table*}[!ht]
	\raggedright
	%	\centering
	\setlength{\arrayrulewidth}{1.2pt}
	\caption{Two-harmonic cumulants up to order eight}
	\scriptsize
	\begin{tabular}[t]{l p{0.7cm} p{15.8cm}}
		\hline
		$\{m_1,m_2\}$ & $\{\alpha\}$ & $c_{2,3}^{\{\alpha\}}\{m_1,m_2\}$\\
		\hline
		$\{2,2\}$ & $\{0\}$ & $  \la  v_2^2 v_3^2\ra  - \la  v_2^2\ra   \la  v_3^2\ra  $\\
		\Xhline{0.3\arrayrulewidth}
		$\{3,2\}$ & $\{6\}$   & $ \la  v_2^3 v_3^2 \cos \left(6 \left(\psi _2-\psi _3\right)\right)\ra  $\\
		\Xhline{0.3\arrayrulewidth}
		$\{2,4\}$ & $\{0\}$   & $ \la  v_2^2 v_3^4\ra -4 \la v_3^2\ra  \la v_2^2
		v_3^2\ra -\la v_2^2\ra  \la v_3^4\ra +4 \la v_2^2\ra  \la v_3^2\ra^2$ \\
		\Xhline{0.3\arrayrulewidth}
		$\{4,2\}$ & $\{0\}$   & $ \la v_2^4 v_3^2\ra -\la v_2^4\ra  \la v_3^2\ra -4 \la v_2^2\ra  \la v_2^2 v_3^2\ra +4 \la v_2^2\ra ^2\la v_3^2 \ra $\\
		\Xhline{0.3\arrayrulewidth}
		$\{3,4\}$ & $\{6\}$   & $ \la  v_2^3 v_3^4 \cos \left(6 \left(\psi _2-\psi _3\right)\right)\ra  -3 \la 
		v_3^2\ra   \la  v_2^3 v_3^2 \cos \left(6 \left(\psi _2-\psi
		_3\right)\right)\ra  $ \\
		\Xhline{0.3\arrayrulewidth}
		$\{5,2\}$ & $\{6\}$   & $ \la  v_2^5 v_3^2 \cos \left(6 \left(\psi _2-\psi _3\right)\right)\ra  -4 \la 
		v_2^2\ra   \la  v_2^3 v_3^2 \cos \left(6 \left(\psi _2-\psi
		_3\right)\right)\ra   $\\	
		\Xhline{0.3\arrayrulewidth}	
		$\{2,6\}$ & $\{0\}$   & $ \la  v_2^2 v_3^6\ra  -9 \la  v_2^2 v_3^2\ra   \la  v_3^4\ra  -9 \la 
		v_3^2\ra   \la  v_2^2 v_3^4\ra  -\la  v_2^2\ra   \la  v_3^6\ra  +36
		\la  v_3^2\ra ^2 \la  v_2^2 v_3^2\ra  +18 \la  v_2^2\ra   \la 
		v_3^2\ra   \la  v_3^4\ra  -36 \la  v_2^2\ra   \la  v_3^2\ra ^3 $ \\
\hline
	\end{tabular}
	\label{two-harmonics}
\end{table*}%
\FloatBarrier
\begin{table*}[!h]
	\raggedright
	%	\centering
	\setlength{\arrayrulewidth}{1.2pt}
	%\caption{two-harmonics cumulants}
	\scriptsize	
	\begin{tabular}[t]{l p{0.7cm} p{15.8cm}}
		\hline
		$\{m_1,m_2\}$ & $\{\alpha\}$  & $c_{2,4}^{\{\alpha\}}\{m_1,m_2\}$\\
		\hline
		$\{2,1\}$ & $\{4\}$   & $  \la  v_2^2 v_4 \cos \left(4 \left(\psi _2-\psi _4\right)\right)\ra  $\\
		\Xhline{0.3\arrayrulewidth}	
		$\{2,2\}$ & $\{0\}$    & $ \la  v_2^2 v_4^2\ra  -\la  v_2^2\ra   \la 
		v_4^2\ra   $\\
		\Xhline{0.3\arrayrulewidth}	
		$\{2,3\}$ & $\{4\}$   & $ \la  v_2^2 v_4^3 \cos \left(4 \left(\psi _2-\psi _4\right)\right)\ra  -2 \la 
		v_4^2\ra   \la  v_2^2 v_4 \cos \left(4 \left(\psi _2-\psi
		_4\right)\right)\ra   $\\	
		\Xhline{0.3\arrayrulewidth}	
		$\{4,1\}$ & $\{4\}$    & $ \la  v_2^4 v_4 \cos \left(4 \left(\psi _2-\psi _4\right)\right)\ra  -3 \la 
		v_2^2\ra   \la  v_2^2 v_4 \cos \left(4 \left(\psi _2-\psi
		_4\right)\right)\ra   $\\	
		\Xhline{0.3\arrayrulewidth}	
		$\{4,2\}$ & $\{8\}$   & $ \la  v_2^4 v_4^2 \cos \left(8 \left(\psi _2-\psi _4\right)\right)\ra  -6 \la 
		v_2^2 v_4 \cos \left(4 \left(\psi _2-\psi _4\right)\right)\ra ^2  $\\	
		\Xhline{0.3\arrayrulewidth}	
		$\{2,4\}$ & $\{0\}$   & $ \la  v_2^2 v_4^4\ra  -4 \la  v_4^2\ra   \la  v_2^2
		v_4^2\ra  -\la  v_2^2\ra   \la  v_4^4\ra  +4
		\la  v_2^2\ra   \la  v_4^2\ra ^2 $\\
		\Xhline{0.3\arrayrulewidth}		
		$\{4,2\}$ & $\{0\}$   & $  \la  v_2^4 v_4^2\ra  -\la  v_2^2 v_4 \cos \left(4 \left(\psi _2-\psi
		_4\right)\right)\ra^2-\la  v_2^4\ra   \la  v_4^2\ra 
		-4 \la  v_2^2\ra   \la  v_2^2 v_4^2\ra  +4 \la 
		v_2^2\ra^2 \la  v_4^2\ra  $\\
\hline
	\end{tabular}
\end{table*}%
\FloatBarrier
\begin{table*}[h!]
	\raggedright
	%	\centering
	\setlength{\arrayrulewidth}{1.2pt}
	%	\caption{two-harmonics cumulants}
	\scriptsize
	\begin{tabular}[t]{l p{0.7cm} p{15.8cm}}
		\hline
		$\{m_1,m_2\}$ & $\{\alpha\}$  & $c_{2,5}^{\{\alpha\}}\{m_1,m_2\}$\\
		\hline
		$\{2,2\}$ & $\{0\}$   & $ \la  v_2^2 v_5^2\ra  -\la  v_2^2\ra   \la  v_5^2\ra  $\\
		\Xhline{0.3\arrayrulewidth}	
		$\{2,4\}$ & $\{0\}$    & $ \la  v_2^2 v_5^4\ra  -4 \la  v_5^2\ra   \la  v_2^2 v_5^2\ra  -\la 
		v_2^2\ra   \la  v_5^4\ra  +4 \la  v_2^2\ra   \la  v_5^2\ra ^2 $\\
		\Xhline{0.3\arrayrulewidth}	
		$\{4,2\}$ & $\{0\}$   & $ \la  v_2^4 v_5^2\ra  -\la  v_2^4\ra   \la  v_5^2\ra  -4 \la  v_2^2\ra  
		\la  v_2^2 v_5^2\ra  +4 \la  v_2^2\ra ^2 \la  v_5^2\ra  $\\
		\Xhline{0.3\arrayrulewidth}	
		$\{5,2\}$ & $\{10\}$    & $ \la  v_2^5 v_5^2 \cos \left(10 \left(\psi _2-\psi _5\right)\right)\ra  $\\
		\Xhline{0.3\arrayrulewidth}	
		$\{2,6\}$ & $\{0\}$   & $ \la  v_2^2 v_5^6\ra  -9 \la  v_2^2 v_5^2\ra   \la  v_5^4\ra  -9 \la 
		v_5^2\ra   \la  v_2^2 v_5^4\ra  -\la  v_2^2\ra   \la  v_5^6\ra  +36
		\la  v_5^2\ra ^2 \la  v_2^2 v_5^2\ra  +18 \la  v_2^2\ra   \la 
		v_5^2\ra   \la  v_5^4\ra  -36 \la  v_2^2\ra   \la  v_5^2\ra ^3 $\\
		\Xhline{0.3\arrayrulewidth}	
		$\{4,4\}$ & $\{0\}$   &  $ \la  v_2^4 v_5^4\ra  -8 \la  v_2^2 v_5^2\ra ^2-4 \la  v_5^2\ra   \la  v_2^4
		v_5^2\ra  -\la  v_2^4\ra   \la  v_5^4\ra  -4 \la  v_2^2\ra   \la 
		v_2^2 v_5^4\ra  +4 \la  v_2^4\ra   \la  v_5^2\ra ^2+32 \la  v_2^2\ra  
		\la  v_5^2\ra   \la  v_2^2 v_5^2\ra  +4 \la  v_2^2\ra ^2 \la 
		v_5^4\ra  -24 \la  v_2^2\ra ^2 \la  v_5^2\ra ^2 $\\	
		\Xhline{0.3\arrayrulewidth}	
		$\{6,2\}$ & $\{0\}$   & $ \la  v_2^6 v_5^2\ra  -\la  v_2^6\ra   \la  v_5^2\ra  -9 \la  v_2^4\ra  
		\la  v_2^2 v_5^2\ra  -9 \la  v_2^2\ra   \la  v_2^4 v_5^2\ra  +18 \la 
		v_2^2\ra   \la  v_2^4\ra   \la  v_5^2\ra  +36 \la  v_2^2\ra ^2 \la 
		v_2^2 v_5^2\ra  -36 \la  v_2^2\ra ^3 \la  v_5^2\ra  $\\			
		\Xhline{0.3\arrayrulewidth}	
		\hline
	\end{tabular}
\end{table*}%
\FloatBarrier
\begin{table*}[h!]
	\raggedright
	\setlength{\arrayrulewidth}{1.2pt}
	%	\caption{two-harmonics cumulants}
	\scriptsize
	\begin{tabular}[t]{l p{0.7cm} p{15.8cm}}
		\hline
		$\{m_1,m_2\}$ & $\{\alpha\}$  & $c_{2,6}^{\{\alpha\}}\{m_1,m_2\}$\\
		\hline
		$\{3,1\}$ & $\{6\}$   & $ \la  v_2^3 v_6 \cos \left(6 \left(\psi _2-\psi _6\right)\right)\ra  $ \\
		\Xhline{0.3\arrayrulewidth}	
		$\{2,2\}$ & $\{0\}$   & $ \la  v_2^2 v_6^2\ra  -\la  v_2^2\ra   \la  v_6^2\ra  $ \\
		\Xhline{0.3\arrayrulewidth}	
		$\{3,3\}$ & $\{6\}$   & $ \la  v_6^3 v_2^3 \cos \left(6 \left(\psi _2-\psi _6\right)\right)\ra  -2 \la  v_2^3 v_6 \cos \left(6 \left(\psi
		_2-\psi _6\right)\right)\ra   \la  v_6^2\ra  $ \\	
		\Xhline{0.3\arrayrulewidth}	
		$\{5,1\}$ & $\{6\}$   & $ \la  v_2^5 v_6 \cos \left(6 \left(\psi _2-\psi _6\right)\right)\ra  -4 \la  v_2^2\ra   \la 
		v_2^3 v_6 \cos \left(6 \left(\psi _2-\psi _6\right)\right)\ra  $ \\	
		\Xhline{0.3\arrayrulewidth}	
		$\{2,4\}$ & $\{0\}$   & $ \la  v_2^2 v_6^4\ra  -4 \la  v_6^2\ra   \la  v_2^2 v_6^2\ra  -\la 
		v_2^2\ra   \la  v_6^4\ra  +4 \la  v_2^2\ra   \la  v_6^2\ra ^2$ \\
		\Xhline{0.3\arrayrulewidth}	
		$\{4,2\}$ & $\{0\}$   & $ \la  v_2^4 v_6^2\ra  -\la  v_2^4\ra   \la  v_6^2\ra  -4 \la 
		v_2^2\ra   \la  v_2^2 v_6^2\ra  +4 \la  v_2^2\ra ^2 \la  v_6^2\ra  $ \\	
		\Xhline{0.3\arrayrulewidth}	
		$\{6,2\}$ & $\{12\}$   & $ \la  v_2^6 v_6^2 \cos \left(12 \left(\psi _2-\psi _6\right)\right)\ra -20 \la  v_2^3 v_6 \cos \left(6 \left(\psi _2-\psi
		_6\right)\right)\ra^2   $ \\
\hline
	\end{tabular}
\end{table*}%
\FloatBarrier
\begin{table*}[h!]
	\raggedright
	\setlength{\arrayrulewidth}{1.2pt}
	%\caption{two-harmonics cumulants}
	\scriptsize
	\begin{tabular}[t]{l p{0.7cm} p{15.8cm}}
		\hline
		$\{m_1,m_2\}$ & $\{\alpha\}$  & $c_{3,4}^{\{\alpha\}}\{m_1,m_2\}$\\
		\hline
		$\{2,2\}$ & $\{0\}$   & $ \la  v_3^2 v_4^2\ra  -\la  v_3^2\ra   \la  v_4^2\ra  $ \\
		\Xhline{0.3\arrayrulewidth}	
		$\{2,4\}$ & $\{0\}$   & $ \la  v_3^2 v_4^4\ra  -4 \la  v_4^2\ra   \la  v_3^2 v_4^2\ra  -\la 
		v_3^2\ra   \la  v_4^4\ra  +4 \la  v_3^2\ra   \la  v_4^2\ra^2 $ \\
		\Xhline{0.3\arrayrulewidth}	
		$\{4,2\}$ & $\{0\}$   & $ \la  v_3^4 v_4^2\ra  -\la  v_3^4\ra   \la  v_4^2\ra  -4 \la 
		v_3^2\ra   \la  v_3^2 v_4^2\ra  +4 \la  v_3^2\ra^2 \la  v_4^2\ra  $ \\	
		\Xhline{0.3\arrayrulewidth}	
		$\{4,3\}$ & $\{12\}$   & $ \la  v_3^4 v_4^3 \cos \left(12 \left(\psi _3-\psi _4\right)\right)\ra  $ \\	
		\Xhline{0.3\arrayrulewidth}	
		$\{2,6\}$ & $\{0\}$   & $ \la  v_3^2 v_4^6\ra  -9 \la  v_3^2 v_4^2\ra   \la  v_4^4\ra  -9 \la 
		v_4^2\ra   \la  v_3^2 v_4^4\ra  -\la  v_3^2\ra   \la  v_4^6\ra  +36
		\la  v_4^2\ra^2 \la  v_3^2 v_4^2\ra  +18 \la  v_3^2\ra   \la 
		v_4^2\ra   \la  v_4^4\ra  -36 \la  v_3^2\ra   \la  v_4^2\ra^3 $ \\
		\Xhline{0.3\arrayrulewidth}	
		$\{4,4\}$ & $\{0\}$   & $\la  v_3^4 v_4^4\ra  -8 \la  v_3^2 v_4^2\ra^2-4 \la  v_4^2\ra   \la 
		v_3^4 v_4^2\ra  -\la  v_3^4\ra   \la  v_4^4\ra  -4 \la  v_3^2\ra  
		\la  v_3^2 v_4^4\ra  +4 \la  v_3^4\ra   \la  v_4^2\ra^2+32 \la 
		v_3^2\ra   \la  v_4^2\ra   \la  v_3^2 v_4^2\ra  +4 \la  v_3^2\ra^2
		\la  v_4^4\ra  -24 \la  v_3^2\ra^2 \la  v_4^2\ra^2 $ \\	
		\Xhline{0.3\arrayrulewidth}	
		$\{6,2\}$ & $\{0\}$   & $ \la  v_3^6 v_4^2\ra  -\la  v_3^6\ra   \la  v_4^2\ra  -9 \la 
		v_3^4\ra   \la  v_3^2 v_4^2\ra  -9 \la  v_3^2\ra   \la  v_3^4 v_4^2\ra 
		+18 \la  v_3^2\ra   \la  v_3^4\ra   \la  v_4^2\ra  +36 \la 
		v_3^2\ra^2 \la  v_3^2 v_4^2\ra  -36 \la  v_3^2\ra^3 \la 
		v_4^2\ra  $ \\			
		\hline
	\end{tabular}
\end{table*}%
\FloatBarrier
\begin{table*}[h!]
	\raggedright
	\setlength{\arrayrulewidth}{1.2pt}
	%	\caption{two-harmonics cumulants}
	\scriptsize
	\begin{tabular}[t]{l p{0.7cm} p{15.8cm}}
		\hline
		$\{m_1,m_2\}$ & $\{\alpha\}$  & $c_{3,5}^{\{\alpha\}}\{m_1,m_2\}$\\
		\hline
		$\{2,2\}$ & $\{0\}$   & $ \la  v_3^2 v_5^2\ra  -\la  v_3^2\ra   \la  v_5^2\ra  $ \\
		\Xhline{0.3\arrayrulewidth}	
		$\{2,4\}$ & $\{0\}$   & $ \la  v_3^2 v_5^4\ra  -4 \la  v_5^2\ra   \la  v_3^2 v_5^2\ra  -\la  v_3^2\ra  
		\la  v_5^4\ra  +4 \la  v_3^2\ra   \la  v_5^2\ra^2$ \\
		\Xhline{0.3\arrayrulewidth}	
		$\{4,2\}$ & $\{0\}$   & $ \la  v_3^4 v_5^2\ra  -\la  v_3^4\ra   \la  v_5^2\ra  -4 \la  v_3^2\ra   \la 
		v_3^2 v_5^2\ra  +4 \la  v_3^2\ra^2 \la  v_5^2\ra  $ \\
		\Xhline{0.3\arrayrulewidth}		
		$\{5,3\}$ & $\{15\}$   & $ \la  v_3^5 v_5^3 \cos \left(15 \left(\psi _3-\psi _5\right)\right)\ra  $ \\
		\Xhline{0.3\arrayrulewidth}		
		$\{2,6\}$ & $\{0\}$   & $ \la  v_3^2 v_5^6\ra  -9 \la  v_3^2 v_5^2\ra   \la  v_5^4\ra  -9 \la  v_5^2\ra  
		\la  v_3^2 v_5^4\ra  -\la  v_3^2\ra   \la  v_5^6\ra  +36 \la  v_5^2\ra^2
		\la  v_3^2 v_5^2\ra  +18 \la  v_3^2\ra   \la  v_5^2\ra   \la  v_5^4\ra  -36
		\la  v_3^2\ra   \la  v_5^2\ra  $ \\
		\Xhline{0.3\arrayrulewidth}	
		$\{4,4\}$ & $\{0\}$   & $\la  v_3^4 v_5^4\ra  -8 \la  v_3^2 v_5^2\ra^2-4 \la  v_5^2\ra   \la  v_3^4 v_5^2\ra 
		-\la  v_3^4\ra   \la  v_5^4\ra  -4 \la  v_3^2\ra   \la  v_3^2 v_5^4\ra  +4
		\la  v_3^4\ra   \la  v_5^2\ra^2+32 \la  v_3^2\ra   \la  v_5^2\ra   \la 
		v_3^2 v_5^2\ra  +4 \la  v_3^2\ra^2 \la  v_5^4\ra  -24 \la  v_3^2\ra^2 \la 
		v_5^2\ra^2 $ \\
		\Xhline{0.3\arrayrulewidth}		
		$\{6,2\}$ & $\{0\}$   & $ \la  v_3^6 v_5^2\ra  -\la  v_3^6\ra   \la  v_5^2\ra  -9 \la  v_3^4\ra   \la 
		v_3^2 v_5^2\ra  -9 \la  v_3^2\ra   \la  v_3^4 v_5^2\ra  +18 \la  v_3^2\ra   \la 
		v_3^4\ra   \la  v_5^2\ra  +36 \la  v_3^2\ra^2 \la  v_3^2 v_5^2\ra  -36 \la 
		v_3^2\ra^3 \la  v_5^2\ra $ \\			
		\hline
	\end{tabular}
\end{table*}%
\FloatBarrier
\begin{table*}[h!]
	\raggedright
	\setlength{\arrayrulewidth}{1.2pt}
	%	\caption{two-harmonics cumulants}
	\scriptsize
	\begin{tabular}[t]{l p{0.7cm} p{15.8cm}}
		\hline
		$\{m_1,m_2\}$ & $\{\alpha\}$  & $c_{3,6}^{\{\alpha\}}\{m_1,m_2\}$\\
		\hline
		$\{2,1\}$ & $\{6\}$   & $ \la  v_3^2 v_6 \cos \left(6 \left(\psi _3-\psi _6\right)\right)\ra  $ \\
		\Xhline{0.3\arrayrulewidth}			
		$\{2,2\}$ & $\{0\}$   & $\la  v_3^2 v_6^2\ra  -\la  v_3^2\ra   \la  v_6^2\ra  $ \\
		\Xhline{0.3\arrayrulewidth}			
		$\{2,3\}$ & $\{6\}$   & $ \la  v_3^2 v_6^3 \cos \left(6 \left(\psi _3-\psi _6\right)\right)\ra  -2 \la  v_3^2 v_6 \cos \left(6 \left(\psi _3-\psi
		_6\right)\right)\ra   \la  v_6^2\ra  $ \\	
		\Xhline{0.3\arrayrulewidth}			
		$\{4,1\}$ & $\{6\}$   & $ \la  v_3^4 v_6 \cos \left(6 \left(\psi _3-\psi _6\right)\right)\ra  -3 \la  v_3^2\ra   \la  v_3^2 v_6 \cos \left(6
		\left(\psi _3-\psi _6\right)\right)\ra  $ \\
		\Xhline{0.3\arrayrulewidth}				
		$\{4,2\}$ & $\{12\}$   & $ \la  v_3^4 v_6^2 \cos \left(12 \left(\psi _3-\psi _6\right)\right)\ra  -6 \la  v_3^2 v_6 \cos \left(6 \left(\psi _3-\psi
		_6\right)\right)\ra ^2 $ \\
		\Xhline{0.3\arrayrulewidth}			
		$\{2,4\}$ & $\{0\}$   & $ \la  v_3^2 v_6^4\ra  -4 \la  v_6^2\ra   \la  v_3^2 v_6^2\ra  -\la  v_3^2\ra  
		\la  v_6^4\ra  +4 \la  v_3^2\ra   \la  v_6^2\ra ^2 $ \\
		\Xhline{0.3\arrayrulewidth}				
		$\{4,2\}$ & $\{0\}$   & $ \la  v_3^4 v_6^2\ra  -\la  v_3^2 v_6 \cos \left(6 \left(\psi _3-\psi _6\right)\right)\ra ^2-\la 
		v_3^4\ra   \la  v_6^2\ra  -4 \la  v_3^2\ra   \la  v_3^2 v_6^2\ra  +4 \la 
		v_3^2\ra ^2 \la  v_6^2\ra  $ \\
\hline
	\end{tabular}
\end{table*}%
\FloatBarrier
\begin{table*}[h!]
	\raggedright
	\setlength{\arrayrulewidth}{1.2pt}
	%	\caption{two-harmonics cumulants}
	\scriptsize
	\begin{tabular}[t]{l p{0.7cm} p{15.8cm}}
		\hline
		$\{m_1,m_2\}$ & $\{\alpha\}$  & $c_{4,5}^{\{\alpha\}}\{m_1,m_2\}$\\
		\hline
		$\{2,2\}$ & $\{0\}$   & $ \la  v_4^2 v_5^2\ra  -\la  v_4^2\ra   \la  v_5^2\ra  $ \\
		\Xhline{0.3\arrayrulewidth}			
		$\{2,4\}$ & $\{0\}$   & $ \la  v_4^2 v_5^4\ra  -4 \la  v_5^2\ra   \la  v_4^2 v_5^2\ra  -\la  v_4^2\ra  
		\la  v_5^4\ra  +4 \la  v_4^2\ra   \la  v_5^2\ra ^2 $ \\
		\Xhline{0.3\arrayrulewidth}			
		$\{4,2\}$ & $\{0\}$   & $ \la  v_4^4 v_5^2\ra  -\la  v_4^4\ra   \la  v_5^2\ra  -4 \la  v_4^2\ra   \la 
		v_4^2 v_5^2\ra  +4 \la  v_4^2\ra ^2 \la  v_5^2\ra  $ \\	
		\Xhline{0.3\arrayrulewidth}			
		$\{2,6\}$ & $\{0\}$   & $ \la  v_4^2 v_5^6\ra  -9 \la  v_4^2 v_5^2\ra   \la  v_5^4\ra  -9 \la  v_5^2\ra  
		\la  v_4^2 v_5^4\ra  -\la  v_4^2\ra   \la  v_5^6\ra  +36 \la  v_5^2\ra ^2
		\la  v_4^2 v_5^2\ra  +18 \la  v_4^2\ra   \la  v_5^2\ra   \la  v_5^4\ra  -36
		\la  v_4^2\ra   \la  v_5^2\ra ^3 $ \\
		\Xhline{0.3\arrayrulewidth}				
		$\{4,4\}$ & $\{0\}$   & $ \la  v_4^4 v_5^4\ra  -8 \la  v_4^2 v_5^2\ra ^2-4 \la  v_5^2\ra   \la  v_4^4 v_5^2\ra 
		-\la  v_4^4\ra   \la  v_5^4\ra  -4 \la  v_4^2\ra   \la  v_4^2 v_5^4\ra  +4
		\la  v_4^4\ra   \la  v_5^2\ra ^2+32 \la  v_4^2\ra   \la  v_5^2\ra   \la 
		v_4^2 v_5^2\ra  +4 \la  v_4^2\ra ^2 \la  v_5^4\ra  -24 \la  v_4^2\ra ^2 \la 
		v_5^2\ra ^2 $ \\
		\Xhline{0.3\arrayrulewidth}				
		$\{6,2\}$ & $\{0\}$   & $ \la  v_4^6 v_5^2\ra  -\la  v_4^6\ra   \la  v_5^2\ra  -9 \la  v_4^4\ra   \la 
		v_4^2 v_5^2\ra  -9 \la  v_4^2\ra   \la  v_4^4 v_5^2\ra  +18 \la  v_4^2\ra   \la 
		v_4^4\ra   \la  v_5^2\ra  +36 \la  v_4^2\ra ^2 \la  v_4^2 v_5^2\ra  -36 \la 
		v_4^2\ra ^3 \la  v_5^2\ra  $ \\			
		\hline
	\end{tabular}
\end{table*}%
\FloatBarrier
\begin{table*}[h!]
	\raggedright
	\setlength{\arrayrulewidth}{1.2pt}
	%	\caption{two-harmonics cumulants}
	\scriptsize
	\begin{tabular}[t]{l p{0.7cm} p{15.8cm}}
		\hline
		$\{m_1,m_2\}$ & $\{\alpha\}$  & $c_{4,6}^{\{\alpha\}}\{m_1,m_2\}$\\
		\hline
		$\{2,2\}$ & $\{0\}$   & $ \la  v_4^2 v_6^2\ra  -\la  v_4^2\ra   \la  v_6^2\ra  $ \\
		\Xhline{0.3\arrayrulewidth}			
		$\{3,2\}$ & $\{12\}$   & $ \la  v_4^3 v_6^2 \cos \left(12 \left(\psi _4-\psi _6\right)\right)\ra  $ \\
		\Xhline{0.3\arrayrulewidth}			
		$\{2,4\}$ & $\{0\}$   & $ \la  v_4^2 v_6^4\ra  -4 \la  v_6^2\ra   \la  v_4^2 v_6^2\ra  -\la  v_4^2\ra  
		\la  v_6^4\ra  +4 \la  v_4^2\ra   \la  v_6^2\ra ^2 $ \\	
		\Xhline{0.3\arrayrulewidth}			
		$\{4,2\}$ & $\{0\}$   & $ \la  v_4^4 v_6^2\ra  -\la  v_4^4\ra   \la  v_6^2\ra  -4 \la  v_4^2\ra   \la 
		v_4^2 v_6^2\ra  +4 \la  v_4^2\ra ^2 \la  v_6^2\ra  $ \\
		\Xhline{0.3\arrayrulewidth}				
		$\{3,4\}$ & $\{12\}$   & $ \la  v_4^3 v_6^4 \cos \left(12 \left(\psi _4-\psi _6\right)\right)\ra  -3 \la  v_6^2\ra   \la  v_4^3 v_6^2 \cos
		\left(12 \left(\psi _4-\psi _6\right)\right)\ra  $ \\
		\Xhline{0.3\arrayrulewidth}			
		$\{5,2\}$ & $\{12\}$   & $ \la  v_4^5 v_6^2 \cos \left(12 \left(\psi _4-\psi _6\right)\right)\ra  -4 \la  v_4^2\ra   \la  v_4^3 v_6^2 \cos
		\left(12 \left(\psi _4-\psi _6\right)\right)\ra  $ \\
		\Xhline{0.3\arrayrulewidth}				
		$\{2,6\}$ & $\{0\}$   & $ \la  v_4^2 v_6^6\ra  -9 \la  v_4^2 v_6^2\ra   \la  v_6^4\ra  -9 \la  v_6^2\ra  
		\la  v_4^2 v_6^4\ra  -\la  v_4^2\ra   \la  v_6^6\ra  +36 \la  v_6^2\ra ^2
		\la  v_4^2 v_6^2\ra  +18 \la  v_4^2\ra   \la  v_6^2\ra   \la  v_6^4\ra  -36
		\la  v_4^2\ra   \la  v_6^2\ra ^3 $ \\
\hline
	\end{tabular}
\end{table*}%
\FloatBarrier
\begin{table*}[h!]
	\raggedright
	\setlength{\arrayrulewidth}{1.2pt}
	%	\caption{two-harmonics cumulants}
	\scriptsize
	\begin{tabular}[t]{l p{0.7cm} p{15.8cm}}
		\hline
		$\{m_1,m_2\}$ & $\{\alpha\}$  & $c_{5,6}^{\{\alpha\}}\{m_1,m_2\}$\\
		\hline
		$\{2,2\}$ & $\{0\}$   & $ \la  v_5^2 v_6^2\ra  -\la  v_5^2\ra   \la  v_6^2\ra  $ \\
		\Xhline{0.3\arrayrulewidth}			
		$\{2,4\}$ & $\{0\}$   & $ \la  v_5^2 v_6^4\ra  -4 \la  v_6^2\ra   \la  v_5^2 v_6^2\ra  -\la  v_5^2\ra  
		\la  v_6^4\ra  +4 \la  v_5^2\ra   \la  v_6^2\ra^2 $ \\
		\Xhline{0.3\arrayrulewidth}			
		$\{4,2\}$ & $\{0\}$   & $ \la  v_5^4 v_6^2\ra  -\la  v_5^4\ra   \la  v_6^2\ra  -4 \la  v_5^2\ra   \la 
		v_5^2 v_6^2\ra  +4 \la  v_5^2\ra^2 \la  v_6^2\ra  $ \\	
		\Xhline{0.3\arrayrulewidth}			
		$\{2,6\}$ & $\{0\}$   & $ \la  v_5^2 v_6^6\ra  -9 \la  v_5^2 v_6^2\ra   \la  v_6^4\ra  -9 \la  v_6^2\ra  
		\la  v_5^2 v_6^4\ra  -\la  v_5^2\ra   \la  v_6^6\ra  +36 \la  v_6^2\ra^2
		\la  v_5^2 v_6^2\ra  +18 \la  v_5^2\ra   \la  v_6^2\ra   \la  v_6^4\ra  -36
		\la  v_5^2\ra   \la  v_6^2\ra^3 $ \\
		\Xhline{0.3\arrayrulewidth}				
		$\{4,4\}$ & $\{0\}$   & $ \la  v_5^4 v_6^4\ra  -8 \la  v_5^2 v_6^2\ra^2-4 \la  v_6^2\ra   \la  v_5^4 v_6^2\ra 
		-\la  v_5^4\ra   \la  v_6^4\ra  -4 \la  v_5^2\ra   \la  v_5^2 v_6^4\ra  +4
		\la  v_5^4\ra   \la  v_6^2\ra^2+32 \la  v_5^2\ra   \la  v_6^2\ra   \la 
		v_5^2 v_6^2\ra  +4 \la  v_5^2\ra^2 \la  v_6^4\ra  -24 \la  v_5^2\ra^2 \la 
		v_6^2\ra^2 $ \\
		\Xhline{0.3\arrayrulewidth}			
		$\{6,2\}$ & $\{0\}$   & $ \la  v_5^6 v_6^2\ra  -\la  v_5^6\ra   \la  v_6^2\ra  -9 \la  v_5^4\ra   \la 
		v_5^2 v_6^2\ra  -9 \la  v_5^2\ra   \la  v_5^4 v_6^2\ra  +18 \la  v_5^2\ra   \la 
		v_5^4\ra   \la  v_6^2\ra  +36 \la  v_5^2\ra^2 \la  v_5^2 v_6^2\ra  -36 \la 
		v_5^2\ra^3 \la  v_6^2\ra  $ \\			
		\hline
	\end{tabular}
\end{table*}%
\FloatBarrier
\begin{table*}[h!]
	\raggedright
	\setlength{\arrayrulewidth}{1.2pt}
	\caption{Three-harmonic cumulants up to order six}
	\scriptsize
	\begin{tabular}[t]{l p{1.3cm} p{14.6cm}}
		\hline
		$\{m_1,m_2,m_3\}$ & $\{\alpha_1,\alpha_2\}$  & $c_{2,3,4}^{\{\alpha_1,\alpha_2\}}\{m_1,m_2,m_3\}$\\
		\hline
		$\{1,2,1\}$ & $\{6,-4\}$   & $ \la  v_3^2 v_2 v_4 \cos \left(2 \left(\psi _2-3 \psi _3+2 \psi _4\right)\right)\ra  $ \\
		\Xhline{0.3\arrayrulewidth}			
		$\{1,2,2\}$ & $\{-6,8\}$   & $ \la  v_4^2 v_3^2 v_2 \cos \left(2 \left(\psi _2+3 \psi _3-4 \psi _4\right)\right)\ra  $ \\
		\Xhline{0.3\arrayrulewidth}			
		$\{2,2,1\}$ & $\{0,4\}$   & $ \la  v_3^2 v_2^2 v_4 \cos \left(4 \left(\psi _2-\psi _4\right)\right)\ra  -\la  v_3^2\ra   \la  v_2^2 v_4 \cos
		\left(4 \left(\psi _2-\psi _4\right)\right)\ra  $ \\	
		\Xhline{0.3\arrayrulewidth}			
		$\{1,2,3\}$ & $\{6,-4\}$   & $ \la  v_3^2 v_4^3 v_2 \cos \left(2 \left(\psi _2-3 \psi _3+2 \psi _4\right)\right)\ra  -2 \la  v_3^2 v_2 v_4 \cos \left(2
		\left(\psi _2-3 \psi _3+2 \psi _4\right)\right)\ra   \la  v_4^2\ra  $ \\
		\Xhline{0.3\arrayrulewidth}				
		$\{1,4,1\}$ & $\{6,-4\}$   & $ \la  v_3^4 v_2 v_4 \cos \left(2 \left(\psi _2-3 \psi _3+2 \psi _4\right)\right)\ra  -3 \la  v_3^2\ra   \la  v_3^2
		v_2 v_4 \cos \left(2 \left(\psi _2-3 \psi _3+2 \psi _4\right)\right)\ra  $ \\
		\Xhline{0.3\arrayrulewidth}			
		$\{2,2,2\}$ & $\{0,0\}$   & $ \la  v_2^2 v_3^2 v_4^2\ra  -\la  v_2^2 v_3^2\ra   \la  v_4^2\ra  -\la  v_3^2\ra  
		\la  v_2^2 v_4^2\ra  -\la  v_2^2\ra   \la  v_3^2 v_4^2\ra  +2 \la  v_2^2\ra  
		\la  v_3^2\ra   \la  v_4^2\ra  $ \\
		\Xhline{0.3\arrayrulewidth}				
		$\{3,2,1\}$ & $\{6,-4\}$   & $ \la  v_2^3 v_3^2 v_4 \cos \left(2 \left(\psi _2-3 \psi _3+2 \psi _4\right)\right)\ra  -2 \la  v_2^2\ra   \la 
		v_3^2 v_2 v_4 \cos \left(2 \left(\psi _2-3 \psi _3+2 \psi _4\right)\right)\ra  $ \\
\hline
	\end{tabular}
	\label{three-harmonics}
\end{table*}%
\FloatBarrier
\begin{table*}[h!]
	\raggedright
	\setlength{\arrayrulewidth}{1.2pt}
	%\caption{three-harmonics cumulants}
	\scriptsize
	\begin{tabular}[t]{l p{1.3cm} p{14.6cm}}
		\hline
		$\{m_1,m_2,m_3\}$ & $\{\alpha_1,\alpha_2\}$  & $c_{2,3,5}^{\{\alpha_1,\alpha_2\}}\{m_1,m_2,m_3\}$\\
		\hline
		$\{1,1,1\}$ & $\{-3,5\}$   & $ \la  v_2 v_3 v_5 \cos \left(2 \psi _2+3 \psi _3-5 \psi _5\right)\ra  $ \\
		\Xhline{0.3\arrayrulewidth}			
		$\{1,1,3\}$ & $\{-3,5\}$   & $ \la  v_5^3 v_2 v_3 \cos \left(2 \psi _2+3 \psi _3-5 \psi _5\right)\ra  -2 \la  v_5^2\ra  \la  v_2 v_3 v_5 \cos \left(2 \psi _2+3 \psi _3-5 \psi
		_5\right)\ra    $ \\
		\Xhline{0.3\arrayrulewidth}			
		$\{1,3,1\}$ & $\{-3,5\}$   & $ \la  v_3^3 v_2 v_5 \cos \left(2 \psi _2+3 \psi _3-5 \psi _5\right)\ra  -2 \la  v_3^2\ra   \la  v_2 v_3 v_5 \cos
		\left(2 \psi _2+3 \psi _3-5 \psi _5\right)\ra  $ \\	
		\Xhline{0.3\arrayrulewidth}			
		$\{3,1,1\}$ & $\{-3,5\}$   & $ \la  v_2^3 v_3 v_5 \cos \left(2 \psi _2+3 \psi _3-5 \psi _5\right)\ra  -2 \la  v_2^2\ra   \la  v_2 v_3 v_5 \cos
		\left(2 \psi _2+3 \psi _3-5 \psi _5\right)\ra  $ \\
		\Xhline{0.3\arrayrulewidth}				
		$\{2,2,2\}$ & $\{-6,10\}$   & $ \la  v_5^2 v_3^2 v_2^2 \cos \left(4 \psi _2+6 \psi _3-10 \psi _5\right)\ra  -4 \la  v_2 v_3 v_5 \cos \left(2 \psi _2+3 \psi _3-5
		\psi _5\right)\ra^2 $ \\
		\Xhline{0.3\arrayrulewidth}			
		$\{4,1,1\}$ & $\{3,5\}$   & $ \la  v_2^4 v_3 v_5 \cos \left(8 \psi _2-3 \psi _3-5 \psi _5\right)\ra  $ \\
		\Xhline{0.3\arrayrulewidth}				
		$\{2,3,1\}$ & $\{9,-5\}$   & $ \la  v_2^2 v_3^3 v_5 \cos \left(4 \psi _2-9 \psi _3+5 \psi _5\right)\ra  $ \\			
		\Xhline{0.3\arrayrulewidth}				
		$\{2,2,2\}$ & $\{0,0\}$   & $ \la  v_2^2 v_3^2 v_5^2\ra  -\la  v_2 v_3 v_5 \cos \left(2 \psi _2+3 \psi _3-5 \psi _5\right)\ra^2-\la  v_2^2
		v_3^2\ra   \la  v_5^2\ra  -\la  v_3^2\ra   \la  v_2^2 v_5^2\ra  -\la 
		v_2^2\ra   \la  v_3^2 v_5^2\ra  +2 \la  v_2^2\ra   \la  v_3^2\ra   \la 
		v_5^2\ra  $ \\		
		\hline
	\end{tabular}
\end{table*}%
\FloatBarrier
\begin{table*}[h!]
	\raggedright
	\setlength{\arrayrulewidth}{1.2pt}
	%\caption{three-harmonics cumulants}
	\scriptsize
	\begin{tabular}[t]{l p{1.3cm} p{14.6cm}}
		\hline
		$\{m_1,m_2,m_3\}$ & $\{\alpha_1,\alpha_2\}$  & $c_{2,3,6}^{\{\alpha_1,\alpha_2\}}\{m_1,m_2,m_3\}$\\
		\hline
		$\{2,2,1\}$ & $\{-6,6\}$   & $ \la  v_3^2 v_2^2 v_6 \cos \left(6 \left(\psi _3-\psi _6\right)\right)\ra  -\la  v_2^2\ra   \la  v_3^2 v_6 \cos
		\left(6 \left(\psi _3-\psi _6\right)\right)\ra   $ \\
		\Xhline{0.3\arrayrulewidth}			
		$\{3,2,1\}$ & $\{0,6\}$   & $ \la  v_2^3 v_3^2 v_6 \cos \left(6 \left(\psi _2-\psi _6\right)\right)\ra  -\la  v_3^2\ra   \la  v_2^3 v_6 \cos
		\left(6 \left(\psi _2-\psi _6\right)\right)\ra  $ \\
		\Xhline{0.3\arrayrulewidth}			
		$\{2,2,2\}$ & $\{0,0\}$   & $ \la  v_2^2 v_3^2 v_6^2\ra  -\la  v_2^2 v_3^2\ra   \la  v_6^2\ra  -\la  v_3^2\ra  
		\la  v_2^2 v_6^2\ra  -\la  v_2^2\ra   \la  v_3^2 v_6^2\ra  +2 \la  v_2^2\ra  
		\la  v_3^2\ra   \la  v_6^2\ra  $ \\		
		\hline
	\end{tabular}
\end{table*}%
\FloatBarrier
\begin{table*}[h!]
	\raggedright
	\setlength{\arrayrulewidth}{1.2pt}
	%\caption{three-harmonics cumulants}
	\scriptsize
	\begin{tabular}[t]{l p{1.3cm} p{14.6cm}}
		\hline
		$\{m_1,m_2,m_3\}$ & $\{\alpha_1,\alpha_2\}$  & $c_{2,4,5}^{\{\alpha_1,\alpha_2\}}\{m_1,m_2,m_3\}$\\
		\hline
		$\{1,2,2\}$ & $\{-8,10\}$   & $ \la  v_5^2 v_4^2 v_2 \cos \left(2 \left(\psi _2+4 \psi _4-5 \psi _5\right)\right)\ra  $ \\
		\Xhline{0.3\arrayrulewidth}			
		$\{2,1,2\}$ & $\{4,0\}$   & $ \la  v_5^2 v_2^2 v_4 \cos \left(4 \left(\psi _2-\psi _4\right)\right)\ra  -\la  v_2^2 v_4 \cos \left(4 \left(\psi _2-\psi
		_4\right)\right)\ra   \la  v_5^2\ra  $ \\
		\Xhline{0.3\arrayrulewidth}			
		$\{3,1,2\}$ & $\{-4,10\}$   & $ \la  v_5^2 v_2^3 v_4 \cos \left(6 \psi _2+4 \psi _4-10 \psi _5\right)\ra  $ \\	
		\Xhline{0.3\arrayrulewidth}			
		$\{1,3,2\}$ & $\{12,-10\}$   & $ \la  v_5^2 v_4^3 v_2 \cos \left(2 \left(\psi _2-6 \psi _4+5 \psi _5\right)\right)\ra  $ \\
		\Xhline{0.3\arrayrulewidth}				
		$\{2,2,2\}$ & $\{0,0\}$   & $ \la  v_2^2 v_4^2 v_5^2\ra  -\la  v_2^2 v_4^2\ra   \la  v_5^2\ra  -\la  v_4^2\ra  
		\la  v_2^2 v_5^2\ra  -\la  v_2^2\ra   \la  v_4^2 v_5^2\ra  +2 \la  v_2^2\ra  
		\la  v_4^2\ra   \la  v_5^2\ra  $ \\		
		\hline
	\end{tabular}
\end{table*}%
\FloatBarrier
\begin{table*}[h!]
	\raggedright
	\setlength{\arrayrulewidth}{1.2pt}
	%\caption{three-harmonics cumulants}
	\scriptsize
	\begin{tabular}[t]{l p{1.3cm} p{14.6cm}}
		\hline
		$\{m_1,m_2,m_3\}$ & $\{\alpha_1,\alpha_2\}$  & $c_{2,4,6}^{\{\alpha_1,\alpha_2\}}\{m_1,m_2,m_3\}$\\
		\hline
		$\{1,1,1\}$ & $\{-4,6\}$   & $ \la  v_2 v_4 v_6 \cos \left(2 \left(\psi _2+2 \psi _4-3 \psi _6\right)\right)\ra  $ \\
		\Xhline{0.3\arrayrulewidth}			
		$\{1,2,1\}$ & $\{8,-6\}$   & $ \la  v_4^2 v_2 v_6 \cos \left(2 \left(\psi _2-4 \psi _4+3 \psi _6\right)\right)\ra  $ \\
		\Xhline{0.3\arrayrulewidth}			
		$\{1,1,3\}$ & $\{-4,6\}$   & $ \la  v_6^3 v_2 v_4 \cos \left(2 \left(\psi _2+2 \psi _4-3 \psi _6\right)\right)\ra  -2 \la  v_2 v_4 v_6 \cos \left(2 \left(\psi
		_2+2 \psi _4-3 \psi _6\right)\right)\ra   \la  v_6^2\ra  $ \\	
		\Xhline{0.3\arrayrulewidth}			
		$\{1,3,1\}$ & $\{-4,6\}$   & $ \la  v_4^3 v_2 v_6 \cos \left(2 \left(\psi _2+2 \psi _4-3 \psi _6\right)\right)\ra  -2 \la  v_4^2\ra   \la  v_2
		v_4 v_6 \cos \left(2 \left(\psi _2+2 \psi _4-3 \psi _6\right)\right)\ra  $ \\
		\Xhline{0.3\arrayrulewidth}				
		$\{2,1,2\}$ & $\{4,0\}$   & $ \la  v_6^2 v_2^2 v_4 \cos \left(4 \left(\psi _2-\psi _4\right)\right)\ra  -\la  v_2^2 v_4 \cos \left(4 \left(\psi _2-\psi
		_4\right)\right)\ra   \la  v_6^2\ra  $ \\
		\Xhline{0.3\arrayrulewidth}			
		$\{3,1,1\}$ & $\{-4,6\}$   & $ \la  v_2^3 v_4 v_6 \cos \left(2 \left(\psi _2+2 \psi _4-3 \psi _6\right)\right)\ra  -2 \la  v_2^2\ra   \la  v_2
		v_4 v_6 \cos \left(2 \left(\psi _2+2 \psi _4-3 \psi _6\right)\right)\ra  $ \\
		\Xhline{0.3\arrayrulewidth}				
		$\{2,2,2\}$ & $\{-8,12\}$   & $ \la  v_6^2 v_4^2 v_2^2 \cos \left(4 \left(\psi _2+2 \psi _4-3 \psi _6\right)\right)\ra  -4 \la  v_2 v_4 v_6 \cos \left(2
		\left(\psi _2+2 \psi _4-3 \psi _6\right)\right)\ra^2 $ \\
\hline
	\end{tabular}
\end{table*}%
\FloatBarrier
\begin{table*}[h!]
	\raggedright
	\setlength{\arrayrulewidth}{1.2pt}
	%\caption{three-harmonics cumulants}
	\scriptsize
	\begin{tabular}[t]{l p{1.5cm} p{14.4cm}}
		\hline
		$\{m_1,m_2,m_3\}$ & $\{\alpha_1,\alpha_2\}$  & $c_{2,5,6}^{\{\alpha_1,\alpha_2\}}\{m_1,m_2,m_3\}$\\
		\hline
		$\{1,2,2\}$ & $\{-10,12\}$   & $ \la  v_6^2 v_5^2 v_2 \cos \left(2 \left(\psi _2+5 \psi _5-6 \psi _6\right)\right)\ra  $ \\
		\Xhline{0.3\arrayrulewidth}			
		$\{2,2,1\}$ & $\{10,-6\}$   & $ \la  v_5^2 v_2^2 v_6 \cos \left(4 \psi _2-10 \psi _5+6 \psi _6\right)\ra  $ \\
		\Xhline{0.3\arrayrulewidth}			
		$\{3,2,1\}$ & $\{0,6\}$   & $ \la  v_2^3 v_5^2 v_6 \cos \left(6 \left(\psi _2-\psi _6\right)\right)\ra  -\la  v_5^2\ra   \la  v_2^3 v_6 \cos
		\left(6 \left(\psi _2-\psi _6\right)\right)\ra  $ \\	
		\Xhline{0.3\arrayrulewidth}			
		$\{2,2,2\}$ & $\{0,0\}$   & $ \la  v_2^2 v_5^2 v_6^2\ra  -\la  v_2^2 v_5^2\ra   \la  v_6^2\ra  -\la  v_5^2\ra  
		\la  v_2^2 v_6^2\ra  -\la  v_2^2\ra   \la  v_5^2 v_6^2\ra  +2 \la  v_2^2\ra  
		\la  v_5^2\ra   \la  v_6^2\ra  $ \\
\hline
	\end{tabular}
\end{table*}%
\FloatBarrier
\begin{table*}[h!]
	\raggedright
	\setlength{\arrayrulewidth}{1.2pt}
	%\caption{three-harmonics cumulants}
	\scriptsize
	\begin{tabular}[t]{l p{1.5cm} p{14.4cm}}
		\hline
		$\{m_1,m_2,m_3\}$ & $\{\alpha_1,\alpha_2\}$  & $c_{3,4,5}^{\{\alpha_1,\alpha_2\}}\{m_1,m_2,m_3\}$\\
		\hline
		$\{1,2,1\}$ & $\{8,-5\}$   & $ \la  v_4^2 v_3 v_5 \cos \left(3 \psi _3-8 \psi _4+5 \psi _5\right)\ra  $ \\
		\Xhline{0.3\arrayrulewidth}			
		$\{2,1,2\}$ & $\{-4,10\}$   & $ \la  v_5^2 v_3^2 v_4 \cos \left(6 \psi _3+4 \psi _4-10 \psi _5\right)\ra  $ \\
		\Xhline{0.3\arrayrulewidth}			
		$\{3,1,1\}$ & $\{4,5\}$   & $ \la  v_3^3 v_4 v_5 \cos \left(9 \psi _3-4 \psi _4-5 \psi _5\right)\ra  $ \\	
		\Xhline{0.3\arrayrulewidth}			
		$\{1,2,3\}$ & $\{8,-5\}$   & $ \la  v_4^2 v_5^3 v_3 \cos \left(3 \psi _3-8 \psi _4+5 \psi _5\right)\ra  -2 \la  v_4^2 v_3 v_5 \cos \left(3 \psi _3-8 \psi _4+5
		\psi _5\right)\ra   \la  v_5^2\ra  $ \\
		\Xhline{0.3\arrayrulewidth}				
		$\{1,4,1\}$ & $\{8,-5\}$   & $ \la  v_4^4 v_3 v_5 \cos \left(3 \psi _3-8 \psi _4+5 \psi _5\right)\ra  -3 \la  v_4^2\ra   \la  v_4^2 v_3 v_5 \cos
		\left(3 \psi _3-8 \psi _4+5 \psi _5\right)\ra  $ \\
		\Xhline{0.3\arrayrulewidth}			
		$\{2,2,2\}$ & $\{0,0\}$   & $ \la  v_3^2 v_4^2 v_5^2\ra  -\la  v_3^2 v_4^2\ra   \la  v_5^2\ra  -\la  v_4^2\ra  
		\la  v_3^2 v_5^2\ra  -\la  v_3^2\ra   \la  v_4^2 v_5^2\ra  +2 \la  v_3^2\ra  
		\la  v_4^2\ra   \la  v_5^2\ra  $ \\
		\Xhline{0.3\arrayrulewidth}				
		$\{3,2,1\}$ & $\{8,-5\}$   & $ \la  v_3^3 v_4^2 v_5 \cos \left(3 \psi _3-8 \psi _4+5 \psi _5\right)\ra  -2 \la  v_3^2\ra   \la  v_4^2 v_3 v_5
		\cos \left(3 \psi _3-8 \psi _4+5 \psi _5\right)\ra  $ \\					
		\hline
	\end{tabular}
\end{table*}%
\FloatBarrier
\begin{table*}[h!]
	\raggedright
	\setlength{\arrayrulewidth}{1.2pt}
	%\caption{three-harmonics cumulants}
	\scriptsize
	\begin{tabular}[t]{l p{1.5cm} p{14.4cm}}
		\hline
		$\{m_1,m_2,m_3\}$ & $\{\alpha_1,\alpha_2\}$  & $c_{3,4,6}^{\{\alpha_1,\alpha_2\}}\{m_1,m_2,m_3\}$\\
		\hline
		$\{2,2,1\}$ & $\{0,6\}$   & $ \la  v_4^2 v_3^2 v_6 \cos \left(6 \left(\psi _3-\psi _6\right)\right)\ra  -\la  v_4^2\ra   \la  v_3^2 v_6 \cos
		\left(6 \left(\psi _3-\psi _6\right)\right)\ra  $ \\
		\Xhline{0.3\arrayrulewidth}			
		$\{2,3,1\}$ & $\{12,-6\}$   & $ \la  v_3^2 v_4^3 v_6 \cos \left(6 \left(\psi _3-2 \psi _4+\psi _6\right)\right)\ra  $ \\
		\Xhline{0.3\arrayrulewidth}			
		$\{2,2,2\}$ & $\{0,0\}$   & $ \la  v_3^2 v_4^2 v_6^2\ra  -\la  v_3^2 v_4^2\ra   \la  v_6^2\ra  -\la  v_4^2\ra  
		\la  v_3^2 v_6^2\ra  -\la  v_3^2\ra   \la  v_4^2 v_6^2\ra  +2 \la  v_3^2\ra  
		\la  v_4^2\ra   \la  v_6^2\ra  $ \\	
		\hline
	\end{tabular}
\end{table*}%
\FloatBarrier
\begin{table*}[h!]
	\raggedright
	\setlength{\arrayrulewidth}{1.2pt}
	%\caption{three-harmonics cumulants}
	\scriptsize
	\begin{tabular}[t]{l p{1.5cm} p{14.4cm}}
		\hline
		$\{m_1,m_2,m_3\}$ & $\{\alpha_1,\alpha_2\}$  & $c_{3,5,6}^{\{\alpha_1,\alpha_2\}}\{m_1,m_2,m_3\}$\\
		\hline
		$\{2,2,1\}$ & $\{0,6\}$   & $ \la  v_5^2 v_3^2 v_6 \cos \left(6 \left(\psi _3-\psi _6\right)\right)\ra  -\la  v_5^2\ra   \la  v_3^2 v_6 \cos
		\left(6 \left(\psi _3-\psi _6\right)\right)\ra  $ \\
		\Xhline{0.3\arrayrulewidth}			
		$\{1,3,2\}$ & $\{15,-12\}$   & $ \la  v_6^2 v_5^3 v_3 \cos \left(3 \left(\psi _3-5 \psi _5+4 \psi _6\right)\right)\ra  $ \\
		\Xhline{0.3\arrayrulewidth}			
		$\{2,2,2\}$ & $\{0,0\}$   & $ \la  v_3^2 v_5^2 v_6^2\ra  -\la  v_3^2 v_5^2\ra   \la  v_6^2\ra  -\la  v_5^2\ra  
		\la  v_3^2 v_6^2\ra  -\la  v_3^2\ra   \la  v_5^2 v_6^2\ra  +2 \la  v_3^2\ra  
		\la  v_5^2\ra   \la  v_6^2\ra  $ \\	
		\hline
	\end{tabular}
\end{table*}%
\FloatBarrier
\begin{table*}[h!]
	\raggedright
	\setlength{\arrayrulewidth}{1.2pt}
	%\caption{three-harmonics cumulants}
	\scriptsize
	\begin{tabular}[t]{l p{1.5cm} p{14.4cm}}
		\hline
		$\{m_1,m_2,m_3\}$ & $\{\alpha_1,\alpha_2\}$  & $c_{4,5,6}^{\{\alpha_1,\alpha_2\}}\{m_1,m_2,m_3\}$\\
		\hline
		$\{1,2,1\}$ & $\{10,-6\}$   & $ \la  v_5^2 v_4 v_6 \cos \left(4 \psi _4-10 \psi _5+6 \psi _6\right)\ra   $ \\
		\Xhline{0.3\arrayrulewidth}			
		$\{1,2,3\}$ & $\{10,-6\}$   & $ \la  v_5^2 v_6^3 v_4 \cos \left(4 \psi _4-10 \psi _5+6 \psi _6\right)\ra  -2 \la  v_5^2 v_4 v_6 \cos \left(4 \psi _4-10 \psi _5+6
		\psi _6\right)\ra   \la  v_6^2\ra  $ \\
		\Xhline{0.3\arrayrulewidth}			
		$\{1,4,1\}$ & $\{10,-6\}$   & $ \la  v_5^4 v_4 v_6 \cos \left(4 \psi _4-10 \psi _5+6 \psi _6\right)\ra  -3 \la  v_5^2\ra   \la  v_5^2 v_4 v_6
		\cos \left(4 \psi _4-10 \psi _5+6 \psi _6\right)\ra  $ \\	
		\Xhline{0.3\arrayrulewidth}			
		$\{2,2,2\}$ & $\{0,0\}$   & $ \la  v_4^2 v_5^2 v_6^2\ra  -\la  v_4^2 v_5^2\ra   \la  v_6^2\ra  -\la  v_5^2\ra  
		\la  v_4^2 v_6^2\ra  -\la  v_4^2\ra   \la  v_5^2 v_6^2\ra  +2 \la  v_4^2\ra  
		\la  v_5^2\ra   \la  v_6^2\ra  $ \\
		\Xhline{0.3\arrayrulewidth}			
		$\{3,2,1\}$ & $\{10,-6\}$   & $ \la  v_4^3 v_5^2 v_6 \cos \left(4 \psi _4-10 \psi _5+6 \psi _6\right)\ra  -2 \la  v_4^2\ra   \la  v_5^2 v_4 v_6
		\cos \left(4 \psi _4-10 \psi _5+6 \psi _6\right)\ra  $ \\		
		\Xhline{0.3\arrayrulewidth}			
		$\{2,2,3\}$ & $\{-10,18\}$   & $ \la  v_6^3 v_5^2 v_4^2 \cos \left(2 \left(4 \psi _4+5 \psi _5-9 \psi _6\right)\right)\ra  $ \\
		\Xhline{0.3\arrayrulewidth}			
		$\{3,2,2\}$ & $\{0,12\}$   & $ \la  v_6^2 v_5^2 v_4^3 \cos \left(12 \left(\psi _4-\psi _6\right)\right)\ra  -\la  v_5^2\ra   \la  v_4^3 v_6^2
		\cos \left(12 \left(\psi _4-\psi _6\right)\right)\ra  $ \\
		\Xhline{0.3\arrayrulewidth}	
		$\{4,2,1\}$ & $\{10,6\}$   & $ \la  v_4^4 v_5^2 v_6 \cos \left(2 \left(8 \psi _4-5 \psi _5-3 \psi _6\right)\right)\ra  $ \\		
		\hline
	\end{tabular}
\end{table*}%
\FloatBarrier
\begin{table*}[h!]
	\raggedright
	\setlength{\arrayrulewidth}{1.2pt}
	\caption{Four-harmonic cumulants up to order six}
	\scriptsize
	\begin{tabular}[t]{l p{1.9cm} p{13.1cm}}
		\hline
		$\{m_1,m_2,m_3,m_4\}$ & $\{\alpha_1,\alpha_2,\alpha_3\}$  & $c_{2,3,4,5}^{\{\alpha_1,\alpha_2,\alpha_3\}}\{m_1,m_2,m_3,m_4\}$\\
		\hline
		$\{1,1,1,1\}$ & $\{3,4,-5\}$   & $ \la  v_2 v_3 v_4 v_5 \cos \left(2 \psi _2-3 \psi _3-4 \psi _4+5 \psi _5\right)\ra  $ \\
		\Xhline{0.3\arrayrulewidth}			
		$\{1,1,2,1\}$ & $\{-3,0,5\}$   & $ \la  v_4^2 v_2 v_5 v_3 \cos \left(2 \psi _2+3 \psi _3-5 \psi _5\right)\ra  -\la  v_4^2\ra   \la  v_2 v_3 v_5 \cos
		\left(2 \psi _2+3 \psi _3-5 \psi _5\right)\ra  $ \\
		\Xhline{0.3\arrayrulewidth}			
		$\{2,1,1,1\}$ & $\{3,-4,5\}$   & $ \la  v_2^2 v_3 v_4 v_5 \cos \left(4 \psi _2-3 \psi _3+4 \psi _4-5 \psi _5\right)\ra  $ \\	
		\Xhline{0.3\arrayrulewidth}			
		$\{3,1,1,1\}$ & $\{-3,4,5\}$   & \scriptsize  $ \la  v_2^3 v_3 v_4 v_5 \cos \left(6 \psi _2+3 \psi _3-4 \psi _4-5 \psi _5\right)\ra  -3 \la  v_2^2 v_4 \cos \left(4 \left(\psi
		_2-\psi _4\right)\right)\ra   \la  v_2 v_3 v_5 \cos \left(2 \psi _2+3 \psi _3-5 \psi _5\right)\ra  $ \\
		\Xhline{0.3\arrayrulewidth}			
		$\{1,1,1,3\}$ & $\{3,4,-5\}$   & $ \la  v_5^3 v_2 v_4 v_3 \cos \left(2 \psi _2-3 \psi _3-4 \psi _4+5 \psi _5\right)\ra  -2 \la  v_2 v_3 v_4 v_5 \cos \left(2 \psi
		_2-3 \psi _3-4 \psi _4+5 \psi _5\right)\ra   \la  v_5^2\ra  $ \\		
		\Xhline{0.3\arrayrulewidth}			
		$\{1,1,3,1\}$ & $\{3,4,-5\}$   & $ \la  v_4^3 v_2 v_5 v_3 \cos \left(2 \psi _2-3 \psi _3-4 \psi _4+5 \psi _5\right)\ra  -2 \la  v_4^2\ra   \la  v_2
		v_3 v_4 v_5 \cos \left(2 \psi _2-3 \psi _3-4 \psi _4+5 \psi _5\right)\ra  $ \\
		\Xhline{0.3\arrayrulewidth}			
		$\{1,2,1,2\}$ & $\{6,-4,0\}$   & $ \la  v_5^2 v_3^2 v_2 v_4 \cos \left(2 \left(\psi _2-3 \psi _3+2 \psi _4\right)\right)\ra  -\la  v_3^2 v_2 v_4 \cos \left(2
		\left(\psi _2-3 \psi _3+2 \psi _4\right)\right)\ra   \la  v_5^2\ra  $ \\
\hline
	\end{tabular}
	\label{four-harmonics}
\end{table*}%
\FloatBarrier
\begin{table*}[h!]
	\raggedright
	\setlength{\arrayrulewidth}{1.2pt}
	\scriptsize
	\begin{tabular}[t]{l p{1.9cm} p{13.1cm}}
		\hline
		$\{m_1,m_2,m_3,m_4\}$ & $\{\alpha_1,\alpha_2,\alpha_3\}$  & $c_{2,3,4,6}^{\{\alpha_1,\alpha_2,\alpha_3\}}\{m_1,m_2,m_3,m_4\}$\\
		\hline
		$\{1,2,1,1\}$ & $\{0,-4,6\}$   & $ \la  v_3^2 v_2 v_6 v_4 \cos \left(2 \left(\psi _2+2 \psi _4-3 \psi _6\right)\right)\ra  -\la  v_3^2\ra   \la  v_2
		v_4 v_6 \cos \left(2 \left(\psi _2+2 \psi _4-3 \psi _6\right)\right)\ra  $ \\
		\Xhline{0.3\arrayrulewidth}			
		$\{1,2,1,2\}$ & $\{-6,-4,12\}$   & \scriptsize $ \la  v_6^2 v_3^2 v_2 v_4 \cos \left(2 \left(\psi _2+3 \psi _3+2 \psi _4-6 \psi _6\right)\right)\ra  -2 \la  v_3^2 v_6 \cos \left(6
		\left(\psi _3-\psi _6\right)\right)\ra   \la  v_2 v_4 v_6 \cos \left(2 \left(\psi _2+2 \psi _4-3 \psi _6\right)\right)\ra  $ \\
		\Xhline{0.3\arrayrulewidth}			
		$\{2,2,1,1\}$ & $\{-6,4,6\}$   & $ \la  v_3^2 v_2^2 v_6 v_4 \cos \left(4 \psi _2+6 \psi _3-4 \psi _4-6 \psi _6\right)\ra  -\la  v_2^2 v_4 \cos \left(4 \left(\psi
		_2-\psi _4\right)\right)\ra   \la  v_3^2 v_6 \cos \left(6 \left(\psi _3-\psi _6\right)\right)\ra  $ \\	
		\Xhline{0.3\arrayrulewidth}			
		$\{1,2,2,1\}$ & $\{0,8,-6\}$   & $ \la  v_4^2 v_3^2 v_2 v_6 \cos \left(2 \left(\psi _2-4 \psi _4+3 \psi _6\right)\right)\ra  -\la  v_3^2\ra   \la 
		v_4^2 v_2 v_6 \cos \left(2 \left(\psi _2-4 \psi _4+3 \psi _6\right)\right)\ra  $ \\
		\Xhline{0.3\arrayrulewidth}	
		\rule[0.62cm]{0pt}{0pt}				
		\hspace*{-0.1cm	}$\{1,2,1,2\}$ & $\{6,-4,0\}$   & $ \begin{aligned} \la  v_6^2 v_3^2 v_2 v_4 \cos \left(2 \left(\psi _2-3 \psi _3+2 \psi _4\right)\right)\ra -\la  v_3^2 v_6 \cos &\left(6 \left(\psi
		_3-\psi _6\right)\right)\ra   \la  v_2 v_4 v_6 \cos \left(2 \left(\psi _2+2 \psi _4-3 \psi _6\right)\right)\ra   \\
		& -\la 
		v_3^2 v_2 v_4 \cos \left(2 \left(\psi _2-3 \psi _3+2 \psi _4\right)\right)\ra   \la  v_6^2\ra 
		\end{aligned} $ \\
\hline
	\end{tabular}
\end{table*}%
\FloatBarrier
\begin{table*}[h!]
	\raggedright
	\setlength{\arrayrulewidth}{1.2pt}
	%	\caption{four-harmonics cumulants}
	\scriptsize
	\begin{tabular}[t]{l p{1.9cm} p{13.1cm}}
		\hline
		$\{m_1,m_2,m_3,m_4\}$ & $\{\alpha_1,\alpha_2,\alpha_3\}$  & $c_{2,3,5,6}^{\{\alpha_1,\alpha_2,\alpha_3\}}\{m_1,m_2,m_3,m_4\}$\\
		\hline
		$\{1,1,1,1\}$ & $\{3,5,-6\}$   & $ \la  v_2 v_3 v_5 v_6 \cos \left(2 \psi _2-3 \psi _3-5 \psi _5+6 \psi _6\right)\ra  $ \\
		\Xhline{0.3\arrayrulewidth}			
		$\{1,1,1,2\}$ & $\{-3,5,0\}$   & $ \la  v_6^2 v_2 v_5 v_3 \cos \left(2 \psi _2+3 \psi _3-5 \psi _5\right)\ra  -\la  v_2 v_3 v_5 \cos \left(2 \psi _2+3 \psi _3-5 \psi
		_5\right)\ra   \la  v_6^2\ra  $ \\
		\Xhline{0.3\arrayrulewidth}			
		$\{2,1,1,1\}$ & $\{3,-5,6\}$   & $ \la  v_2^2 v_3 v_5 v_6 \cos \left(4 \psi _2-3 \psi _3+5 \psi _5-6 \psi _6\right)\ra  $ \\	
		\Xhline{0.3\arrayrulewidth}			
		$\{2,1,1,2\}$ & $\{-3,-5,12\}$   & $ \la  v_6^2 v_2^2 v_5 v_3 \cos \left(4 \psi _2+3 \psi _3+5 \psi _5-12 \psi _6\right)\ra  $ \\
		\Xhline{0.3\arrayrulewidth}			
		$\{1,3,1,1\}$ & $\{-9,5,6\}$   & \scriptsize $ \la  v_3^3 v_2 v_6 v_5 \cos \left(2 \psi _2+9 \psi _3-5 \psi _5-6 \psi _6\right)\ra  -3 \la  v_2 v_3 v_5 \cos \left(2 \psi _2+3
		\psi _3-5 \psi _5\right)\ra   \la  v_3^2 v_6 \cos \left(6 \left(\psi _3-\psi _6\right)\right)\ra  $ \\		
		\Xhline{0.3\arrayrulewidth}			
		$\{1,1,1,3\}$ & $\{3,5,-6\}$   & $ \la  v_6^3 v_2 v_5 v_3 \cos \left(2 \psi _2-3 \psi _3-5 \psi _5+6 \psi _6\right)\ra  -2 \la  v_2 v_3 v_5 v_6 \cos \left(2 \psi
		_2-3 \psi _3-5 \psi _5+6 \psi _6\right)\ra   \la  v_6^2\ra  $ \\
		\Xhline{0.3\arrayrulewidth}			
		$\{1,1,3,1\}$ & $\{3,5,-6\}$   & $ \la  v_5^3 v_2 v_6 v_3 \cos \left(2 \psi _2-3 \psi _3-5 \psi _5+6 \psi _6\right)\ra  -2 \la  v_5^2\ra   \la  v_2
		v_3 v_5 v_6 \cos \left(2 \psi _2-3 \psi _3-5 \psi _5+6 \psi _6\right)\ra  $ \\
\hline
	\end{tabular}
\end{table*}%
\FloatBarrier
\begin{table*}[h!]
	\raggedright
	\setlength{\arrayrulewidth}{1.2pt}
	%	\caption{four-harmonics cumulants}
	\scriptsize
	\begin{tabular}[t]{l p{1.9cm} p{13.1cm}}
		\hline
		$\{m_1,m_2,m_3,m_4\}$ & $\{\alpha_1,\alpha_2,\alpha_3\}$  & $c_{2,4,5,6}^{\{\alpha_1,\alpha_2,\alpha_3\}}\{m_1,m_2,m_3,m_4\}$\\
		\hline
		$\{1,1,2,1\}$ & $\{-4,0,6\}$   & $ \la  v_5^2 v_2 v_6 v_4 \cos \left(2 \left(\psi _2+2 \psi _4-3 \psi _6\right)\right)\ra  -\la  v_5^2\ra   \la  v_2
		v_4 v_6 \cos \left(2 \left(\psi _2+2 \psi _4-3 \psi _6\right)\right)\ra  $ \\
		\Xhline{0.3\arrayrulewidth}			
		$\{1,1,2,2\}$ & $\{4,10,-12\}$   & $ \la  v_6^2 v_5^2 v_2 v_4 \cos \left(2 \left(\psi _2-2 \psi _4-5 \psi _5+6 \psi _6\right)\right)\ra  $ \\
		\Xhline{0.3\arrayrulewidth}			
		$\{1,2,2,1\}$ & $\{8,0,-6\}$   & $ \la  v_5^2 v_4^2 v_2 v_6 \cos \left(2 \left(\psi _2-4 \psi _4+3 \psi _6\right)\right)\ra  -\la  v_5^2\ra   \la 
		v_4^2 v_2 v_6 \cos \left(2 \left(\psi _2-4 \psi _4+3 \psi _6\right)\right)\ra  $ \\	
		\Xhline{0.3\arrayrulewidth}			
		$\{2,1,2,1\}$ & $\{-4,10,-6\}$   & $ \la  v_5^2 v_2^2 v_6 v_4 \cos \left(4 \psi _4-10 \psi _5+6 \psi _6\right)\ra  -\la  v_2^2\ra   \la  v_5^2 v_4 v_6
		\cos \left(4 \psi _4-10 \psi _5+6 \psi _6\right)\ra  $ \\			
		\hline
	\end{tabular}
\end{table*}%
\FloatBarrier
\begin{table*}[h!]
	\raggedright
	\setlength{\arrayrulewidth}{1.2pt}
	%	\caption{four-harmonics cumulants}
	\scriptsize
	\begin{tabular}[t]{l p{1.9cm} p{13.1cm}}
		\hline
		$\{m_1,m_2,m_3,m_4\}$ & $\{\alpha_1,\alpha_2,\alpha_3\}$  & $c_{3,4,5,6}^{\{\alpha_1,\alpha_2,\alpha_3\}}\{m_1,m_2,m_3,m_4\}$\\
		\hline
		$\{1,1,1,1\}$ & $\{4,5,-6\}$   & $ \la  v_3 v_4 v_5 v_6 \cos \left(3 \psi _3-4 \psi _4-5 \psi _5+6 \psi _6\right)\ra  $ \\
		\Xhline{0.3\arrayrulewidth}			
		$\{1,1,1,2\}$ & $\{-4,-5,12\}$   & $ \la  v_6^2 v_3 v_5 v_4 \cos \left(3 \psi _3+4 \psi _4+5 \psi _5-12 \psi _6\right)\ra  $ \\
		\Xhline{0.3\arrayrulewidth}			
		$\{1,2,1,1\}$ & $\{-8,5,6\}$   & $ \la  v_4^2 v_3 v_6 v_5 \cos \left(3 \psi _3+8 \psi _4-5 \psi _5-6 \psi _6\right)\ra  $ \\	
		\Xhline{0.3\arrayrulewidth}			
		$\{1,1,1,3\}$ & $\{4,5,-6\}$   & $ \la  v_6^3 v_3 v_5 v_4 \cos \left(3 \psi _3-4 \psi _4-5 \psi _5+6 \psi _6\right)\ra  -2 \la  v_3 v_4 v_5 v_6 \cos \left(3 \psi
		_3-4 \psi _4-5 \psi _5+6 \psi _6\right)\ra   \la  v_6^2\ra  $ \\
		\Xhline{0.3\arrayrulewidth}			
		$\{1,1,3,1\}$ & $\{4,5,-6\}$   & $ \la  v_5^3 v_3 v_6 v_4 \cos \left(3 \psi _3-4 \psi _4-5 \psi _5+6 \psi _6\right)\ra  -2 \la  v_5^2\ra   \la  v_3
		v_4 v_5 v_6 \cos \left(3 \psi _3-4 \psi _4-5 \psi _5+6 \psi _6\right)\ra  $ \\		
		\Xhline{0.3\arrayrulewidth}			
		$\{1,2,1,2\}$ & $\{8,-5,0\}$   & $ \la  v_6^2 v_4^2 v_3 v_5 \cos \left(3 \psi _3-8 \psi _4+5 \psi _5\right)\ra  -\la  v_4^2 v_3 v_5 \cos \left(3 \psi _3-8 \psi _4+5
		\psi _5\right)\ra   \la  v_6^2\ra  $ \\
		\Xhline{0.3\arrayrulewidth}			
		$\{1,3,1,1\}$ & $\{4,5,-6\}$   & $ \la  v_4^3 v_3 v_6 v_5 \cos \left(3 \psi _3-4 \psi _4-5 \psi _5+6 \psi _6\right)\ra  -2 \la  v_4^2\ra   \la  v_3
		v_4 v_5 v_6 \cos \left(3 \psi _3-4 \psi _4-5 \psi _5+6 \psi _6\right)\ra  $ \\
\hline
	\end{tabular}
\end{table*}%
\FloatBarrier

\twocolumn
\section{A comprehensive study for multiparticle correlations at the presence of flow fluctuation}\label{appB}

This appendix reviews the role of multiparticle correlations in Eq.~\eqref{QVecsIIp} in extracting information about flow fluctuations with more mathematical details. 
Although the whole picture is already well understood, few intermediate steps are not mentioned explicitly over the past years. A consequence of this study is a direct algorithm to compute the statistical uncertainty of a generic multiparticle correlation. The final result is equivalent to the computations in section \ref{statisticalSec}. This algorithm is implemented into the function $\verb|Nsigma2P|$ in the {\tt Mathematica} package.

Before studying the statistical properties of flow fluctuation, we start with a simple one-dimensional example which typically happens in many experimental measurements (see Ref.~\cite{kendallBook}). Assume we have an experiment with a measurable variable $x$, and suppose a theory provides us a p.d.f. for the variable $x$, namely  $p(x)$. We aim to test our theoretical prediction by comparing the moments of $p(x)$, 
\bea\label{intMoment}
\la x^n \ra =\int dx \, x^n \,p(x),
\eea
with the measurements. Consider we setup several measurements of the variable $x$ in which the first measurement leads to an outcome $x_1$, the second measurement to $x_2$, and so on. If we repeat this process $N$  times, and assume all measurements are independent, the p.d.f. of the measured sequence  $(x_1,\ldots,x_N)$ is given by
\bea\label{1DF}
P(x_1,\ldots,x_N)=p(x_1)\cdots p(x_N).
\eea
Now, we show that the quantities $\mu_n$,    
\bea\label{muSum}
\mu_n =\frac{1}{N}\sum_{i=1}^{N} x_i^n,
\eea
is an estimator for $\la x^n\ra$. To this end, we compute the mean value of $\mu_n$ using p.d.f. in Eq.~\eqref{1DF},
\begin{equation} \label{eq1p}
\begin{split}
\overline{\mu_n}  &=\int dx_1\cdots dx_N \left(\frac{1}{N}\sum_{i=1}^{N} x_i^n\right)P(x_1,\ldots,x_N).
\end{split}
\end{equation}
By substituting Eq.~\eqref{1DF} into the above relation, one finds $\overline{\mu_n} =\la x^n \ra$. The equality is exact  if we redo the series of measurements $(x_1,\ldots,x_N)$ infinite times and at each time compute the summation in Eq.~\eqref{muSum}, and finally compute the average over the results. Given that measuring a quantity infinite times is not practical, we would like to estimate the actual $\la x^n \ra$ by doing the summation in Eq.~\eqref{muSum} once with a reasonable number of measurements $N$. With the finite number of measurements, we get the following estimation $\la x^n \ra=\mu_n+\sigma_{\mu_n}$, where $\sigma_{\mu_n}$ is the statistical uncertainty, sourced by the finite number of measurements. To calculate the statistical error of $\mu_n$, we compute $ \overline{\mu_n^2 }$ as follows:
\begin{equation} \label{eq2p}
\begin{split}
\overline{\mu_n^2 } &=\int dx_1\cdots dx_N \left(\frac{1}{N}\sum_{i=1}^{N} x_i^n\right)^2P(x_1,\ldots,x_N)\\
&=\frac{1}{N^2}\sum_{i=1}^N\left[\int dx x^{2n} p(x)\right]+\frac{1}{N^2}\sum_{i\neq j} \left[\int dx x^n p(x)\right]^2\\
&=\frac{N}{N^2} \la x^{2n} \ra+\frac{N(N-1)}{N^2} \la x^n \ra^2.\\ 
\end{split}
\end{equation}
Consequently, we obtain the variance of the random variable as
\begin{equation}\label{sigmaMu}
\begin{split}
\sigma_{\mu_n}^2&= \overline{\mu_n^2 }- \overline{\mu_n }^2=\frac{1}{N}\left(\la x^{2n} \ra-\la x^n \ra^2\right)\\
&\simeq\frac{1}{N}\left(\mu_{2n}-\mu_n^2\right),
\end{split}
\end{equation}
where in the last line, we substitute $\la x^n \ra$ with the estimation $\mu_n$.
In summary, by doing $N$ measurements, we estimate the true value of $\la x^n \ra$ with statistical uncertainty $\sigma_{\mu_n}$. There are similarities and differences between the above simple example and that we are looking for in heavy-ion physics, as we will explain in the following. 

The features of the collective models in heavy-ion physics can be categorized into two different parts. First one is the collective evolution part, which leads to a deterministic one-particle distribution function $f(\varphi)$ (Eq.~\eqref{Fourier}) for a given initial state.\footnote{The function $f(\varphi)d\varphi$ returns the probability of finding one particle in the interval $(\varphi,\varphi+d\varphi)$ in the azimuthal direction.} And the second one is the initial state, which is fundamentally stochastic due to the quantum nature of the nucleus wave function. For that reason, we rewrite the Eq.~\eqref{Fourier} as follows:
\bea\label{b1}
f_{\{\hat{v}\}}(\varphi)=\frac{1}{2\pi}\sum_{n=-\infty}^{\infty}\hat{v}_n e^{-in\varphi},
\eea
where $\hat{v}_n= v_ne^{i n\psi_n}$, $\hat{v}_0=1$, $\{\hat{v}\}\equiv \{\hat{v}_1,\hat{v}_2,\ldots\}$, and $\hat{v}_{-n}=\hat{v}_n^*$. The main difference between the above equation and that in Eq.~\eqref{Fourier} is the subscript $\{\hat{v}\}$ in $f_{\{\hat{v}\}}(\varphi)$ to label the one-particle distribution function for different events with different flow harmonics in the final state.

Our physics of interest is encoded in $f_{\{\hat{v}\}}(\varphi)$,  and we would like to measure its Fourier coefficients $\hat{v}_n$ experimentally. Theoretically, the Fourier coefficients can be obtained by
\bea\label{b8}
\la e^{in\varphi} \ra_f=\int f_{\{\hat{v}\}}(\varphi) e^{in\varphi}=\hat{v}_n.
\eea
Experimentally, we have a distribution of $M$  particles ($M$ is the multiplicity) in the azimuthal direction, and we need to estimate the true value of $\hat{v}_n$ from the finite number of particles. This can be done by replacing the  integral in Eq~\eqref{b8} with the following summation,
\bea\label{qn}
q_n=\frac{1}{M}\sum_{i=1}^{M} e^{in\varphi_i},
\eea
where $q_n$ is called normalized flow vector. Focusing on a single event, we can skip subscript $\{\hat{v}\}$ for the moment. Then the probability of finding a specific configuration of particles $(\varphi_1,\ldots,\varphi_M)$ in the final state is given by
\bea\label{Ffi}
F(\varphi_1,\ldots,\varphi_M)=f(\varphi_1)\cdots f(\varphi_M),
\eea
similar to Eq.~\eqref{1DF}.  In an event, the multiplicity $M$ is a parameter related to our model's physical parameters. For instance, in hydrodynamic models, the multiplicity is related to hydrodynamic initial time and the initial energy density deposited into a given region in the transverse direction. As a result, there are cases with low multiplicity and interesting physics which normalized flow vectors are not helpful because of high statistical uncertainty.

We could overcome this problem, by using many event averages. However, as we explained earlier, two different events have two different one-particle distribution function $f_{\{\hat{v}\}}(\varphi)$, and consequently two different values for $\hat{v}_n$. It means we cannot simply increase the statistics by collecting many events. However, we are still able to extract statistically stable information about flow harmonics fluctuation. Referring to Eq.~\eqref{b1}, the function $f_{\{\hat{v}\}}(\varphi)$ is fully characterized by its Fourier coefficients $\hat{v}_n$. As a result, the fluctuating $f_{\{\hat{v}\}}(\varphi)$ can be encoded in an infinite dimensional p.d.f. shown by $p(\hat{v}_{\pm1},\hat{v}_{\pm 2},\ldots)$. Besides, the multiplicity $M$ is different from one event to the other generally. To consider this point, we generalize the flow harmonic p.d.f. to $p(M,\hat{v}_{\pm1},\hat{v}_{\pm 2},\ldots)$ where
\bea
\int_{(M,\hat{v})} p(M,\hat{v}_{\pm1},\hat{v}_{\pm 2},\ldots)=1.
\eea
In the above, we have used the following short notation,
\bea
\int_{(M,\hat{v})}\equiv \sum_{M=0}^{\infty} \,\int\left[\prod_{n=-\infty}^{\infty}d\hat{v}_{n}\right].
\eea
Considering the flow and multiplicity fluctuations, we should modify the joint p.d.f. in Eq.~\eqref{Ffi} into the following form:
\begin{equation} \label{eq1Ap}
\begin{split}
F(\{\varphi\})&=\int_{(M,\hat{v})} p(M,\hat{v}_{\pm1},\hat{v}_{\pm 2},\ldots)\\
&\times f_{\{\hat{v}\}}(\varphi_1)\cdots f_{\{\hat{v}\}}(\varphi_M)f_{\{\hat{v}\}}(0)f_{\{\hat{v}\}}(0)\cdots,
\end{split}
\end{equation}
where $F(\{\varphi\})$ is an infinite dimensional p.d.f. of particle azimuthal angles $(\varphi_1,\varphi_2,\ldots)$. Note that with present interpretation, each event has infinite multiplicity. In practice, however, the events have finite values of multiplicity, which means the p.d.f. $p(M,\hat{v}_{\pm1},\hat{v}_{\pm 2},\ldots)$ is (approach to) zero at large $M$. In case the multiplicity is not fluctuating, we can rewrite $p(M,\hat{v}_{\pm1},\hat{v}_{\pm 2},\ldots)=\delta_{M,M_0}p(M_0,\hat{v}_{\pm1},\hat{v}_{\pm 2},\ldots)$, which consequently leads to
\begin{equation} \label{eq1ApII}
\begin{split}
F(\varphi_1,&\ldots,\varphi_{M_0})=\int \left[\prod_{n=-\infty}^{\infty}d\hat{v}_{n}\right]\\
&\times p(M_0,\hat{v}_{\pm1},\hat{v}_{\pm 2},\ldots) f_{\{\hat{v}\}}(\varphi_1)\cdots f_{\{\hat{v}\}}(\varphi_{M_0}).
\end{split}
\end{equation}
The infinite dimensional function $ p(M_0,\hat{v}_{\pm1},\hat{v}_{\pm 2},\ldots)$ in the Cartesian coordinate reads as $p(v_{1,x},v_{1,y},v_{2,x},v_{2,y},\ldots)$. By keeping $k$ harmonics in the argument, we reach the flow harmonics fluctuation p.d.f.~\eqref{pdf}. If we ignore the flow fluctuations, $p(M_0,\hat{v}_{\pm1},\hat{v}_{\pm 2},\ldots)$ should be replaced by delta functions as well. Then we obtain the simpler case for $F$ mentioned in Eq.~\eqref{Ffi}. As we will see in the following subsection, although the statistically stable values for flow harmonics $\hat{v}_n$ are not accessible in a single event, we can still find accurate values for the moments of the p.d.f. $p(M,\hat{v}_{\pm1},\hat{v}_{\pm 2},\ldots)$ by using many events.

\subsection{Statistical uncertainty of multiparticle correlations }

In the present subsection, we elaborate on the algorithm's details behind the functions $\verb|Nsigma2P|$. The experimental estimator of the moment $\la \hat{v}_{a_1}\cdots \hat{v}_{a_k} \ra$, as we will find out, is related to the normalized flow vectors' product  $q_{a_1} \cdots q_{a_k}$. By referring to Eq.~\eqref{qn}, we start with splitting this product as the following:
\begin{equation} \label{eq2Ap}
\begin{split}
& M^k q_{a_1} \cdots q_{a_k}=\sum_{i_1,\ldots,i_k}^Me^{i(a_1 \varphi_{i_k}+\cdots+a_k \varphi_{i_1})}\\
&=\sum_{i_1}^Me^{i(a_1+\cdots+a_k) \varphi_{i_1}}+\sum_{i_1\neq i_2}^Me^{i a_1 \varphi_{i_1}+i(a_2+\cdots+a_k) \varphi_{i_2}},\\
&+\cdots+\sum_{i_1\neq i_2}^Me^{i(a_1+\cdots+a_{k-1})\varphi_{i_1}+i a_k \varphi_{i_2} \varphi_{i_2}}\\
&+\cdots+\sum_{i_1\neq \cdots\neq i_k}^Me^{i(a_1 \varphi_{i_k}+\cdots+a_k \varphi_{i_1})},
\end{split}
\end{equation}
where in the above, we have separated summations to the terms that all $k$ particles are auto-correlated, $k-1$ particles are auto-correlated, up to the case that all the auto-correlations are removed.

Before proceeding, let us rewrite the above summation in a more compact form by using set partitions. Consider that the set of all partitions of the set $X$ is shown by $\mathcal{P} X$,
\begin{equation} \label{eq2App}
\begin{split}
&\mathcal{P}\{i_1,\ldots,i_k\}=\Big\{ \{ \{i_1,\ldots,i_k  \} ,\\
&\{ \{i_1\},\{i_2,\ldots,i_k\}  \},\ldots, \{ \{i_1 \},\ldots,\{i_k \}  \}   \Big\}.
\end{split}
\end{equation}
In the analogy of $\mathcal{P}$, we define the operator $\mathcal{I}$ and $\A $ acting on the set $\{i_1,\ldots,i_k  \}$ and $\{a_1,\ldots,a_k  \}$ as 
\begin{equation} \label{B5}
\begin{split}
&\mathcal{I}\equiv\mathcal{I}\{i_1,\ldots,i_k\}=\Big\{ \{ i_1=\cdots=i_k  \} ,\\
&\{ i_1\neq i_2=\cdots=i_k  \},\ldots, \{ i_1 \neq\cdots\neq i_k  \}   \Big\},\\
&=\{ \mathcal{I}_1,\ldots,\mathcal{I}_{B_k}  \},
\end{split}
\end{equation}
and
\begin{equation} \label{B6}
\begin{split}
&\A \equiv\A \{a_1,\ldots,a_k\}=\Big\{ \{ a_1+\ldots+a_k  \},\\
&\{ a_1, a_2+\ldots+a_k  \},\ldots, \{ a_1,\ldots, a_k  \}   \Big\},\\
&=\{ \A _1,\ldots,\A _{B_k} \},
\end{split}
\end{equation}
where $B_k$ is the Bell number corresponds to the number of all partitions of a set with $k$ elements. Although these definitions look complicated, they help us to rewrite the Eq.~\eqref{eq2Ap} more compactly,
\begin{equation} \label{eq2Apppp}
\begin{split}
& M^k q_{a_1} \cdots q_{a_k}=\sum_{i =1}^{B_k}\sum_{\mathcal{I}_i}e^{i(a_1 \varphi_{i_1}+\cdots+a_k \varphi_{i_k})}.
\end{split}
\end{equation}

Like what has been done in Eq.~\eqref{eq1p}, we perform the average $\la q_{a_1} \cdots q_{a_k} \ra$  with respect to $F(\{\varphi\})$. To this end, several integrations have to be done. For example
\begin{equation} \label{eq4Ap}
\begin{split}
\int  &d\varphi_1d\varphi_2\cdots \,f_{\{\hat{v}\}}(\varphi_1)\cdots f_{\{\hat{v}\}}(\varphi_M)f_{\{\hat{v}\}}(0)f_{\{\hat{v}\}}(0)\cdots \\
&\times \sum_{i_1\neq i_2}^Me^{i a_1 \varphi_{i_1}+i(a_2+\cdots+a_k) \varphi_{i_2}}= M(M-1)\hat{v}_{a_1}\hat{v}_{a_2+\cdots+a_k}.
\end{split}
\end{equation}
Similarly, and by employing notations in Eqs.~\eqref{B5} and \eqref{B6}, we obtain
\begin{equation} \label{B9}
\begin{split}
\la q_{a_1} &\cdots q_{a_k} \ra=\int d\varphi_1d\varphi_2\cdots \, (q_{a_1} \cdots q_{a_k})\, F(\{\varphi\}),\\
&=\int_{(M,\hat{v})} \int d\varphi_1d\varphi_2\cdots p(M,\hat{v}_{\pm1},\hat{v}_{\pm 2},\ldots)\\
&\times f_{\{\hat{v}\}}(\varphi_1)\cdots f_{\{\hat{v}\}}(\varphi_M)f_{\{\hat{v}\}}(0)f_{\{\hat{v}\}}(0)\cdots\\
&\times \frac{1}{M^k}\sum_{j =1}^{B_k}\sum_{\mathcal{I}_j}e^{i(a_1 \varphi_{i_k}+\cdots+a_k \varphi_{i_1})},\\
&=\left\la\frac{1}{M^k} \sum_{j =1}^{B_k} \sum_{\mathcal{I}_j}  \prod_{z\in\A _j} \hat{v}_z \right\ra.
\end{split}
\end{equation}
In the above, the summation on $\mathcal{I}_j$ is trivial. To find out the final expression, we need to calculate the number of terms in this summation. In fact, by noting that each $i_j$ runs from $1$ to $M$,  it is easy to see that
\bea
\sum_{\mathcal{I}_j} 1=(M)_{||\mathcal{I}_j||},
\eea 
where the $(M)_i=M(M-1)\cdots(M-i+1)$ is the falling factorial, and $||\mathcal{I}_i||$ is the number of elements in the set $\mathcal{I}_j$. As a result, we eventually find
\bea\label{B11p}
\la q_{a_1} \cdots q_{a_k} \ra= \left\la \frac{1}{M^k} \sum_{j =1}^{B_k} (M)_{||\A _j||}  \prod_{z\in\A _j} \hat{v}_z \right\ra, 
\eea
where we have used the fact that $||\mathcal{I}_j||=||\A _j||$.

The last expression is rather simple. To make it more clear, we elaborate on it with two examples in the following. First consider $a_1=n$ and $a_2=-n$. In this case,
\bea
\A =\{ \{ 0 \},\{n,-n \}  \},
\eea
which leads to
\begin{equation}\label{77}
\begin{split}
\la q_n q_{-n} \ra &=\left\la\frac{1}{M^2}\Big[M+M(M-1) (\hat{v}_n  \hat{v}_{-n})  \Big]\right\ra,\\
&=\frac{1}{\la M\ra}+\left\la\frac{M-1}{M}  v_n^2\right\ra. 
\end{split}
\end{equation}
The correlations on the right-hand side are moments of $p(M,\hat{v}_{\pm1},\hat{v}_{\pm 2},\ldots)$. Instead of $\la q_n q_{-n} \ra$, we could have started with $\la q_n q_{-n}(M/M-1) \ra$. In such a case, we would have obtained,
\bea\label{78}
\left\la  \frac{q_n q_{-n} M}{M-1} \right\ra=\frac{1}{\la M-1\ra}+\left\la v_n^2\right\ra.
\eea
By rearranging the above relation and using $q_n=Q_n/M$ (see Eq.~\eqref{cQvec}), we recover Eq.~\eqref{c22}.

For the second example, assume  $a_1=2$, $a_2=3$, and $a_3=-5$. In this special case, we have
\begin{equation} \label{B10}
\begin{split}
\A =\{ \{ 0\},\{ 2,-2 \},\{ 3,-3 \},\{ 5,-5 \},\{ 2,3,-5 \}  \}
\end{split}
\end{equation}
which consequently leads to
\begin{equation} \label{B11}
\begin{split}
\la q_{2} &q_{3} q_{-5} \ra=\\
&\Big\la\frac{1}{M^3}\Big[M+M(M-1)\big(\hat{v}_2\hat{v}_{-2} +\hat{v}_3\hat{v}_{-3} +\hat{v}_5\hat{v}_{-5} \big)\\
&+M(M-1)(M-2) \hat{v}_2 \hat{v}_3 \hat{v}_{-5}  \Big]\Big\ra\\
&=\Big\la\frac{1}{M^2}+\frac{M-1}{M^2}\left( v_2^2 + v_3^2 + v_5^2 \right)\\
&+\frac{(M-1)(M-2)}{M^2}  v_2v_3 v_5 e^{i(2\psi_2+3\psi_3-5\psi_5)} \Big\ra.
\end{split}
\end{equation}
By replacing $\la v_n^2 \ra$ from Eq.~\eqref{78} and starting from $\la q_{2} q_{3} q_{-5} M^2/(M-1)(M-2)\ra$, we find
\begin{equation} 
\begin{split}
\la& v_2 v_3 v_5 e^{i(2\psi_2+3\psi_3-5\psi_5)} \ra=\\
&\Big\la \frac{M^2 q_{2} q_{3} q_{-5}-M(q_{2}q_{-2}+q_{3}q_{-3}+q_{5}q_{-5})+2}{M^2-3M+2}\Big\ra,
\end{split}
\end{equation}
which is coincident with the result of function ${\tt cQvec[\{1,1,1\},\{-3,5\},\{2,3,5\},M,Q]}$ with replacing $Q_n=M q_n$, as we expected. It means our computations so far are compatible with the results of Refs.~\cite{Bilandzic:2013kga,Bilandzic:2010jr}.

Given that the set $\mathcal{I}_{B_k}=\{i_1\neq \cdots\neq n_k\}$ with $||\mathcal{I}_{B_k}||=k$ is the unique most populated set in the set of partitions $\mathcal{I}$, only the term $j=B_k$ survives the limit $M\to\infty $ in Eq.~\eqref{B11p}. As a result, we obtain 
\begin{equation} \label{eq5Ap}
\begin{split}
\lim_{M\to\infty}\,\la q_{a_1} \cdots q_{a_k} \ra=\la \hat{v}_{a_1}\cdots \hat{v}_{a_k} \ra,
\end{split}
\end{equation}
which is manifestly correct because, in the limit $M\to\infty$, there is no statistical fluctuation in a single event. This indicates that the true value of $\la q_{a_1} \cdots q_{a_k} \ra$ with no statistical error is encoded in the sector $\mathcal{I}_{B_k}$. Referring to Eq.~\eqref{B9}, and keep only the term $j=B_k$ from both sides, we obtain,
\begin{equation} \label{eq5App}
\begin{split}
\la \hat{v}_{a_1}\cdots \hat{v}_{a_k} \ra=\left\la \frac{1}{(M)_k}\sum_{i_1\neq \cdots\neq i_k}e^{i(a_1 \varphi_{i_k}+\cdots+a_k \varphi_{i_1})}\right\ra.
\end{split}
\end{equation}
The meaning of Eq.~\eqref{eq5App} is that by removing the auto-correlations and averaging over many events (as it is done in Refs.~\cite{Borghini:2000sa,Borghini:2001vi,Bilandzic:2010jr,Bilandzic:2013kga}) the effect of statistical fluctuations in a single event is removed and we obtain accurate moments of $p(M,\hat{v}_{\pm1},\hat{v}_{\pm 2},\ldots)$ distribution.

The above statement is valid only when the average $\la \cdots \ra$ is performed over infinitely many numbers of events. Similar to computations lead to Eq.~\eqref{sigmaMu}, we compute the statistical fluctuation of the right-hand side of Eq.~\eqref{eq5App} with the finite number of events, $N$. Defining
\bea\label{84}
\la k \ra_{a_1,\ldots,a_k}=\frac{1}{(M)_k}\sum_{i_1\neq \cdots\neq i_k}e^{i(a_1 \varphi_{i_1}+\cdots+a_k \varphi_{i_k})},
\eea 
and $\mathcal{D}\varphi = d\varphi_{1}d\varphi_2\cdots$, one can compute the following integrals,
\begin{equation} \label{eq7Ap}
\begin{split}
\overline{\la k \ra}& \equiv \int \D\varphi_{(1)}F_{(1)}(\{\varphi\})\cdots d\varphi_{(N)}F_{(N)}(\{\varphi\})\\
&\hspace*{2cm}\times \left[\frac{1}{N}\sum_{i=1}^{N} \la k \ra_{(i)}\right],\\
& =\la \la k \ra \ra,\\
\overline{\la k \ra\la \ell \ra}&\equiv \int \D\varphi_{(1)}F_{(1)}(\{\varphi\})\cdots d\varphi_{(N)}F_{(N)}(\{\varphi\}) \\
&\times\left[\frac{1}{N^2}\sum_{i,j=1}^{N} \la k\ra_{(i)} \la \ell\ra_{(j)}\right],\\
&=\frac{1}{N}\left[\la\la k \ra\la \ell \ra\ra+(N-1)\la \la k \ra\ra\la\la \ell \ra\ra\right].
\end{split}
\end{equation}
The quantity $\la \la k \ra\ra$ has been computed in Eq.~\eqref{eq5App} while we need to elaborate $\la \la k \ra\la \ell \ra\ra$.\footnote{We ignored the subscripts $a_i$ in the notation $\la k \ra_{a_1,\ldots,a_k}$ for simplicity. } To this end, we need to study a summation as
\bea\label{B20}
\sum_{i_1\neq \cdots\neq i_k} \sum_{i_{k+1}\neq \cdots\neq i_{k+\ell}}\cdots.
\eea
Considering the set $\{i_1,i_2,\ldots,i_{k+\ell}  \}$, we immediately find that in reorganizing the summation in Eq.~\eqref{B20} no partition with more than two elements is allowed. Assume a set as $\{i_1,x\}$ where the element $x$ can be any $i_j$ with $j=k+1,\ldots,\ell$. Add any third element to $\{i_1,x\}$ contradicts with one of the inequalities $i_1\neq \cdots\neq i_k$ or $i_{k+1}\neq \cdots\neq i_{\ell}$   in Eq.~\eqref{B20}. As a result, the partitions are started from sets as
\bea\label{87}
\{ \{i_1,i_{s_1} \},\ldots,\{i_k,i_{s_k}\},\{i_{s_{k+1}}\},\ldots,\{i_{s_{\ell}}\}   \}.
\eea
Here we conventionally assumed $\ell\geq k$. The $\ell$-tuple $(s_1,\ldots,s_\ell)$ is a permutation of $(k+1,\ldots,k+\ell)$ which are $\ell!$ different kinds. However, $(\ell-k)$ elements of the $\ell$-tuple are placed in a set with a single element. Therefore, the permutation of that subset does not lead to a different partition in Eq.~\eqref{87}. As a result, the number of partitions in Eq.~\eqref{87} is $\ell!/(k-\ell)!$. We are allowed to break each set $\{i_r,i_{s_r}\}$ into two, and the result will be a legal partition. After breaking all sets as $\{i_r,i_{s_r}\}$ we reach to
\bea
\{ \{i_1\},\ldots,\{ i_{k+\ell}\} \}.
\eea

Similar to $i_j$ partitions, the partitions of harmonics $a_i$ quantities start from
\bea\label{89}
\{ a_1+a_{s'_1},\ldots,a_k+a_{s'_k},a_{k+1},\ldots,a_\ell \},
\eea
where $(s'_1,\ldots,s'_\ell)=\text{perm}(1,\ldots,\ell)$ and can be continued by splitting each summation $a_i+a_{s'_i}$ into two. Defining the set
\begin{equation} \label{B5p}
\begin{split}
\A '\equiv\{ \A '_1,\ldots,\A '_{f_{k,\ell}} \},
\end{split}
\end{equation}
as set of all $f_{k,\ell}$ partitions of  set \eqref{89}, we obtain a relation similar to Eq.~\eqref{B11p}, 
\begin{equation} \label{eq9Ap}
\begin{split}
&\la \la k \ra\la \ell \ra \ra=\\
&\left\la \frac{1}{(M)_k(M)_\ell}\sum_{j =1}^{f_{k,\ell}} (M)_{||\A '_j||}   \prod_{z\in\A '_j} \hat{v}_z \right\ra.
\end{split}
\end{equation}
Now, we can define the covariance matrix as follows
\begin{equation} \label{eq8ApI}
\begin{split}
\text{cov}(\la k \ra,\la \ell \ra)&=\overline{\la k \ra\la \ell \ra}-\overline{\la k \ra}\,\overline{\la \ell \ra}\\
&=\frac{1}{N}\left[\la\la k \ra\la \ell \ra\ra-\la\la k \ra\ra\la\la \ell \ra\ra\right].
\end{split}
\end{equation}
The statistical fluctuation of $\la \la k \ra \ra$ is given by $\text{cov}(\la k \ra,\la k \ra)$. 

For clarity, let us elaborate on a simple example with $a_1=n$ and $a_2=-n$. The partition set is given by
\begin{equation} \label{eq8App}
\begin{split}
\A '&=\{  \{2n,-2n \},\{ 0,0\},\{ n,n,-2n \},\{2n,-n,-n\} \\
&, \{n,-n,0 \},\{ n,n,-n,-n \}  \}.
\end{split}
\end{equation}
As a result, Eqs.~\eqref{eq9Ap} and \eqref{eq8ApI} lead to the following relation for the statistical uncertainty of $\la\la 2 \ra_{-n,n}\ra$,
\begin{equation} \label{eq8Appp}
\begin{split}
\sigma_{\la\la 2 \ra_{-n,n}\ra}^2&=\frac{1}{N}\Bigg[\Bigg\la\frac{ \hat{v}_{-2n}\hat{v}_{2n} +1}{M(M-1)}+\\
&\frac{(M-2)}{M(M-1)}\bigg[\hat{v}_{-2n}\hat{v}^2_n+\hat{v}_{-n}^2\hat{v}_{2n}+\hat{v}_{-n}\hat{v}_{n}\bigg]\\
&+\frac{(M-2)(M-3)}{M(M-1)}\hat{v}_{-n}^2\hat{v}_{n}^2\Bigg\ra-\la \hat{v}_{-n}\hat{v}_n\ra^2\Bigg].
\end{split}
\end{equation}
Referring to Eqs.~\eqref{eq5App} and \eqref{84}, one notes that $\hat{v}_{a_1}\cdots \hat{v}_{a_k}=\la k \ra_{a_1,\ldots,a_k}$. Substituting this relation into the above equation, we obtain Eq.~\eqref{46}, compatible with \cite{Bilandzic:2013kga,Bilandzic:2014qga} if we ignore the flow fluctuation and set $\psi_n=\psi_{2n}$. 

\section{Flow fluctuation and particle distribution decomposition}\label{appBp}

This appendix introduces a generic procedure to connect $q$th order cumulants to the $q$PC. We elaborate 3PC for $n=2,\ldots,5$ explicitly at the end of the appendix.

\subsection{Nonflow correlations vs. flow-induced correlations}

One of the main motivations of using cumulants in studying the flow harmonic fluctuations is removing the nonflow effects. Moreover, cumulants systematically classify the deviation of the flow harmonic p.d.f.~\eqref{pdf} from Gaussianity. In this part, we would like to show how much the observed features of these two concepts are entangled.

Ignoring the flow fluctuation, Eq.~\eqref{Ffi} should be replaced by \cite{Borghini:2000sa} 
\begin{equation} \label{FfiNonFlow}
\begin{split}
F(\varphi_1,\ldots,\varphi_M)=f(\varphi_1)&\cdots f(\varphi_M)\\
&+F_\text{nonflow}(\varphi_1,\ldots,\varphi_M),
\end{split}
\end{equation}
where $F_\text{nonflow}(\varphi_1,\ldots,\varphi_M)$ is correlation developed by nonflow effects in a single event. One notes that $F_\text{nonflow}(\varphi_1,\ldots,\varphi_M)$ still contains
products of two-particles or more joint distributions. For instance
\begin{equation}
\begin{split}
F_{\text{nonflow}}&(\varphi_1,\ldots,\varphi_M)=\\
&F_{\text{nonflow}}^{(2)}(\varphi_1,\varphi_2)f(\varphi_3)\cdots f(\varphi_M)\\
&\;\,\vdots\\
&+F_{\text{nonflow}}^{(M)}(\varphi_1,\ldots,\varphi_M).
\end{split}
\end{equation} 
A standard way to suppress the nonflow effects is using cumulants. By measuring $m$th order cumulant we practically measure a genuine correlation among $m$th particles. Therefore by measuring the $m$th order cumulant of the distribution, we automatically remove any $n$-particle nonflow correlations with $n<m$ \cite{Borghini:2000sa,Borghini:2001vi}. 

In the presence of flow fluctuation, the same structure as Eq.~\eqref{FfiNonFlow} appears even if we ignore nonflow effects. In such a case, Eq.~\eqref{eq1ApII} is written as the following (for the detail see appendix \ref{appBp}) 
\begin{equation} \label{FfiFlowFlu}
\begin{split}
F(\varphi_1,\ldots,\varphi_M)=\overline{f}(\varphi_1)&\cdots \overline{f}(\varphi_M)\\
&+F_\text{flow-fluc}(\varphi_1,\ldots,\varphi_M),
\end{split}
\end{equation}
where $\overline{f}(\varphi)$ is the many events average of the single-particle distribution Eq.~\eqref{Fourier} and $F_\text{flow-fluc}(\varphi_1,\ldots,\varphi_M)$ is the joint correlation part appears because of averaging over many fluctuating events. As a result, there are two entangled parts in genuine multiparticle correlations: nonflow effects in a single event develop the first part, and the second part fictitiously appears because of averaging over many fluctuating events. To study lower-order cumulants, which contain nontrivial and important information about flow fluctuations, we need to use other methods to reduce the nonflow effects \cite{Jia:2017hbm,Adler:2002pu}.

\subsection{Genuine $q$-particle correlation functions}\label{AppC2}

 By inserting Eq.~\eqref{b1} into Eq.~\eqref{eq1ApII}, we find the following expression,\footnote{Here, we ignore the multiplicity fluctuation.} 
\begin{equation} \label{c1}
\begin{split}
F(\varphi_1,\ldots,\varphi_M)=\frac{1}{(2\pi)^M}&\sum_{a_1,\ldots,a_M} \la \hat{v}_{a_1} \cdots \hat{v}_{a_M} \ra \\
&\times e^{-ia_1 \varphi_1-\cdots-ia_M \varphi_M}.
%&\int \left[\prod_{n=1}^{\infty}d\hat{v}_{n}\right]p(\hat{v}_{1},\hat{v}_{2},\ldots) f_{\{\hat{v}\}}(\varphi_1)\cdots f_{\{\hat{v}\}}(\varphi_M),
\end{split}
\end{equation}
The above relation cannot be decomposed into the product of $M$ distinct single-particle distribution. However, we are still able to rewrite the moment $ \la \hat{v}_{a_1} \cdots \hat{v}_{a_M} \ra$ as a summation of all possible genuine correlations between $\hat{v}_{a_i}$s  (cumulants $\la \cdots \ra_c$),
\begin{equation} \label{c2}
\begin{split}
\la \hat{v}_{a_1} \cdots \hat{v}_{a_M} \ra&=\la \hat{v}_{a_1}\ra_c\cdots \la \hat{v}_{a_M}\ra_c\\
&+\la \hat{v}_{a_1} \hat{v}_{a_2}\ra_c \la \hat{v}_{a_3}\ra_c \cdots \la \hat{v}_{a_M} \ra+\cdots\\
&\;\;\vdots\\
&+\la \hat{v}_{a_1} \hat{v}_{a_2}\ra_c \cdots \la \hat{v}_{a_{M-1}}\hat{v}_{a_M}\ra_c +\cdots \\
&\;\;\vdots\\
&+\la \hat{v}_{a_1} \cdots \hat{v}_{a_M} \ra_c.
\end{split}
\end{equation}
Using the above equation, we can rewrite Eq.~\eqref{c1} as
\begin{equation} \label{c3}
\begin{split}
F(\varphi_1,&\ldots,\varphi_M)= f_{\{\la \hat{v} \ra_c\}}(\varphi_1)\cdots f_{\{\la \hat{v} \ra_c\}}(\varphi_M)\\
&+F_{\text{flow-fluc}}(\varphi_1,\ldots,\varphi_M)
\end{split}
\end{equation}
where
\begin{equation} \label{c4}
\begin{split}
F_{\text{flow-fluc}}&(\varphi_1,\ldots,\varphi_M)=\\
&f_{\la \hat{v}_1 \hat{v}_2 \ra_c}(\varphi_1,\varphi_2)f_{\{\la \hat{v} \ra_c\}}(\varphi_3)\cdots f_{\{\la \hat{v} \ra_c\}}(\varphi_M)\\
&\;\,\vdots\\
&+f_{\la \hat{v}_1 \cdots\hat{v}_M \ra_c}(\varphi_1,\ldots,\varphi_M),
\end{split}
\end{equation}
and
\begin{equation} \label{c5}
\begin{split}
f_{\la \hat{v}_1\cdots \hat{v}_q \ra_c}&(\varphi_1,\ldots,\varphi_q)=\\
&\frac{1}{(2\pi)^q}\sum_{a_1,\ldots,a_q}\la \hat{v}_{a_1} \cdots \hat{v}_{a_q} \ra_c e^{-i a_1\varphi_1-\cdots-i a_q\varphi_q}.
\end{split}
\end{equation}
We see from above that the apparent correlation induced by fluctuation depends on the event by event fluctuation.

Considering the rotational symmetry, the constraint $a_1+\cdots+a_k=0$ should be included into the above summation. By this constraint, the cumulant $\la \hat{v}_{a_1} \cdots \hat{v}_{a_q} \ra_c$  is equivalent to that we have already studied in section \ref{generatingFunction}. Defining $\Delta\varphi_k=\varphi_{k+1}-\varphi_1$
and replacing the notation $f_{\la \hat{v}_1\cdots \hat{v}_q \ra_c}$ with $C_q(\Delta\varphi_1,\ldots,\Delta\varphi_{q-1})$, the $q$PC reads as
\begin{equation} \label{c5p}
\begin{split}
&C_q(\Delta\varphi_1,\ldots,\Delta\varphi_{q-1})=\\
&\frac{1}{(2\pi)^q}\Bigg[1+2\sum_{k=1}^{\infty}\;\sum_{\substack{n_1,\ldots,n_k\\\alpha_1,\ldots,\alpha_{k-1}}}\,\sum_{m_1+\cdots+m_k=q}\\
&c_{n_1,\ldots,n_k}^{\{\alpha_1,\ldots\alpha_{k-1}\}}\{m_1,\ldots,m_k\} \Phi_{n_1,\ldots,n_k}^{\{\alpha_1,\ldots\alpha_{k-1}\}}(\Delta\varphi_1,\ldots,\Delta\varphi_{k-1})\Bigg].
\end{split}
\end{equation}
The function $\Phi_{n_1,\ldots,n_k}^{\{\alpha_1,\ldots\alpha_{k-1}\}}(\Delta\varphi_1,\ldots,\Delta\varphi_{k-1})$ can be obtained by comparing Eqs.~\eqref{c5} and \eqref{c5p} straightforwardly. For instance
\begin{equation} \label{c5pp}
\begin{split}
&\Phi_n(\Delta\varphi)=\cos(n\Delta\varphi),\\
\\
&\Phi_{2,4}^{\{4\}}(\Delta\varphi_1,\Delta\varphi_2)=\cos(2\Delta\varphi_1-4\Delta\varphi_2)\\
&+\cos(-4\Delta\varphi_1+2\Delta\varphi_2)+\cos(2\Delta\varphi_1+2\Delta\varphi_2),\\
\\
&\Phi_{2,3,5}^{\{-3,5\}}(\Delta\varphi_1,\Delta\varphi_2)=\cos(3\Delta\varphi_1-5\Delta\varphi_2)\\
&+\cos(-5\Delta\varphi_1+3\Delta\varphi_2)+\cos(2\Delta\varphi_1-5\Delta\varphi_2)\\
&+\cos(-5\Delta\varphi_1+2\Delta\varphi_2)+\cos(2\Delta\varphi_1+3\Delta\varphi_2)\\
&+\cos(3\Delta\varphi_1+2\Delta\varphi_2),
\end{split}
\end{equation}
and so on. The above expression for two-particle distribution leads to the following well-known result,
\bea\label{c6}
C_2(\Delta \varphi)=\frac{1}{(2\pi)^2}\left[1+2\sum_{n=1}^{\infty} c_n\{2\}\cos(n\Delta\varphi)\right].
\eea
Given that for harmonics $n=2,\ldots,5$, there are only two  third-order cumulants $c_{2,4}^{\{4\}}\{2,1\}$ and $c_{2,3,5}^{\{-3,5\}}\{1,1,1\}$, 3PC is written as
\begin{equation} \label{c9}
\begin{split}
&C_3(\Delta \varphi_1,\Delta \varphi_2)=\\
&\frac{1}{(2\pi)^3}\bigg[1+2c_{2,4}^{\{4\}}\{2,1\}\,\Phi_{2,4}^{\{4\}}(\Delta\varphi_1,\Delta\varphi_2)\\
&\hspace*{1cm}+2c_{2,3,5}^{\{-3,5\}}\{1,1,1\}\,\Phi_{2,3,5}^{\{-3,5\}}(\Delta\varphi_1,\Delta\varphi_2)\bigg].
\end{split}
\end{equation}
The other correlation functions can be obtained accordingly.

%\vspace*{0.4cm}
\section{Rotationally symmetric generating function}\label{appC}

We elaborate on the derivation of Eq.~\eqref{J2} here. We first rewrite the distribution $p(v_{n_1,x},v_{n_1,y},\ldots,v_{n_k,x},,v_{n_k,y})$ as follows:
\begin{equation} \label{collectiveNote}
\begin{split}
p(\vec{X}_{n_1,\ldots,n_k},\vec{Y}_{n_1,\ldots,n_k}),
\end{split}
\end{equation}
where we have used the following notation:
\begin{equation} \label{notation}
\begin{split}
\vec{X}&=(v_{n_1,x},\ldots,v_{n_k,x}), \\
\vec{Y}&=(v_{n_1,y},\ldots,v_{n_k,y}).
\end{split}
\end{equation}
The generating function of the above distribution is written as $G(\vec{P},\vec{Q})=\la e^{i\,L} \ra$, where
\bea\label{D4}
L=i\vec{X}\cdot\vec{P}+i\vec{Y}\cdot\vec{Q}.
\eea
Similar to Eq.~\eqref{notation}, we have used the notation
\begin{equation} \label{notation2}
\begin{split}
\vec{P}&=(k_{n_1,x},\ldots,k_{n_k,x}), \\
\vec{Q}&=(k_{n_1,y},\ldots,k_{n_k,y}).
\end{split}
\end{equation}
Presenting Eq.~\eqref{D4} in polar coordinate and  in an arbitrary reaction plane angle $\Phi_{\text{RP}}$, we find
\begin{equation} \label{GinPolar1}
\begin{split}
L=i\sum_{i=1}^{k} v_{n_i}\, k_{n_i}\cos\left[n_i\left(\psi_{n_i}-\phi_{n_i}+\Phi_{\text{RP}}\right)\right],
\end{split}
\end{equation}
where we have used polar coordinate $v_n e^{in\psi_n}=v_{n,x}+iv_{n,y}$ together with
\bea\label{Kpolar}
k_{n,x}=k_n\cos n\phi_n,\quad k_{n,y}=k_n\sin n\phi_n.
\eea
After changing variables $\delta\psi_{n_i}=\psi_{n_{i+1}}-\psi_{n_1}$ and $\delta\phi_{n_i}=\phi_{n_{i+1}}-\phi_{n_1}$, we obtain
\begin{equation}\label{D7}
\begin{split}
&L=i\,v_{n_1} k_{n_1}\cos n_1(\psi_{n_1}-\phi_{n_1}+\Phi_{\text{RP}})\\
&+i\sum_{i=2}^k v_{n_i} k_{n_i}\\
&\times\cos \left[n_i\left(\psi_{n_1}+\delta\psi_{n_{i-1}}-\delta\phi_{n_{i-1}}-\phi_{n_1}+\Phi_{\text{RP}}\right)\right].
\end{split}
\end{equation}
We eventually fix our reference frame such that $\phi_{n_1}=\Phi_{\text{RP}}$ to obtain Eq.~\eqref{J2}.

\end{document}